\documentclass[11pt]{article}

\usepackage[letterpaper,margin=1in]{geometry}

\usepackage{hyperref}

\usepackage{graphicx} 
\usepackage{mathtools}  
\usepackage{bbm}
\usepackage{bbold}
\usepackage{amsthm}
\usepackage{xcolor}
\usepackage{acronym}
\usepackage{algorithm}
\usepackage{algpseudocode}
\usepackage{diagbox}

\usepackage{hyperref}
\usepackage{cleveref}

\newacro{iid}[i.i.d.]{Independent and Identically Distributed}
\newacro{bdp}[BDP]{Backward Dynamic Programming}

\newcommand{\bb}[1]{\mathbb{#1}}

\newcommand{\eqdef}{\vcentcolon =}
\newcommand{\alg}{\operatorname{ALG}}
\newcommand{\opt}{\operatorname{OPT}}
\newcommand{\comp}{\operatorname{CR}}
\newcommand{\val}[2]{\operatorname{V}_{#1}^{#2}}
\newcommand{\gap}[2]{\operatorname{\Delta}_{#1}^{#2}}

\newcommand{\bx}{\mathbf{X}}
\newcommand{\be}{\mathbb{E}}

\newcounter{dummy} \numberwithin{dummy}{section}

\newtheorem{theorem}[dummy]{Theorem}

\newtheorem{proposition}[dummy]{Proposition}

\newtheorem{lemma}[dummy]{Lemma}

\newtheorem{corollary}[dummy]{Corollary}

\theoremstyle{remark}
\newtheorem*{remark}{Remark}

\theoremstyle{definition}
\newtheorem{definition}[dummy]{Definition}

\numberwithin{equation}{section}

\title{Prophet Inequalities: Competing with the Top $\ell$ Items is Easy}
\author{
  Mathieu Molina\thanks{Inria, FairPlay joint team and 
  CREST, ENSAE, mathieu.molina@inria.fr}
  \and
  Nicolas Gast\thanks{Univ. Grenoble Alpes, Inria, CNRS, 
  Grenoble INP and LIG, nicolas.gast@inria.fr}
  \and
  Patrick Loiseau\thanks{Inria, FairPlay joint team, patrick.loiseau@inria.fr}
  \and
  Vianney Perchet\thanks{CREST, ENSAE, and Criteo AI Lab,
  FairPlay joint team, vianney.perchet@normalesup.org}
}

\date{}

\begin{document}

\allowdisplaybreaks

\maketitle

\begin{abstract}  We explore a prophet inequality problem, where the values of a sequence of items are drawn i.i.d. from some distribution, and an online decision maker must select one item irrevocably. We establish that $\mathrm{CR}_{\ell}$ the worst-case competitive ratio between the expected optimal performance of an online decision maker compared to that of a prophet who uses the average of the top $\ell$ items is exactly the solution to an integral equation. This quantity $\mathrm{CR}_{\ell}$ is larger than $1-e^{-\ell}$. This implies that the bound converges exponentially fast to $1$ as $\ell$ grows. In particular for $\ell=2$, $\mathrm{CR}_{2} \approx 0.966$ which is much closer to $1$ than the classical bound of $0.745$ for $\ell=1$. Additionally, we prove asymptotic lower bounds for the competitive ratio of a more general scenario, where the decision maker is permitted to select $k$ items. This subsumes the $k$ multi-unit i.i.d. prophet problem and provides the current best asymptotic guarantees, as well as enables broader understanding in the more general framework. Finally, we prove a tight asymptotic competitive ratio when only static threshold policies are allowed.\end{abstract}

\section{Introduction}

Decision makers are frequently confronted with the arduous task of making crucial decisions with limited information. For example, when a seller wants to sell a limited number of items to a stream of customers, potential future customers with a high willingness to pay must be taken into account. How long should the seller wait before finally lowering its expectations, and to what price? Is the current customer likely to be the best we can hope to interact with? This common challenge is at the heart of many online selection problems \cite{Borodin1998}. Two of the simplest and most famous versions of this online selection problem are specific instances of optimal stopping problems \cite{Chow71}: the secretary problem where inputs are adversarial \cite{Ferguson1989}, and prophet inequalities where inputs are random \cite{Correa2019}. In this work, we focus on the prophet inequality problem.

A classical way of measuring the performance of an online decision problem is to consider the so-called competitive ratio, the worst-case ratio between the performance of an online algorithm and the performance of an oracle that usually has access to more information than the decision maker. This has been the focus of many works, in online matching \cite{Mehta2013}, scheduling \cite{Motwani1994}, or metrical task systems \cite{Bubeck2018} to name but a few, where explicit upper and lower bounds on the competitive ratio were provided. This metric makes it possible to design robust algorithms, that are able to always perform approximately well in any circumstances.

The classical prophet inequality dates back to the $1970s$, with \cite{Krengel1977} famously showing that a gambler allowed to select a single item using an optimal online algorithm can always recover at least half of the item value chosen by an omniscient prophet able to see the future, this $1/2$ factor being the best possible. Following this, \cite{SamuelCahn1984} proved that a simple threshold algorithm can actually achieve this competitive ratio of $1/2$. A refined prophet inequality was proved in \cite{Hill83}, who showed that if $V$ is the maximum expected performance of an online algorithm, then the expected maximum value is always smaller than $2V-V^2$. 

\medskip

Following these works, variations with different assumptions on the item distributions were considered. If there are no assumptions on the joint distribution of the sequence of values, then \cite{Hill83b} showed that the worst-case comparison between the gambler and the prophet can be arbitrarily bad in the number of items. Conversely, assumptions on the joint distribution can be strengthened: \cite{Kertz1982} are the first to consider the \ac{iid} setting, in which the values of all items are independently drawn from the same distribution. An implicit upper bound of approximately $0.745$ on the competitive ratio was proposed by \cite{Kertz1986} who reduces the computation of the worst-case competitive ratio to a finite dimensional optimization problem. This upper bound was proven to be tight by \cite{correa2017} through the construction of an explicit adaptive quantile algorithm that achieves this bound, showing that the worst-case competitive ratio in the \ac{iid} setting is exactly the solution to an integral equation (with a numerical value of around $0.745$). Some other works have further investigated the \ac{iid} case: \cite{PerezSalazar2022} show that this optimal competitive ratio can be approached with fewer different thresholds, and \cite{Jiang2022} show that the competitive ratio for a finite number of items can be computed as the solution to a linear program.

\medskip

One important observation is that the worst-case instances tend to involve distributions that depend on the number of items and have a particularly heavy tail. This worst-case distribution does not correspond to the most commonly encountered distributions. In particular, for most distributions, the optimal online algorithm tends to perform better than what the worst-case instance suggests. As a result, some authors propose to use a different benchmark. \cite{Kennedy1985} and \cite{Kertz1986b}, for instance, studied the competitive ratio when the comparison is made with respect to a weaker prophet that receives the average reward of the top $\ell$ items. They prove that, in the case where valuations are independent but not necessarily identically distributed, the competitive ratio of any online algorithm cannot be larger than $1-1/(\ell+1)$, and that this bound is attained.  In this work, we consider the same benchmark of \cite{Kennedy1985,Kertz1986b}, but with \ac{iid} valuations  as in \cite{Kertz1982,Kertz1986,correa2017}. We prove a lower bound on the competitive ratio and provide a quantile algorithm to solve the problem.\\

\paragraph*{Contributions}

We consider a setting with $n \in \bb{N}$ items whose valuations are $(X_1,\dots,X_n)$. The variables $(X_i)_{i\ge1}$ are \ac{iid} non-negative random variables with finite expectation drawn according to some distribution $F$, and we denote by $X_{(1)}\geq \dots \geq X_{(n)}$ their order statistics. The finite expectation assumption implies that the expectations of all order statistics are also finite. We consider online algorithms that observe valuations sequentially, and upon seeing the valuation $X_i$ of an item $i$ irrevocably decide to select the item and receive the value, or move to the next item.
As a result, a valid algorithm induces a stopping time $\tau$ that corresponds to the selected item, and the expected performance of the algorithm is $E[X_{\tau}]$. We will compare the performance of an algorithm to the average valuation of the top $\ell$ items, $\be_{X \sim F}[\ell^{-1}\sum_{i \in [\ell]} X_{(i)}]$.
For $n$ and $\ell$, we define the competitive ratio as:
\begin{equation}
    \comp_{\ell}(n)=\inf_{F} \frac{  \sup_{\tau} \be_{X \sim F}[X_{\tau}]}{\be_{X \sim F}\left[\frac{1}{\ell}\sum_{i \in [\ell]} X_{(i)}\right]}.
    \label{eq:CR_nl}
\end{equation}
This corresponds to the competitive ratio achievable by an algorithm that knows the distribution $F$. Note that we always have $\comp_{\ell}(n)\leq 1$ by taking a constant distribution $X_i=1$. We also define $\comp_{\ell}\eqdef \inf_{n \geq \ell} \comp_{\ell}(n)$ as the worst-case competitive ratio over all possible number of items.  To better illustrate the practical implications of the model, consider the following two concrete interpretations. This competitive ratio can correspond to competing against an \emph{imperfect prophet} who makes mistakes with some probability: the prophet can identify the top-$\ell$ items, but instead of selecting the maximum she selects one of the top-$\ell$ items uniformly at random. This competitive ratio could also correspond to a multi-unit setting with a budget of $\ell$ items, where a \emph{stronger agent} competes against a regular prophet: the agent is allowed to re-select the same item multiple times. At least one optimal policy ends up selecting $\ell$ times the same item. This is immediate from the linearity of expectation: $\sup_{\tau_1,\dots,\tau_{\ell}} \be[ \sum_{i \in [\ell]} X_{\tau_i}]= \sum_{i \in [\ell]} \sup_{\tau_i} \be[X_{\tau_i}]= \ell \sup_{\tau} \be[X_{\tau}]$.

Our first result is the following exact characterization of the worst-case competitive ratio: 
\begin{theorem}
    \label{thm:competitive}
    For all positive integers $\ell$, we have that $\comp_{\ell}$ is the unique solution in $[0,1)$ to the integral equation
    \begin{equation}
        \frac{1}{\ell !}\int_0^{\infty} \frac{\nu^{{\ell}-1}}{e^{\nu}\left( \frac{1}{\comp_{\ell}} -1  \right)+\sum_{i=0}^{{\ell}} \frac{\nu^{i}}{i!}} d\nu=1.
        \label{eq:integral-eq}
    \end{equation}
\end{theorem}

Our analysis is based on a non-trivial generalization of the quantile algorithm presented in \cite{correa2017}, where this algorithm provides a lower bound on the competitive ratio. We show that maximizing the parameters of the proposed quantile algorithm (\Cref{alg:algorithm}) is equivalent to solving a non-linear discrete boundary value problem. We then show that this discrete boundary value problem corresponds to a continuous boundary value problem in the limit where $n$ goes to infinity. This limiting competitive ratio lower bound yields the integral equation \eqref{eq:integral-eq}. Finally, we prove that the competitive ratio for a finite $n$ is lower bounded by its limit as $n$ grows, hence the limit performance of the algorithm is a lower bound on $\comp_{\ell}$. We obtain the matching upper bound by adapting a worst-case instance from \cite{Allen2021} to our setting.

 \begin{table}[t]
    \begin{center}
        \begin{tabular}{|c|c|c|c|c|c|}
            \hline
            $\ell$ & 1 & 2 & 3 & 4 & 5 \\ \hline
            $\comp_{\ell}$& 0.745 & 0.966 & 0.997 & 0.9998 & 0.999993 \\ \hline
        \end{tabular}
        \caption{First digits of $\comp_{\ell}$.}\label{tab:c_values}
    \end{center}
\end{table}

There are no explicit solutions to the integral equation~\eqref{eq:integral-eq}, but this equation can be easily solved numerically. We report the first values of the competitive ratio $\comp_{\ell}$ in \Cref{tab:c_values}. In the special case $\ell=1$, we recover the integral equation of \cite{Kertz1986,correa2017} that yields the competitive ratio of $0.745$ for the classical \ac{iid} prophet.  What is striking is that for $\ell\ge2$, the competitive ratio increases extremely fast towards $1$. In particular, for $\ell=2$, the competitive ratio is larger than $0.966$, which is much closer to $1$ than $0.745$. For any distribution $F$, one has:
\begin{align*}
    \sup_{\tau\text{ stopping time}}\be[X_{\tau}]&\geq 0.745~\be[X_{(1)}] & \text{(result of \cite{correa2017})}\\
    \sup_{\tau\text{ stopping time}}\be[X_{\tau}]&\geq 0.966~\be[\frac12(X_{(1)}+X_{(2)})] & \text{(our bound).}
\end{align*}
The main reason for this difference is that the worst-case instances for $\ell=1$ rely on the first and second maximum being very different. In fact, such an instance is relatively easy when using the benchmark $(X_{(1)}+X_{(2)})/2$. We observe numerically that $\comp_{\ell}$ grows exponentially fast to $1$ (roughly of the order of $1-10^{-\ell}$). Our second main result is to show that indeed the competitive ratio provably converges exponentially fast to $1$ as $\ell$ grows:
\begin{theorem}
    \label{th:expo-bound}
    For all positive integers $\ell$, we have: 
    \begin{equation}
        \comp_{\ell} > 1-e^{-\ell}.
    \end{equation}    
\end{theorem}
Note that this bound is loose as can be seen with the values in \Cref{tab:c_values} but still provides the exponential convergence rate to $1$. Compared to the tight competitive ratio of $1-1/(\ell+1)$ in \cite{Kennedy1985} without the \ac{iid} assumption, the convergence towards $1$ is noticeably faster. This exponential convergence rate explains why the jump from $\ell=1$ to $\ell=2$ was so marked. 

All of our results are obtained through a quantile algorithm which is known to have additional benefits, such as being usable in a learning setting \cite{Rubinstein2020}. Moreover, the algorithm only needs point-wise access to the quantile function, compared to the backward dynamic programming method that needs to compute an integral for each threshold. \\

There are multiple variants of the original setting of Theorem \ref{thm:competitive} that can be studied. A first possible extension is to consider a prophet that does not use the average of the top $\ell$ items, but instead uses a convex combination of the top $\ell$ items. Another possible variant is to compute the worst-case competitive ratio for the prophet. We prove some results for both of these extensions.

We also show that our algorithm can be extended with provable guarantees to a more general setting introduced by \cite{Kennedy1987} for the non \ac{iid} case, where the decision maker is allowed to select $k$ items. This setting encompasses the $k$ multi-unit \cite{Alaei2011,Jiang2022} \ac{iid} prophet problem when $\ell=k$. An algorithm that sequentially selects $k$ items will induce a sequence of stopping times $(\tau_{i})_{i \in [k]}$, for which if $\tau_i<\infty$ then $\tau_i<\tau_{i+1}$. We thus consider the following competitive ratio:
\begin{equation}
\comp_{k,\ell}(n)= \inf_{F}  \frac{\sup_{\tau_1<\dots <\tau_k} \be_{X \sim F}\left[ \frac{1}{k}\sum_{i \in [k]} X_{\tau_{i}}\right]}{\be_{X \sim F}\left[ \frac{1}{\ell}\sum_{i \in [\ell]} X_{(i)}\right]}.
\end{equation}

We prove asymptotic guarantees on $\comp_{k,\ell}(n)$.
\begin{theorem} \label{thm:general_liminf}
For all positive integers $k$ and $\ell$, we have:
\begin{equation}
\liminf_{n \rightarrow \infty} \comp_{k,\ell}(n) \geq \frac{\comp_{\ell}}{k} \sum_{j \in [k]} \frac{1}{ \prod_{t \in [j]} \theta_{j,\ell}},
\end{equation}
where $\theta_{1,\ell}=1$ and $\theta_{2,\ell},\dots,\theta_{k,\ell}$ are the unique parameters such that the following boundary value problem admits a solution,
\begin{align} \label{eq:continuousBVP}
\frac{db^1(t)}{dt} &= \frac{\ell}{\comp_{\ell}} -\ell \cdot \gamma_{\ell+1}\circ \gamma_{\ell}(b^{j}(t)), \nonumber \\
\frac{db^j(t)}{dt} &= \ell \left( \theta_{j,\ell}  \cdot\gamma_{\ell+1}\circ \gamma_{\ell}^{-1}(b^{j-1}(t))-\gamma_{\ell+1}\circ \gamma_{\ell}(b^{j}(t))\right), \quad && \text{for $2 \leq j \leq k$} \\
b^j(0)&=0, \quad b^j(1)=1, \quad && \text{for $1 \leq j \leq k$} \nonumber,
\end{align}
where $\gamma_{x}(t)$ and $\gamma_{x}^{-1}(t)$ are respectively the cumulative distribution function and the quantile function from a $\mathrm{Gamma}(x,1)$ random variable evaluated at $t$. 
 
\end{theorem}

Remark that the first differential equation admits a solution if and only if $\comp_{\ell}$ is the unique solution to the integral equation of Theorem \ref{thm:competitive}.

Finally, we extend the analysis to the setting where the decision maker is restricted to use static thresholds policies. A static threshold policy $T \in \bb{R}$ accepts items whenever $X_i \geq T$ and less than $k$ items were already selected. More specifically, a static threshold $T$ induces stopping times $\tau_1(T)=\inf \{i\in \bb{N} \mid X_i \geq T \}$ and $\tau_i(T)=\inf \{ i \in \bb{N}  \mid X_i \geq T, i > \tau_{i-1}\}$ for $i \geq 2$. When the decision maker and the prophet must select $k$ and $\ell$ items respectively and the decision maker is restricted to static threshold policies we define the competitive ratio as
\begin{equation}
\comp_{k,\ell}^S(n)=\inf_{F}  \frac{ \sup_{T \in \bb{R}} \be_{X \sim F}\left[ \frac{1}{k}\sum_{i \in [k]} X_{\tau_{i}(T)}\right]}{\be_{X\sim F} \left[ \frac{1}{\ell}\sum_{i \in [\ell]} X_{(i)}\right]}.
\end{equation}

Using intermediary results derived for Theorems \ref{thm:competitive} and \ref{thm:general_liminf} and extending some of the analysis from \cite{Correa2019} and \cite{Arnosti2021}, we nearly characterize the exact competitive ratio for static thresholds:

\begin{theorem} \label{thm:static}
For all positive integers $k$ and $\ell$, and $n \geq \max(k,\ell)$, we have:
\begin{equation}
\left \vert \comp^S_{k,\ell}(n) - \frac{\sum_{j=1}^k \Pr(\mathrm{Gamma}(j,1) \leq \ell)}{k} \right \vert  \leq O\left( \frac{1}{n} \right).
\end{equation}
\end{theorem}

This result recovers a special case of the tight static threshold competitive ratio provided in \cite{Arnosti2021} when $n \rightarrow \infty$ and with \ac{iid} valuations, but is on other aspects more general by allowing for $\ell \neq k$. \\

\paragraph*{Roadmap} The rest of the paper is organized as follows. In \Cref{sec:finite-n}, we present the quantile algorithm and the analysis of the competitive ratio for finite $n$. In \Cref{sec:limit-n} we show how to use the limit performance guarantees (as the number of items $n$ goes to infinity) to construct the ODE and derive Theorem \ref{thm:competitive} and Theorem \ref{th:expo-bound}. We prove the matching upper bound of $\ell/c_{\ell}$ on the competitive ratio in \Cref{sec:tight}. In \Cref{sec:general} we show how to extend our algorithm to the selection of $k$ items and construct a corresponding system of $k$ ODE to obtain Theorem \ref{thm:general_liminf}. \Cref{sec:static} deals with the static threshold setting and proves Theorem \ref{thm:static}. \Cref{sec:extension} deals with some direct extensions to the setting considered, namely non uniform distributions for the \emph{imperfect prophet}, and worst-case distribution against a prophet. The detailed proofs of all results are presented in \Cref{sec:proof}. Finally, some additional related works are presented in \Cref{sec:related-work}.

\section{Competitive ratio guarantees for a given number of items}
\label{sec:finite-n}

Because the optimal online algorithm depends in a complicated fashion on the distribution, the direct analysis of the competitive ratio is hard.  In this section, we construct an explicit simpler algorithm, and we derive an analysis of the competitive ratio of this algorithm. This algorithm builds on quantiles of the Beta distribution.\medskip

For the remainder of the paper and for simplicity of notations, we will consider that $F$ is absolutely continuous\footnote{This assumption is standard for this type of analysis and simplifies the exposition. The proof can be adapted to general $F$ by adding randomization between ties when the distribution has atoms, as in \cite{correa2017}.} with respect to the Lebesgue measure and admits a density $f\geq 0$. The function $F$ is the cumulative distribution function $F(x)=\Pr(X\leq x)$, and we denote by $F^{-1}$ its quantile function: for each $p\in[0,1]$,  $F^{-1}(p)=\inf\{ x \in \bb{R} \mid p \leq F(x) \}$. Most quantities depend on $n$ and this will be omitted, except punctually when this makes the understanding clearer and will be denoted as $x(n)$ for some quantity $x$.

\subsection{The quantile algorithm}

We define the quantile algorithm \Cref{alg:algorithm}, which takes as an input the known distribution $F$, and an increasing sequence $0=\epsilon_{0} < \dots < \epsilon_{n}=1$. For each item $i$, this algorithm samples a quantile $q_i$ from a Beta($\ell,n-\ell$) distribution truncated to $\epsilon_{i-1}$ and $\epsilon_i$ and selects item $i$ if and only if $F(X_i)\ge 1-q_i$.  We denote by $\alg_n$ the algorithm's expected performance for a sequence of $n$ items. This algorithm generalizes the quantile algorithm described in \cite{correa2017} for the special case $\ell=1$, and no mentions of Beta distributions were made.

\begin{algorithm}
    \caption{Quantile algorithm for $\ell$}\label{alg:algorithm}
    \begin{algorithmic}
    \State \textbf{Input:} Partition $(\epsilon_i)_{0\leq i \leq n}$ of $[0,1]$, distribution $F$ of the $X_i$.
    \For{$i \in [n]$:}
        \State Draw $q_i$ from $\mathrm{Beta}(\ell,n-\ell)$ truncated between $\epsilon_{i-1}$ and $\epsilon_i$
        \If{$X_i \geq F^{-1}(1-q_i)$:}
            \State Accept item $i$ and stop
        \EndIf
    \EndFor 
    \end{algorithmic}
\end{algorithm}

Before showing a bound on the performance of the algorithm, we introduce some notations regarding the Beta distribution. We recall that the density of a $\mathrm{Beta}(\ell,n-\ell)$ random variable is equal to
\begin{equation}
    \psi_{\ell,n-\ell}(x)=\frac{x^{\ell}(1-x)^{n-\ell}}{B(\ell,n-\ell)},
\end{equation}
where $B(\ell, n-\ell)\eqdef (\ell-1)!(n-\ell-1)!/(n-1)!$ is the normalization constant.  As $q_i$ is drawn from a distribution truncated between $\epsilon_{i-1}$ and $\epsilon_i$, we denote the normalizing factor of this $\psi_{\ell,n-\ell}$ truncated distribution by $\alpha_i=\int_{\epsilon_{i-1}}^{\epsilon_i} \psi_{\ell,n-\ell}(x)dx$. Finally, we define $a_i\eqdef\alpha_i\be[(1-q_i)]$, where $\be[(1-q_i)]$ is the expected probability of not selecting item $i$ when observing it.

Our first result provides a bound on the performance of Algorithm~\ref{alg:algorithm} (valid for any sequence of $\epsilon_i$s), as a function of the quantities $\alpha_i$ and $a_i$ defined above. 
\begin{proposition}\label{prop:alg_perf}
    For $\opt_{\ell,n}=\be[ \sum_{i \in [\ell]} X_{(i)}]$ and $\rho_i=\alpha_i^{-1}\prod_{j=1}^{i-1} a_j/\alpha_j$, we have the inequality
    \begin{equation}
    \label{eq:min-max-rho}
        \frac{\min_{i \in [n]}\rho_i}{n} \opt_{\ell,n} \le \alg_n \le \frac{\max_{i \in [n]}\rho_i}{n} \opt_{\ell,n}.
    \end{equation}
\end{proposition}

\begin{proof}[Sketch of proof]
    The proof decomposes in two steps. First, we compute an expression of $\alg_n$: remarking that the quantiles $q_i$ are independent, we can show that the performance of $\alg_n$ is equal to $\sum_{i \in [n]} \rho_i \int_{\epsilon_{i-1}}^{\epsilon_i} \psi_{\ell,n-\ell}(q) dq$. Second, we derive an expression for $\opt_{\ell,n}$ as an expectation of the function $R(Q)$, where $R(q)$ is the expected reward of accepting an item with threshold $F^{-1}(1-q)$ and $Q$ is some random variable. In \cite{correa2017}, the distribution of $Q$ was shown to be the density $(n-1)(1-q)^{n-2}$ in the special case $\ell=1$. We prove by using the density for a general order statistic, that the right density in the case $\ell>1$ is $\psi_{\ell,n-\ell}$. Because $\opt_{\ell,n}=n\be_{q \sim \psi_{\ell,n-\ell}}[R(q)]$, we can take the minimum or the maximum over the $\rho_i$ to obtain \eqref{eq:min-max-rho}. A full proof is provided in \Cref{app:alg_perf}. 
\end{proof}

To obtain bounds on the competitive ratio of the algorithm, we can divide the above inequality by $\opt_{\ell,n}/\ell$.

\subsection{Optimizing the parameters of the algorithm} \label{subsec:maximize}

Looking at \Cref{prop:alg_perf}, we see that the $\rho_i$s are functions of $\epsilon_i$. A lower bound on the performance of the algorithm is therefore obtained if we can find the sequence $\epsilon_i$ which maximizes $\min_{i \in [n]} \rho_i$. A natural choice of $\epsilon_i$ would be to find a sequence such that all the $\rho_i$ are equal, this would lead to an algorithm whose performance is \emph{exactly} $(\max_{i}\rho_i/n)\opt_n$. This would entail a lower bound on $\comp_{\ell}(n)$. It is, however, not clear whether such a sequence of $\epsilon_i$ exists. Moreover we need this sequence of $\epsilon_i$s to be increasing (\emph{i.e.}, $\epsilon_{i-1}<\epsilon_i$) for the algorithm to be well defined. 
 
We will first see that finding the $\epsilon_i$ such that $\rho_i=\rho_{i+1}$ is equivalent to solving a discrete boundary value problem on a non-linear transformation of the $\epsilon_i$ by an incomplete beta function. We then use this transformed problem to prove the existence of such an $\epsilon_i$.

We introduce the variables $b_i=\beta_{\ell,n-\ell}(\epsilon_i)$ which is a nonlinear transformation of $\epsilon_i$, where
\begin{equation}
    \beta_{\ell,n-\ell}(x)=\frac{\int_0^{x} t^{\ell-1}(1-t)^{n-\ell-1}dt}{B(\ell,n-\ell)}.
\end{equation}
is the cumulative distribution function of $\psi_{\ell,n-\ell}$, also called the regularized incomplete beta function. 

Because $\beta_{\ell,n-\ell}$ is strictly monotone as a distribution function with an associated positive density, it has an inverse, and we can recover $\epsilon_i$ with $\beta^{-1}_{\ell,n-\ell}(b_i)$. We also define the discrete difference operator $\Delta$, with $\Delta[b_i]=b_{i+1}-b_i$. It is the discrete analogue of the continuous differentiation operator. Similarly, we define $\Delta^2[b_i]=\Delta[\Delta[b_i]]=b_{i+2}-2b_{i+1}-b_i$. 

\begin{lemma} \label{lemma:recurrence_relation}
    All the $\rho_i$ are equal if and only if the following difference equation holds for all $b_i$:
    \begin{equation} \label{eq:recurrence_relation}
        \Delta[b_i]= -\frac{\ell}{n} \beta_{\ell+1,n-\ell}\circ \beta^{-1}_{\ell,n-\ell}(b_i) + b_1.
    \end{equation}
\end{lemma}

\begin{proof}[Sketch of proof]
    The main idea is that we can actually express $\alpha_i$ and $a_i$ in terms of $\beta_{\ell,n-\ell}$. Indeed $\alpha_i=\beta_{\ell,n-\ell}=\beta_{\ell,n-\ell}(\epsilon_i)-\beta_{\ell,n-\ell}(\epsilon_{i-1})=b_i-b_{i-1}$, and after some computations it can be showed that $a_i=(n-\ell)(\beta_{\ell,n+1-\ell}(\epsilon_i)-\beta_{\ell,n+1-\ell}(\epsilon_{i-1}))/n$. Then using that there is an exact recurrence relationship between $\beta_{\ell,n+1-\ell}$ and $\beta_{\ell,n-\ell}$, as well as between $\beta_{\ell+1,n-\ell}$ and $\beta_{\ell,n-\ell}$, we can obtain a non-linear second order difference equation. Then, for every $i\geq 2$, it is sufficient to sum these difference equations for all $j \leq i$, to obtain a recurrence relation directly on the $b_i$. The full proof can be found in \Cref{app:recurrence_relation}. 
\end{proof}

\begin{remark}
Note that obtaining an explicit recurrence relation of $\epsilon_{i+1}$ with `simple' functions of the previous $\epsilon_i$ is difficult: developing the integrals $a_i$ and $\alpha_i$ in $\epsilon_i$ and $\epsilon_{i+1}$ yields an implicit polynomial equation in $\epsilon_{i+1}$ parameterized by $\epsilon_i$. This would entail finding roots of a sequence of polynomials of degree $n-\ell+1$. The approach that we use here is to obtain an explicit recurrence relation on a non-linear transformation of the $\epsilon_i$ (the $b_i$), and not on the $\epsilon_i$ themselves. This non-linear transformation has no inverse expressible with `simple' functions exactly whenever $\ell\geq 2$; this explains why the task is considerably more difficult than for the case $\ell=1$  studied in \cite{correa2017}. This difficulty is one of the core obstacles to an extension of the work of \cite{correa2017}.

It is quite remarkable that the recurrence relation is `almost' linear, in that if $\beta_{\ell+1,n-\ell}$ were to be replaced with $\beta_{\ell,n-\ell}$ we would have recovered the identity when composing with $\beta_{\ell,n-\ell}^{-1}$. We will see below that when $n$ goes to infinity we recover the Gamma distribution in the limit. As an aside when $\ell$ also goes to infinity we can recover the Normal distribution.
\end{remark}

By using Lemma~\ref{lemma:recurrence_relation}, we can therefore focus on proving the existence of the correct constant $b_1$ which will imply the required condition $\rho_{i+1}=\rho_i$.

\begin{proposition} \label{prop:existence}
    There exists an increasing sequence $0=\epsilon_0<\epsilon_1 < \dots < \epsilon_{n-1}<\epsilon_n=1$ and $c_{\ell}(n) \in [\ell,\ell+1]$ such that all the $\rho_i$ are equal to $n/c_{\ell}(n)$. 
\end{proposition}

\begin{proof}[Sketch of proof] Because $\beta_{\ell,n-\ell}$ is a bijection from $[0,1]$ to $[0,1]$, $\beta_{\ell,n-\ell}(0)=0$ and $\beta_{\ell,n-\ell}(1)=1$, finding the right partition $\epsilon_i$ partition is equivalent to finding the right $b_i$ partition. Due to the continuity of the recurrence relation, $b_n$ is a continuous function of $b_1$, and the intermediate value theorem proves the existence. We must ensure that there is a solution in $[\ell/n,(\ell+1)/n]$ to guarantee the monotonicity of the $b_i$ and thus of the $\epsilon_i$. The full proof is provided in \Cref{app:existence}. 
\end{proof}

\Cref{prop:existence} shows that there exists a sequence of $\epsilon_i$s such that our algorithm is well-defined and satisfies that $n/\rho_i=c_\ell(n)$ for all $i$. In particular, such an algorithm has a competitive ratio of $\ell/c_{\ell}(n)$. This shows that for all $n$, $\comp_{\ell}(n)\geq \ell/c_{\ell}(n)$.

In the remainder of the paper, we improve this result in two directions. First, the quantity $c_{\ell}(n)$ depends on $n$. In \Cref{sec:limit-n}, we show how to obtain a guarantee that does not depend on $n$, by studying the limit as $n$ grows. 
 Second, the fact that $c_{\ell}(n) \in [\ell,\ell+1]$ implies a competitive ratio of at least $\ell \rho_1/n=\ell/c_{\ell}(n) \in [\frac{\ell}{\ell+1},1]$. The bound on $\ell/(\ell+1)$ is the same as the result from \cite{Kennedy1985,Kertz1986b} for the non-\ac{iid} case. 
    We then show in \Cref{sec:large-l} that, in the \ac{iid} setting, $\comp_{\ell}$ is actually exponentially close to $1$ when $\ell$ is large.

\section{Competitive ratio guarantees as $n$ grows}
\label{sec:limit-n}

Until now, we have proven how to obtain guarantees that depend on the number of items. 
In this section, we first show that the worst-case for the competitive ratio is for large $n$. Then, we use a limiting ODE to characterize the competitive ratio given by our quantile algorithm when $n$ goes to infinity.

\subsection{$\comp_{\ell}(n)$ is minimized for very large $n$}

By using our analysis of the previous section, we cannot directly conclude that $\comp_{\ell}(n)$ is a monotone function of the number of items $n$, nor that the competitive ratio might be small for large $n$. It might be possible that the value of $\comp_{\ell}$ is actually reached for some $n^*$ such that $\comp_{\ell}(n^*)=\comp_{\ell}$ (we know that this is not the case for $\ell=1$, see \cite{correa2017}). Our \Cref{lemma:doubling} generalizes Lemma $3.2$ from \cite{Allen2021} (that deals with the special case $\ell=1$) and shows that we can always transform an instance with $n$ items $(F,n)$ to an instance with $2n$ items $(\widetilde{F},2n)$ that is at least as difficult as the $(F,n)$ instance.  

\begin{lemma} \label{lemma:doubling}
For any $n\geq {\ell}$, let $(X_1,\dots,X_n)$ \ac{iid} distributed according to $F$, and $(Y_1,\dots,Y_{2n})$ i.i.d distributed according to $\sqrt{F}$. We have for $\tau_X$ and $\tau_Y$ the optimal stopping for respectively the $X_i$ and $Y_i$ that
\begin{equation*}
 \frac{\be[X_{\tau_X}]}{\opt_{\ell,n}(F)} \geq  \frac{\be[Y_{\tau_Y}]}{\opt_{\ell,2n}(\sqrt{F})}.
\end{equation*}
\end{lemma}

\begin{proof}[Sketch of proof]
The original Lemma proves this for ${\ell}=1$. They remark that for each $X_i$, we can simulate a draw of $Y_{2i}$ and $Y_{2i+1}$ conditionally on their maximum being equal to $X_i$. The item $X_i$ is then accepted whenever $Y_{2i}$ or $Y_{2i+1}$ is accepted by $\tau_Y$. This proves that $\be[X_{\tau_X}]$ is at least $\be[X_{\tau_Y}]$. This property holds for $\ell>1$ as we are still only allowed to select a single item. What changes, however, is $\opt_{\ell}$. We show that $\be[Y_{({\ell})}]\geq \be[X_{({\ell})}]$. It is sufficient to prove that $Y_{(\ell)}$ stochastically dominates $X_{(\ell)}$, which will imply the inequality for the expectations. This can be proved by looking at the difference of the respective cumulative distribution functions and looking at the monotonicity of the derivative of the difference. The full proof can be found in \Cref{app:doubling}. \end{proof}

Note that this lemma implies that $\comp_{\ell}(2n)\le \comp_{\ell}(n)$. In particular, this implies that
\begin{equation}\label{eq:lim_inf}
    \comp_{\ell}=\inf_{n \in \bb{N}} \comp_{\ell}(n) = \liminf_{n \in \bb{N}} \comp_{\ell}(n).
\end{equation}
This explains why, in the rest of the section, we focus on the limiting behavior of the quantile algorithm as $n$ goes to infinity. Note that this does not imply that $\comp_{\ell}(n)$ decreases with $n$.

\subsection{Limiting ODE as $n$ goes to infinity}

We can now focus on analyzing the limit (as $n\rightarrow \infty$) of our discrete boundary value problem described by the recurrence relation in \Cref{eq:recurrence_relation}. Let us first recall this difference equation using that $b_1=c_{\ell}(n)/n$:
\begin{equation}
    \label{eq:difference-eq}
    \Delta[b_i]= \frac{-\ell\beta_{\ell+1,n-\ell}\circ \beta^{-1}_{\ell,n-\ell}(b_i) + c_{\ell}(n)}{n}.
\end{equation}
We show below that this difference equation converges to an ordinary differential equation as $n$ goes to infinity, by using the property that the limit of a Beta random variable is a Gamma random variable. Recall that the cumulative distribution function of a $\mathrm{Gamma}(k,1)$ random variable is 
\begin{equation}
    \gamma_{\ell}(z)=\frac{\int_0^{z} t^{\ell-1}e^{-t}dt}{\Gamma(\ell)},
\end{equation}
where $\Gamma(\ell)=(\ell-1)!$ is the normalizing constant.  This function is also called the regularized lower incomplete Gamma function. Because of the integral representation $\Gamma(\ell)=\int_0^{\infty}t^{\ell-1}e^{-t}dt$, it is clear that $\gamma_{\ell}(\infty)=1$ and thus this is a proper distribution. 

\begin{lemma} \label{lemma:unif_conv}
    The sequence of functions  $x\mapsto -{\ell}\beta_{\ell+1,n-{\ell}}\circ \beta_{{\ell},n-{\ell}}^{-1}(x)$, defined on $[0,1]$, converges uniformly towards $-{\ell}\gamma_{\ell+1}\circ \gamma_{\ell}^{-1}(x)$ as $n$ goes to infinity.
\end{lemma}

\begin{proof}[Sketch of proof] We can prove the point-wise convergence by using the property that Beta random variables converge in distribution to Gamma random variables. Then to show that the convergence is uniform, we first show the monotonicity of the sequence of functions through usual formulas for beta functions. We can conclude by Dini's convergence theorem as the input space is compact. The full proof can be found in \Cref{app:unif_conv}. \end{proof}

\begin{remark}
The concise representation of the function of interest through Beta functions enables us to easily prove uniform convergence. Indeed, if we attempt to prove uniform convergence of first $\beta_{\ell,n-\ell}^{-1}$ and then $\beta_{\ell+1,n-\ell}$, this does not suffice as the output of the inverse of $\gamma_{\ell}$ is unbounded, so the input space of the last function is non-compact. Looking directly at the composition enables us to skip this difficulty, thus avoiding tedious technical computations, and piece-wise analysis. 
\end{remark}

Before stating formally the result, we start by giving the intuition on how to construct the limiting ODE. \Cref{lemma:unif_conv} suggests to approximate the difference equation~\eqref{eq:difference-eq} by 
\begin{equation} \label{eq:ode}
    \frac{db(t)}{dt}=-\gamma_{\ell+1} \circ \gamma^{-1}_{\ell}(b(t))+c_\ell,
\end{equation}
where the solution $b$ satisfies the boundary conditions $b(0)=0$ and $b(1)=1$, and where the constant $c_\ell$ is an unknown value that replaces $c_\ell(n)$.

As $c_\ell(n) \geq {\ell}$, we also consider that $c_\ell\ge\ell$. This implies that $b$ is strictly increasing until at least the first $t_1$ for which $b(t_1)=1$. So $b$ is a bijection over $[0,t_1]$, and we can consider the inverse function $t(b)$ with $t(1)=t_1$ and $t(0)=0$. Requiring that $t(1)=1$ leads to the following integral equation:
\begin{equation*}
\int_0^1 \frac{db}{-{\ell} \gamma_{\ell+1}\circ \gamma_{\ell}^{-1}(b)+c_{\ell}}=t(1)-t(0)=1-0=1.
\end{equation*}

Using a change of variable $\nu=\gamma_{{\ell}}^{-1}$, we define $c_{\ell}$ as the constant which satisfies the following integral equation:
\begin{equation}
    \label{eq:integral_form}
    1=\frac{1}{\Gamma({\ell})}\int_0^{\infty} \frac{\nu^{{\ell}-1}e^{-\nu}}{c_{\ell}-{\ell}\gamma_{\ell+1}(\nu)} d\nu =\frac{1}{\Gamma({\ell})}\int_0^{\infty} \frac{\nu^{{\ell}-1}}{e^{\nu}(c_{\ell}-{\ell})+{\ell} \sum_{i=0}^{{\ell}} \frac{\nu^{i}}{i!}} d\nu.
\end{equation}

Note that this last integral equation implies that $c_{\ell}>{\ell}$, as otherwise when $\nu$ goes to $\infty$ the integrand becomes equivalent to $\nu^{{\ell}-1}/\nu^{{\ell}}=1/\nu$ which integrates to $\log(\nu)$ and diverges.

The next proposition formalizes the intuition, and shows the limit of $c_{\ell}(n)$ to indeed be the $c_{\ell}$ defined as the solution to this integral equation. The proof uses the same arguments as that in \cite{Kertz1986,Jiang2022}, with the main difference being the actual value of the limit, and proving the uniform convergence in \Cref{lemma:unif_conv}, which has already been detailed. We defer the actual proof of the convergence to \Cref{app:lim}.

\begin{proposition} \label{thm:lim}
For $c_{\ell}$ the solution to \Cref{eq:integral_form}, we have 
\begin{equation*}
\lim_{n \rightarrow \infty} c_{\ell}(n)=c_{\ell}.
\end{equation*}
\end{proposition}
Combining this result with \Cref{eq:lim_inf} implies $\comp_{\ell} \geq \ell/c_{\ell}$. Remark that if $c_{\ell}$ is solution to \Cref{eq:integral_form}, then $\ell/c_{\ell}$ is solution to \Cref{eq:integral-eq}. This proves the first part of Theorem \ref{thm:competitive}, namely that $\comp_{\ell}$ is greater than the solution to the integral equation~\eqref{eq:integral-eq}.

\subsection{Asymptotic competitive ratio as $\ell$ grows}
\label{sec:large-l}

\begin{figure}[ht]
    \centering
         \includegraphics[width=0.7 \textwidth]{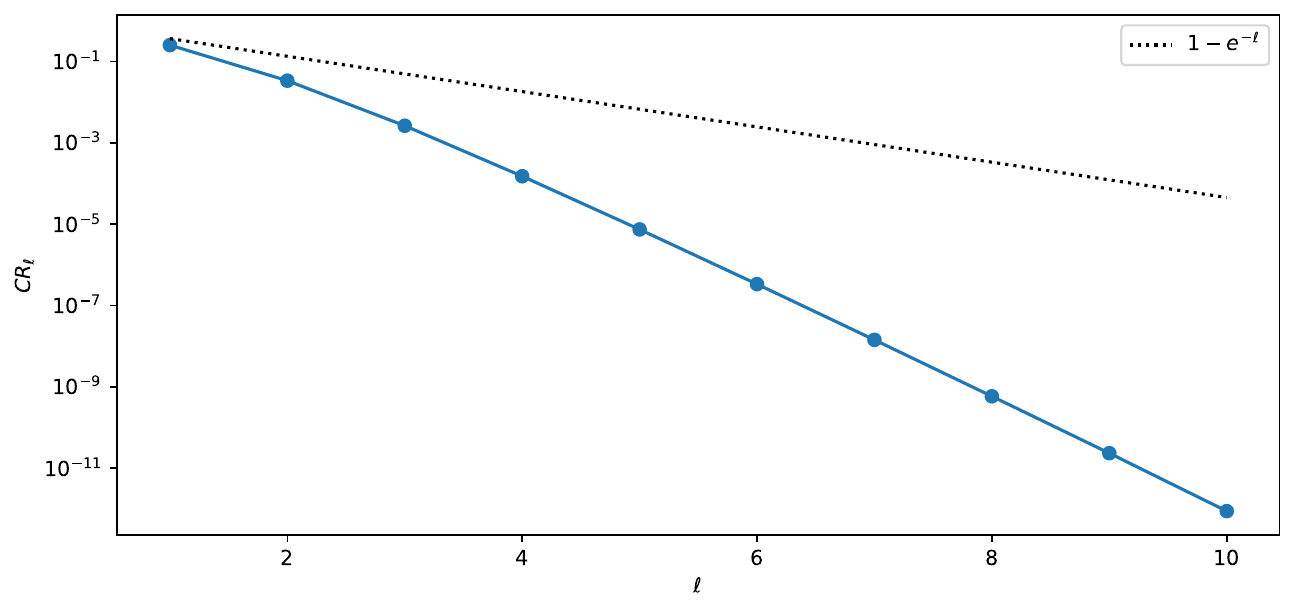} 
    \caption{$\comp_{\ell}$ as a function of $\ell$: the competitive ratio converges exponentially fast to $1$.}
    \label{fig:CR_LB}
\end{figure}

While the above integral equation does not lead to a close form expression for $c_\ell$, it can be used to provide an easy characterization for the behavior of $\comp_\ell$ as $\ell$ grows. Here, we recall and prove Theorem \ref{th:expo-bound}, which states that for all $\ell$:
\begin{equation}
    \comp_{\ell} >  1- e^{-\ell}.
\end{equation}
This result confirms what we observe in \Cref{tab:c_values}: the competitive ratio goes exponentially fast towards $1$.  To push the comparison deeper, we plot in \Cref{fig:CR_LB}(a) the value of the $1-\ell/c_\ell$ as a function of $\ell$ with a $y$-axis in log-scale. We observe that numerically $c_{\ell} \approx 1-10^{-\ell}$. This is closer to $1$ than $1-\exp(-\ell)$ predicted by Theorem \ref{th:expo-bound}, but still of the correct order. This does not tell us whether $\ell/c_{\ell}$ is the best possible lower bound on $\comp_{\ell}$ but shows that the competitive ratio must lie between $1-\exp(-\ell)$ and $1$ for all $\ell$.

\begin{proof}[Proof of Theorem \ref{th:expo-bound}]
    For $c\geq \ell$, let us consider the integral in \Cref{eq:integral_form}:
    \begin{equation*}
    \frac{1}{\Gamma({\ell})}\int_0^{\infty} \frac{\nu^{{\ell}-1}}{e^{\nu}(c-{\ell})+{\ell} \sum_{i=0}^{{\ell}} \frac{\nu^{i}}{i!}} d\nu < \frac{1}{\ell \Gamma({\ell})}\int_0^{\infty} \frac{\nu^{{\ell}-1}e^{-\nu}}{(\frac{c}{\ell}-1)+e^{-\nu}\sum_{i=0}^{{\ell-1}} \frac{\nu^{i}}{i!}} d\nu.
    \end{equation*}
    Let $h(\nu)\eqdef e^{-\nu}\sum_{i=0}^{\ell-1}\nu^{i}/i!$, then $h'(\nu)=-\nu^{\ell-1}e^{-\nu}/\Gamma(\ell)$. Hence 
    \begin{align*}
    \frac{1}{\ell \Gamma({\ell})}\int_0^{\infty} \frac{\nu^{{\ell}-1}e^{-\nu}}{(\frac{c}{\ell}-1)+e^{-\nu}\sum_{i=0}^{{\ell-1}} \frac{\nu^{i}}{i!}} d\nu &= \frac{-1}{\ell}\left[ \log\left(\frac{c}{\ell}-1+h(\nu)\right)\right]_{0}^{\infty}\\
    &=\frac{1}{\ell}\log\left(\frac{c}{c-\ell}\right).
    \end{align*}
    For $\tilde{c}_{\ell}$ such that $\log(\tilde{c}_{\ell}/(\tilde{c}_{\ell}-\ell))/\ell=1$, then $\tilde{c}_{\ell}=\ell / (1-e^{-\ell})$. Moreover, because the integral in \Cref{eq:integral_form} is decreasing in $c$, we have that $c_{\ell} < \tilde{c}_{\ell}$. Finally
    \begin{equation*}
    \comp_{\ell} = \frac{\ell}{c_{\ell}} > \frac{\ell}{\tilde{c}_{\ell}} = 1-e^{-\ell}. \qedhere
    \end{equation*} 
    \end{proof}

\section{Tightness of $\ell/c_{\ell}$} \label{sec:tight}

In this section, we show that the lower bound is actually tight, proving the result of Theorem \ref{thm:competitive}. To do so, we adapt the worst-case instance for $\ell=1$ from \cite{Allen2021} to an arbitrary $\ell$.\\

In \cite{Allen2021}, an equivalent random arrival model is used, where each item is assigned independently a uniform arrival time in $[0,1]$, and the decision maker observes the items in order of increasing arrival times. The optimal online policy $r_n(t)$ then corresponds to accepting an item with value $X$ at time $t$ if $r_n(t)>X$. The limit policy $\lim_{n \rightarrow \infty} r_n(t)$ is then studied, and the competitive ratio on an instance constructed from a differential equation is computed. Here, we similarly construct a worst-case example by leveraging the ODE \Cref{eq:ode}. For $b(t)$ the solution to the boundary value problem \Cref{eq:ode}, let $y(t)=1-b(t)$. The function $y(t)$ satisfies for all $t \in [0,1]$ the ODE
\begin{equation*}
y'(t)= \ell \gamma_{\ell+1} \circ \gamma_{\ell}^{-1}(1-y(t)) -c_{\ell}
\end{equation*}
with initial conditions $y(0)=1$.
Let $q\in (0,1)$ a parameter, $p=\exp(-\gamma_{\ell}^{-1}(1-y(q)))$, and 

\begin{equation*}
H=\frac{1}{y'(q)\log(p)} - \int_q^1 \frac{ds}{y'(s)} .
\end{equation*}
We define a threshold policy $r(t)$ as 
\begin{equation*}
r(t)=-\int_t^1 \frac{ds}{y'(s)} \quad \text{for $t \in [q,1]$ and $r(t)=H$ for $t \in [0,q)$.}
\end{equation*}
Finally, we define $F_q$ the distribution of the maximum as 
\begin{equation}
F_q(x) =
\left\{ 
    \begin{array}{l}
        \exp(-\gamma_{\ell}^{-1}(1-y(r^{-1}(x))) \quad \text{for $t \in [0,r(q)]$} \\
        p \quad \text{for $t \in (r(q),H]$} \\
        1 \quad \text{for $t \in (H,\infty)$}.
    \end{array} 
\right.
\end{equation}
This distribution is constructed such that $r$ is optimal in the limit. The $X_i$ are thus distributed according to $F_q^{1/n}$. 

\begin{proposition} \label{prop:tightness}
\begin{equation}
\lim_{q \rightarrow 0} \lim_{n \rightarrow \infty} \sup_{\tau\text{ stopping time}} \frac{\be_{X \sim F_q^{1/n}}\left[ X_{\tau} \right]}{ \be_{X \sim F_q^{1/n}} \left[ \frac{1}{\ell} \sum_{i=1}^{\ell} X_{(i)} \right]}= \frac{\ell}{c_{\ell}}.
\end{equation}
\end{proposition}

\begin{proof}[Sketch of proof]
The first step is to show that the threshold policy $r(t)$ defined above is indeed the limit online optimal policy as $n \rightarrow \infty$, for this particular instance $F_q$. Then, we compute the value of the competitive ratio as $q$ goes to $0$, and show through integration by parts that this ratio equals $\ell/c_{\ell}$, using the characterization of $c_{\ell}$ with the integral equation~\eqref{eq:integral_form}. The full proof can be found in \Cref{app:tight}
\end{proof}

The above property implies that $\comp_{\ell} \leq \ell/c_{\ell}$, which concludes the proof of Theorem \ref{thm:competitive}.

\section{Extension to the selection of $k$ items} 
\label{sec:general}

Until now we have assumed, that the number of items that can be selected by the decision maker is only $1$. Here, we consider an extension of the problem where the decision maker sequentially selects $k$ items, and where all the items can be selected at most once. This corresponds to the more general setting of \cite{Kennedy1987}, which encloses the previous model.

\subsection{General algorithm and guarantees}

This section focuses on giving guarantees on $\comp_{k,\ell}(n)$ for $k>1$. This time, the algorithm needs to be adaptive not only in $n$ but also in $k$. We define the quantile algorithm \Cref{alg:algorithm2}, which takes as an input the known distribution $F$, and for $j \in [k]$, the increasing sequences $0 = \vcentcolon \varepsilon^j_{j-1} <\varepsilon^j_j<\dots <\varepsilon^{j}_{n}\eqdef 1$. The algorithm is the direct extension of \Cref{alg:algorithm}. The only difference is that, in this new version, the algorithm uses thresholds that depend both on the number of items already observed ($i-1$) and on the number of items already selected ($j-1$).

\begin{algorithm}
    \caption{Quantile algorithm for $(k,\ell)$}\label{alg:algorithm2}
    \begin{algorithmic}
    \State \textbf{Input:} Partition $(\epsilon_i^j)_{j-1 \leq i \leq n}$ of $[0,1]$ for all $j \in [k]$, distribution $F$ of the $X_i$.
    \State $j \gets 1$ \hfill\emph{ // We are currently selecting the $j$th item.}
    \For{$i \in [n]$:}
        \State Draw $q^j_i$ from $\mathrm{Beta}(\ell,n-\ell)$ truncated between $\epsilon^j_{i-1}$ and $\epsilon^j_i$
        \If{$X_i \geq F^{-1}(1-q^j_i)$}
            \State Accept item $i$
            \If{$j=k$}
            \State Stop \hfill\emph{ // because we selected $k$ items}
            \Else
            \State $j\gets j+1$
            \EndIf
        \EndIf
    \EndFor 
    \end{algorithmic}
\end{algorithm}

Similarly to earlier, we define $\alpha^j_i \eqdef \int_{\varepsilon^j_{i-1}}^{\varepsilon^j_i} \psi_{\ell,n-\ell}(x)dx=\beta_{\ell,n-\ell}(\epsilon_i^j)-\beta_{\ell,n-\ell}(\epsilon_{i-1}^j)$, and $a_i^j=\alpha_i^j \be[(1-q_i^j)]$ where $\be[(1-q_i^j)]$ is the expected probability of not selecting item $i$ when we are observing it and have currently already selected $j-1$ items.
Similarly to \Cref{prop:alg_perf}, we can obtain a guarantee on $\alg_{k,n} \eqdef \be[ \sum_{ j \in [k]} X_{\tau_{j}}] $, the performance of \Cref{alg:algorithm2}, with respect to $\opt_{\ell,n}$.

\begin{proposition} \label{prop:alg_perf2}
We have the inequality
\begin{equation}
 \frac{\sum_{j=1}^k \min_{ i \in \{j,\dots, n\}}   \rho_i^j}{n} \opt_{\ell,n} \leq \alg_{k,n}   \leq \frac{\sum_{j=1}^k \max_{ i \in \{j,\dots, n\}}   \rho_i^j}{n} \opt_{\ell,n},
\end{equation}
where 
\begin{equation}
\rho_i^j=\frac{1}{\alpha_i^j}\sum_{1\leq t_1 < \dots <t_{j-1} \leq i-1} \left[ \prod_{s=1}^{j-1} \left(1-\frac{a_{t_s}^s}{\alpha_{t_s}^s}\right) \prod_{r_1=1}^{t_1-1} \frac{a^1_{r_1}}{\alpha^1_{r_1}} \dots \prod_{r_j=t_{j-1}+1}^{i-1} \frac{a^j_{r_j}}{\alpha^j_{r_j}} \right].
\end{equation}
\end{proposition}

\begin{proof}[Sketch of proof]
This is similar to \Cref{prop:alg_perf}, the main difference being that the performance of the algorithm conditionally on the $q_i$ being already drawn is more difficult to express. See \Cref{app:alg_perf2} for the proof. 
\end{proof}

Remark that the $\rho_i^j$ are only defined for $i\geq j$, as time $j$ is the first possible time for the $j$-th item to be selected. If for all $j \in [k]$, we have that the $\rho_i^j$ are all equal in $i$, meaning that $\rho_i^j=\rho_j^j$, then this readily implies from the previous proposition that $\alg_{k,n}= \frac{\sum_{j \in [k]} \rho_j^j}{n} \opt_{\ell,n}$, and that
\begin{equation}
\comp_{k,\ell}(n) \geq \frac{\ell}{k}\frac{\sum_{j \in [k]} \rho_j^j}{n} \opt_{\ell,n},
\end{equation}

For all $j \in [k]$, we want to find the $\epsilon_i^j$ that equalizes all the $\rho_i^j$ across the different $i$. Here, looking at the expression of $\rho_i^j$, finding any meaningful recurrence relationship might seem hopeless. However, remark that the probability of reaching item $i$ while waiting to select item $j$ only depends on the $\epsilon_r^t$ for $t \leq j$, and thus so does $\rho_i^j$. This implies that in order to equalize the $\rho_i^j$ across $i$, it must be done inductively over $j$: first select the $\epsilon_i^1$ such that $\rho_i^j=\rho_1^j$, then select the $\epsilon_i^2$ such that $\rho_i^2=\rho_2^2$, and so on. We show that, this is equivalent to a system of $k$ difference equations in a non-linear transformation of the $\epsilon_i^j$.

\begin{lemma} \label{lemma:recurrence_relation2}
For $b_i^j=\beta_{\ell,n-\ell}(\epsilon_i^j)$, the condition that for all fixed $j$ the $\rho_i^j$ are equalized, is equivalent to the following system of difference equations over the $b_i^j$:
\begin{align} \label{eq:recurrence_relation2}
\Delta[b_i^j]&=-\frac{\ell}{n}\left(\beta_{\ell+1,n-\ell}\circ \beta_{\ell,n-\ell}^{-1}(b_i^j)-\frac{\rho_{j-1}^{j-1}}{\rho_j^{j}} \beta_{\ell+1,n-\ell}\circ \beta_{\ell,n-\ell}^{-1}(b_i^{j-1})\right)+b_j^j,\quad && \text{for $j\geq2$, $i\geq j$}\\
\Delta[b_i^1]&= -\frac{\ell}{n} \left( \beta_{\ell+1,n-\ell}\circ \beta_{\ell,n-\ell}^{-1}(b_i^1)\right)+b_1^1, \quad &&\text{for $i\geq 1$}.
\end{align}
\end{lemma}

\begin{proof}[Sketch of proof]
The most difficult part, is to actually identify the recurrence relationship between the $\alpha_i^j$ and $a_i^j$ that the equality of the $\rho_i^j$ imposes. While it was immediate for $j=1$ as we simply had $\alpha_{i+1}^1=a_i^1$, it is not clear what relationship can be obtained for $j\geq 2$. Fortunately, a simple relation is obtained by incorporating for $j$ the previous constant $\rho_{j-1}^{j-1}$. Indeed, this is equivalent for $j>1$ to $\alpha_{i+1}^j =a_i^j  + (\alpha_i^{j-1}-a_i^{j-1})(\rho_{j-1}^{j-1})/\rho_j^j$. From there, obtaining the recurrence relation is as before based on the properties of the Beta function. For the proof, see \Cref{app:recurrence_relation2}. 
\end{proof}

We define $\theta_{j,\ell}(n) \eqdef \rho_{j-1}^{j-1}/\rho_j^j$. From this recurrence relation, we can prove the existence of the solution to the system of boundary discrete value problem by using the exact same continuity argument as in \Cref{prop:existence}. The proof is however much more technically involved, and requires using several estimates of $b_j^j$ and $\theta_{j,\ell}(n)$ when $n$ grows large.

\begin{proposition} \label{prop:existence2}
There exists some $n_0 \in \mathbb{N}$, such that for $n \geq n_0$, there exist  $k$ increasing sequences $0=\epsilon_{j-1}^{j}< \epsilon_j^j < \dots < \epsilon_n^j =1$ for $j \in [k]$ and $c_{j,\ell}(n)$ such that: for a given $j$ all the $\rho_i^j$ are equal, $b_j^j=c_{j,\ell}(n) \cdot n^{ -  \ell\cdot (((\ell+1)/\ell)^{j}-1)}$ with $c_{j,\ell}(n)$ being bounded between two positive constants independent of $n$.
\end{proposition}

\begin{proof}[Sketch of proof]
The quantities $b_n^j$ are still the composition of continuous functions (yet different functions for each $i \geq j+1$), so the same intermediate value argument can be applied to prove existence. If $\theta_{j,\ell}(n)$ is too big or too small compared to some constants, so is $b_n^j$ compared to $1$. Regarding the exponent in $n$, it can be proved by induction using the relation between $c_{j,\ell}(n)$ and $\theta_{j,\ell}(n)$. The monotonicity comes from the requirement that all the $\rho_i^j$ are equal and thus must be of the same sign. See \cref{app:existence2} for the full proof.
\end{proof}

This proposition suggests that the discrete boundary value problem can be approximated in the limit by the continuous boundary value problem in \Cref{eq:continuousBVP}. By \Cref{thm:lim} and Theorem \ref{thm:competitive} we already have that $\lim_{n\rightarrow \infty} c_{1,\ell}(n) = \ell/\comp_{\ell}$.

The goal is then to find $(\theta_{2,\ell},\dots, \theta_{k,\ell})$ such that this non-linear ODE system admits a solution $b=(b^1,\dots,b^k)$ over $[0,1]$, which will be unique. Note that these constants can be found by sequentially solving the $j$-th ODE and finding the $j$-th relevant constant.
    
Through \Cref{prop:alg_perf2}, solving the discrete boundary value problem for a finite $n$ directly translates to a lower bound on the competitive ratio $\comp_{k,\ell}(n)$. To show that the limiting competitive ratio can also be lower bounded, we must show that the solutions $\theta_j(n)$ to the discrete problem converge, which naturally ends up being the solution to the above continuous boundary value problem. 

\begin{proposition} \label{prop:convergence}
There exists unique $\theta_{2,\ell},\dots, \theta_{k,\ell}$ constants such that the boundary value problem in \Cref{eq:continuousBVP} admits a solution. We also have that $\lim_{n \rightarrow \infty} \theta_{j,\ell}(n)=\theta_{j,\ell}$, and $\theta_{j,\ell} \geq 1$. Moreover, this also implies the convergence of $c_{j,\ell}(n)$ toward a constant $c_{j,\ell}$, and for all $j \geq 2$ the relationship:
\begin{equation}
\theta_{j,\ell}=\frac{(\ell+1)c_{j,\ell}}{\ell(\ell!)^{1/\ell} \cdot c_{j-1,\ell}^{1+1/\ell}}.
\end{equation}
\end{proposition} 

\begin{proof}[Sketch of proof]
The main idea is to couple the convergence of the Euler scheme with the uniform convergence of the drift function, as the difference equations are Euler schemes that use $\beta_{\ell,n-\ell}$ instead of $\gamma_{\ell}$. The sequence $\theta_{j,\ell}(n)$ is bounded, and thus admits at least one accumulation point. If we prove the convergence, on a subsequence if needed, of the discrete solution towards the ODE with $\theta$ an accumulation point of $\theta_{j,\ell}(n)$ then this shows the existence of a constant such that the continuous boundary value problem admits a solution. This constant is then shown to be unique, which proves that there is only a single accumulation point, and the sequence must converge. However, a key technical difficulty is that the Euler method requires Lipschitzness of the drift function, which is not true for $\gamma_{\ell+1} \circ \gamma_{\ell}^{-1}$ over $[0,1]$. Thus we must use refined arguments proving the convergence on $[0,1-\epsilon]$ to then extend the convergence over $[0,1]$. The full proof can be found in \Cref{app:convergence}.
\end{proof}

We now combine the above results to prove Theorem \ref{thm:general_liminf}. Letting $\theta_{1,\ell}=1$ for ease of notation, we have that $\rho_i^j=\rho_j^j= 1/(c_{1,\ell}\cdot \prod_{t \in [k]} \theta_{t,\ell})$, and therefore applying $\liminf$ on the inequality from \Cref{prop:alg_perf2} and using that the $\theta_{j,\ell}$ converge from \Cref{prop:convergence},
\begin{equation} \label{eq:liminf_gen}
\liminf_{n \rightarrow \infty} \comp_{k,\ell}(n) \geq \liminf_{n \rightarrow \infty} \frac{\sum_{j \in [k]} \rho_j^j(n)}{n} = \lim_{n \rightarrow \infty} \frac{\ell}{k} \frac{1}{c_{1,\ell}(n)} \sum_{j \in [k]} \frac{1}{\prod\limits_{t \in [j]} \theta_{t,\ell}(n)}=\frac{\comp_{\ell}}{k}  \sum_{j \in [k]} \frac{1}{\prod\limits_{t \in [j]} \theta_{t,\ell}},
\end{equation}
which concludes the proof.

\subsection{Numerical results for general setting}

\begin{figure}[ht]
    \centering
    \includegraphics[width=0.6 \textwidth]{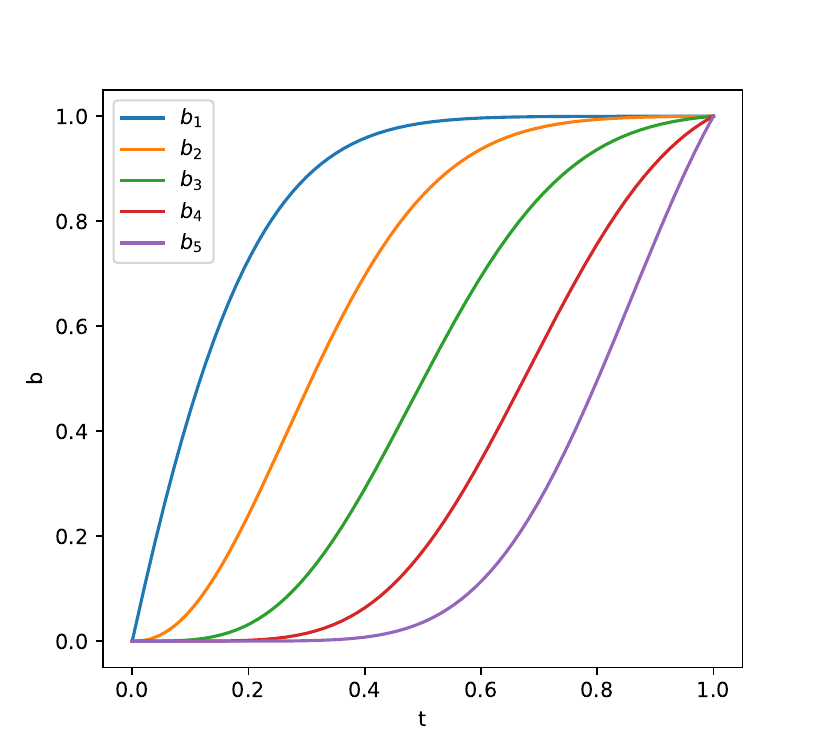}
    \caption{Solution to the continuous boundary value problem for $k=\ell=5$.}
    \label{fig:solution}
\end{figure}

Using numerical optimization, we compute the constants $\theta_{j,\ell}$ and provide in \Cref{tab:c_values_gen} the numerical value of the asymptotic lower bound on $\comp_{k,\ell}(n)$ from Theorem \ref{thm:general_liminf}. We also display the solution to the continuous boundary value problem for $k=\ell=5$ in \Cref{fig:solution}.

 \begin{table}[t]
    \begin{center}
        \begin{tabular}{|c|c|c|c|c|c|}
            \hline
            \diagbox{$k$}{$\ell$} & 1 & 2 & 3 & 4 & 5 \\ \hline
            1 & 0.745 & 0.966 & 0.997 & 0.9998 & 0.999993 \\ 
            2 & 0.486 & 0.829 & 0.964 & 0.995 & 0.9995 \\ 
            3 & 0.332 & 0.645 &  0.864 & 0.964 & 0.993 \\
            4 & 0.24997 & 0.498 & 0.724 & 0.885 & 0.964 \\
            5 & 0.19997 & 0.3998 & 0.596 & 0.772 & 0.898 \\ \hline
        \end{tabular}
        \caption{First digits of $(\comp_{\ell}/k) \sum_{j \in [k]} \prod_{t \in [j]} \theta_t^{-1}$.}\label{tab:c_values_gen}
    \end{center}
\end{table}

We do not have a proof of the asymptotic tightness of \Cref{eq:liminf_gen}, nor do we have a proof that $\comp_{k,\ell}=\liminf_{n \rightarrow \infty} \comp_{k,\ell}(n)$, although we conjecture that both statements are true. The asymptotic tightness is harder to show due to the lack of simple integral characterization, and the same strategy as \Cref{lemma:doubling} cannot be used for a selection budget greater than $1$. Nevertheless, we still show that the infinite-dimensional optimization problem of computing $\comp_{k,\ell}(n)$ is simpler than it appears, and can be reduced to a finite-dimensional optimization by applying the balayage technique from \cite{Kertz1982}. Solving this optimization problem numerically would then provide valid upper bounds on $\comp_{k,\ell}(n)$.

\begin{proposition} \label{prop:reduction2}
The value of $\comp_{k,\ell}(n)$ is attained by a discrete distribution with a support of $2+k(k-1)/2+k(n-k)$ points on $[0,1]$. 
\end{proposition}
See \cref{app:balayage} for the proof. This proposition is actually stronger than a similar result of \cite{Jiang2022} (Lemma $7.2$), who show that for the $(k,k)$ case, using an increasingly finer discretization over values to solve the optimization problem $\comp_{k,k}(n)$ approximates well the optimal value. Here we have shown that not only can this be extended to any general $(k,\ell)$ setting, but mainly that at least one minimum distribution \emph{must} lie in a discretization linear in $n$, and it is unnecessary to make the discretization any finer.

\section{Static thresholds} \label{sec:static}

We now restrict the competitive ratio analysis to the set of static threshold policies. Similarly to previous works  \cite{Arnosti2021}, we allow for random tie breaks when the distributions are discrete. For simplicity, the exposition will use continuous distribution. \\

In the \ac{iid} single item setting, it has been known that the threshold $F^{-1}(1-1/n)$ achieves a competitive ratio of $1-1/e$. A simple alternate proof of this fact was presented in \cite{Correa2019} using the representation of $\be[\max_i X_i]$ as the expectation of $nR(Q)$ for $Q$ distributed according to some distribution, and the Jensen inequality. As we have generalized this result and obtained that $\opt_{\ell,n}$ is equal to $n\be[R(Q)]$ with $Q$ distributed according to $\mathrm{Beta}(\ell,n-\ell)$, we use the same method to prove the following lower bound:

\begin{proposition}\label{prop:static_LB}
The performance of the algorithm that uses static threshold $T=F^{-1}(1-\ell/n)$ is greater than
\begin{equation}
\frac{\sum_{j=1}^k \gamma_j(\ell)}{k}-\frac{1}{n}\left(1-\gamma_j(\ell) - \frac{\ell^{j-1} e^{-\ell}}{(j-1)!}\right) -o\left( \frac{1}{n^2} \right).
\end{equation}
\end{proposition}

For the full proof see \Cref{app:static_LB}. Compared to the proof of $1-1/e$ in \cite{correa2017}, multiple additional algebraic manipulations are necessary. This result is actually even more precise, in the sense that the expected reward of the $j$-th item is up to the error term exactly $\gamma_j(\ell)/\ell$. One aspect of this result that is remarkable, is that the threshold \emph{only depends on $\ell$ and not on $k$}. This is quite surprising as this suggests for the decision maker to target the expected demand of the prophet in order to achieve a good competitive ratio. \\

To obtain an upper bound, we can adapt results from \cite{Arnosti2021} which deals with the $k$ multi-unit static threshold prophet secretary problem. It so happens that their worst-case instance is \ac{iid}. We use the following modified example: Let $F^*$ be the distribution such that $X=1$ with probability $1-1/n^2$, and $X=nW_{k,\ell}$ with probability $1/n^2$ where 
\begin{equation}
W_{k,\ell}= \frac{\ell^2}{k} \frac{\Pr(\mathrm{Poisson}(\ell)<k)}{\Pr(\mathrm{Poisson}(\ell)>k)}.
\end{equation}

This example provides an asymptotically tight upper bound.
\begin{proposition} \label{prop:static_UB}
For $(T^*,p^*)$ the optimal static threshold and random tie-break under $F^*$, we have 
\begin{equation}
\lim_{n \rightarrow \infty} \frac{\be_{X \sim F}\left[ \frac{1}{k}\sum_{i \in [k]} X_{\tau_{i}(T^*)}\right]}{\be_{X\sim F} \left[ \frac{1}{\ell}\sum_{i \in [\ell]} X_{(i)}\right]} =\frac{ \sum_{j=1}^k \gamma_j(\ell)}{k}.
\end{equation}
\end{proposition}

The full proof can be found in \Cref{app:static_UB}. The combination of these two results immediately yields Theorem \ref{thm:static}. 

\begin{remark}
For the special case $k=1$, the same arguments of \Cref{lemma:doubling} can be applied. For any threshold $T$, the expected performance of this threshold over an instance $X_1,\dots,X_n \sim F$ will be greater than for an instance $Y_1,\dots, Y_{2n} \sim \sqrt{F}$, as playing on the first instance with a static threshold is equivalent to playing against $(\max(Y_{2i},Y_{2i+1}))_{i\in[n]}$. Therefore, $\comp_{\ell}^S\eqdef \inf_{n \geq \ell} \comp_{\ell}^S(n)$ and thus $\comp_{\ell}^S=\gamma_1(\ell)=1-e^{-\ell}$.
\end{remark}

A direct consequence of the previous remark and Theorem \ref{th:expo-bound}, is that for $k=1$ there is a positive gap between the worst-case performance of static threshold policies compared to dynamic threshold policies:
\begin{corollary}
For $\ell \in \mathbb{N}$, $\comp_{\ell} >\comp_{\ell}^S $.
\end{corollary}

\begin{figure}[ht]
    \centering
    \includegraphics[width=1 \textwidth]{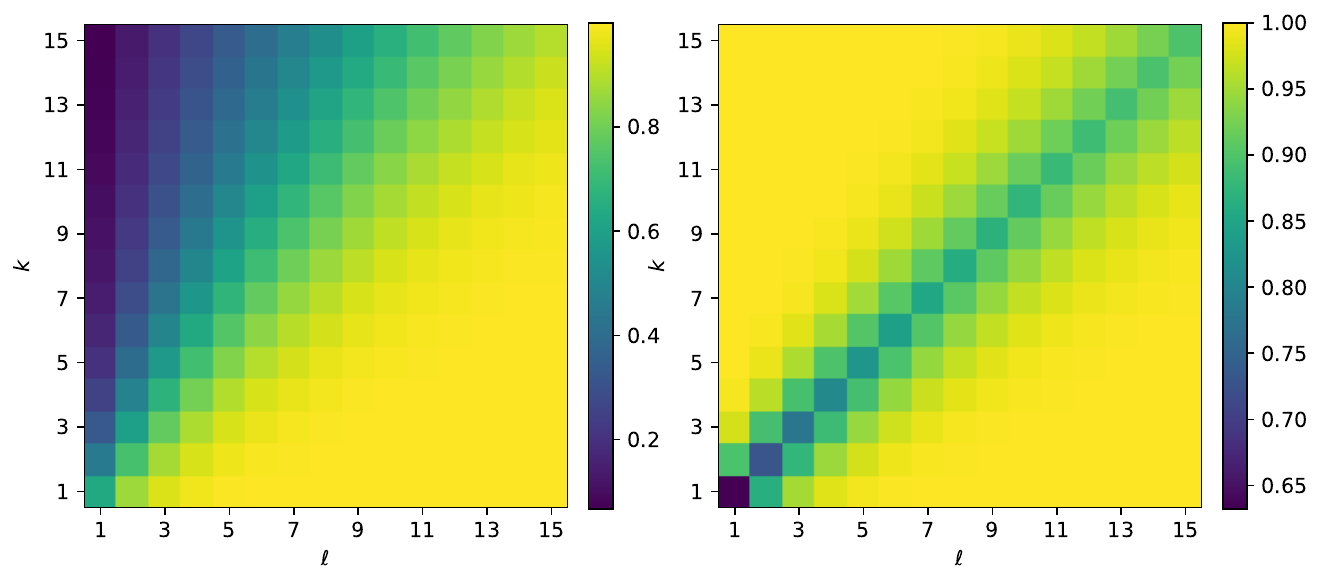}
    \caption{For $k,\ell \in [15]$: on the left $\comp_{k,\ell}^S$, on the right $\comp_{k,\ell}^S \cdot \min(\ell/k,1)$}
    \label{fig:static}
\end{figure}

We compute $\sum_{j=1}^k \gamma_j(\ell)/k$ for different $k$ and $\ell$ and represent them in the left plot of \Cref{fig:static}. We observe that when either $k$ or $\ell$ or both grow large, the competitive ratio goes towards $\min(\ell/k,1)$. In addition, the convergence to $\min(\ell/k,1)$ seems to be the slowest for $k=\ell$ and otherwise exponential, away from $k=\ell$, as can be observed on the right plot of \Cref{fig:static}.

This result is intuitive, and we present a simple explanation for $\ell=1$ and $k$ arbitrary. For the single item \ac{iid} worst case instance, the first maximum is very far from the second maximum, and thus all other order statistics. For this specific distribution, while having a large $k$ allows a greater probability of selecting the actual maximum, all the other selected values will be negligible compared to the maximum. Hence the expected reward of the decision maker will approach $\be[\max_i X_i]$, and the mean reward $\be[\max_i X_i]/k$ which leads to a competitive ratio of order $1/k$.\\

An interesting open question is whether similar guarantees extend to the prophet secretary setting, where distributions are not identical anymore and arrive in random order. In \cite{Arnosti2021}, the $k=\ell$ case is studied, which proves that expected demand policies are not tight for $k \leq 4$, but are tight whenever $k=\ell >4$. The proof of the general $k \neq \ell$ setting presented here relies heavily on the \ac{iid} assumption, but it is likely that the values of the competitive ratio in the prophet secretary setting remain similar to the \ac{iid} setting

\section{Extensions} \label{sec:extension}

\subsection{Selecting order statistics with decreasing distributions}

In this subsection we will only consider the case $k=1$ for exposition purposes, the case $k>1$ can be treated similarly. \\

Up until now, we have assumed as a benchmark $\be[ \sum_{i \in [\ell]} X_{(i)}/\ell]$, which can be reformulated according to our \emph{imperfect prophet} as $\be[X_{(S)}]$ where $S \sim \mathrm{Unif}([\ell])$. Can we say anything when the probability distribution of $S$ is not uniform? We can generalize this benchmark to any decreasing probability mass function $\Pr(S=s)=p_s$ for $s \in \bb{N}$, where $p_1 \geq p_2 \geq \dots \geq  p_{\ell} > p_{\ell+1}=0$ and such that $\sum_{s \in [\ell]} p_s =1$. Remark that this implies $p_1\geq 1/\ell$. Let $\mathbf{p}=(p_1,\dots,p_{\ell}, \dots)$, and $\comp(\mathbf{p})$ be the competitive ratio with the benchmark $\be[X_{(S)}]$ where $S$ is distributed according to $\mathbf{p}$ over $\bb{N}$. Clearly, $\comp_{\ell}=\comp((\mathbf{1}_{\ell},0,\dots)/\ell)$. For simplicity and to avoid convergence issues in $n$, we only allow for finite distributions, i.e. $\sup \{i \mid p_i >0\} <\infty$, but the model could still be defined even for a countably infinite distribution by setting $X_{(s)}=0$ if $s > n$.

\begin{proposition} \label{prop:convex_comb}
For $\mathbf{p}=(p_1,\dots,p_{\ell},\dots) \in \bb{R}_{+}^{\bb{N}}$, with $\mathbf{p}$ non-increasing, $\ell \eqdef \sup \{i \mid p_i >0\} <\infty$ and $\sum_{i \in \bb{N}} p_i=1$, we have 
\begin{equation}
\comp(\mathbf{p}) \geq \frac{1}{p_{1} \cdot c(\mathbf{p})},
\end{equation}
where $c(\mathbf{p})$ is the unique solution in $[1/p_{1}, \infty)$ to the integral equation
\begin{equation*}
1=\int_0^{\infty} \frac{ e^{-\nu} \sum_{s=1}^\ell \frac{p_s-p_{s+1}}{p_1}  \frac{\nu^{s-1}}{(s-1)!}}{c(\mathbf{p}) - \sum_{s \in [\ell]} \frac{p_s-p_{s+1}}{p_1} \cdot s \cdot \gamma_{s+1}(\nu)} d\nu.
\end{equation*}
\end{proposition}

\begin{proof}
The core difference compared to $\comp_{\ell}$, is that $\psi_{\ell,n-\ell}$ is replaced with a mixture of Beta random variables of weights proportional to $p_{s}-p_{s+1}$, which explains the monotonicity condition on the $p_s$. The rest of the proof follows similarly to the previous section. Because the quantile function of mixtures cannot be directly expressed nicely, the final integral equation is messier. Nonetheless, the mixture inherits the monotonicity in $n$ from $\beta_{n,n-s}$, and thus the uniform convergence follows. See the full proof in \Cref{app:convex_comb}.
\end{proof}

\begin{figure} \label{fig:cr_alpha}
    \centering
    \includegraphics[width=0.7\linewidth]{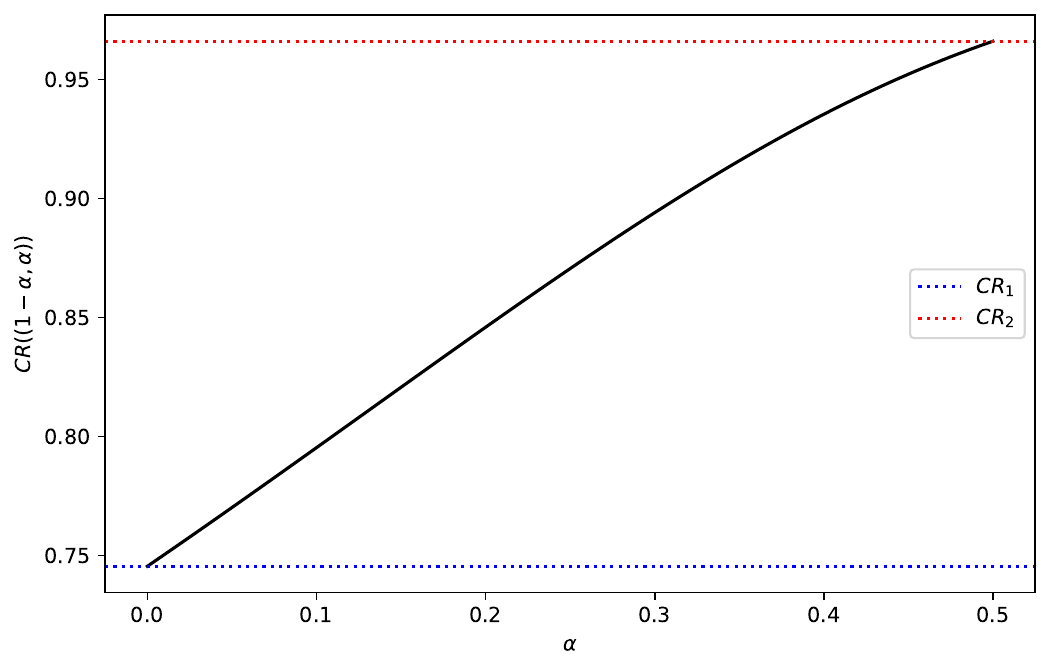}
    \caption{Competitive ratio for $\mathbf{p}=(1-\alpha,\alpha)$, for $\alpha \in [0,1/2]$.}
    \label{fig:enter-label}
\end{figure}

As a special case, when the distribution $\mathbf{p}$ is only supported by $1$ and $2$, we have $\comp_1 \leq \comp(\mathbf{p}) \leq \comp_2$, and $\comp(\mathbf{p})$ interpolates between those two values as a function of $\mathbf{p}$ as can be seen in \Cref{fig:cr_alpha}. This also corresponds in this special case to the comparison against a fractional order statistic.

\subsection{Worst-case for the prophet}

The competitive ratio $\comp_{k,\ell}$ can be interpreted as evaluating the worst-case relative loss of the online agent having less information compared to the prophet. In most settings, the prophet performs always strictly better than the online decision maker. However, here we have a prophet which, while she has more information, can be more limited in taking decisions. Hence, the worst-case distribution \emph{for the prophet} could be of interest. Let 
\begin{equation}
D_{k,\ell} \eqdef  \sup_{n \geq \max(k,\ell)} \sup_{F}  \frac{\sup_{\tau_1<\dots <\tau_k}\be_{X \sim F}\left[ \frac{1}{k}\sum_{i \in [k]} X_{\tau_{i}}\right]}{\be_{X \sim F}\left[ \frac{1}{\ell}\sum_{i \in [\ell]} X_{(i)}\right]}.
\end{equation}
We can exactly compute $D_{k,\ell}$:
\begin{proposition}
For $(k,\ell) \in \bb{N}^2$, we have
\begin{equation}
 D_{k,\ell} = \max \left\{ \frac{\ell}{k}, 1 \right\}.
\end{equation}
\end{proposition}

\begin{proof}
For the upper-bound, if $\ell \leq k$:
\begin{equation*}
\sum_{i=1}^{k} X_{\tau_i} \leq \sum_{i=1}^{\ell} X_{(i)} + (k-\ell) \cdot \frac{1}{\ell} \sum_{i=1}^{\ell} X_{(i)} = \frac{k}{\ell} \sum_{i=1}^{\ell} X_{(i)},
\end{equation*}
which yields $D_{k,\ell} \leq 1$ by taking the expectation. If $k \leq \ell$, then $\sum_{i=1}^{k} X_{\tau_i} \leq \sum_{i=1}^\ell$, hence $D_{k,\ell} \leq \ell/k$. Overall $D_{k,\ell} \leq \max\{ \ell/k,1\}$. For the matching lower bound, take $F$ the distribution such $X=0$ with probability $1-1/n^2$ and $X=1$ otherwise. The probability for at least one $X_i$ to be equal to $1$ is of order $1/n$, and the probability for strictly more than one $X_i$ to be equal to $1$ is of order $1/(2n^2)$. Therefore the chance that a second non-zero variable appears in the sequence is negligible in front of only one appearing. The optimal policy is to select a non-zero value when it appears. Thus $\lim_{n \rightarrow \infty} \be[\sum_{i=1}^k X_{\tau_i} ]/ \be[\sum_{i=1}^\ell X_{(i)}]=1$, and normalizing by $k$ and $\ell$ yields the worst-case instance for the prophet when $k \leq \ell$. For $k \geq \ell$ simply take a constant distribution. 
\end{proof}

\section{Detailed proofs}
\label{sec:proof}

\subsection{Proof of \Cref{prop:alg_perf}} \label{app:alg_perf}

Let $R(q) \eqdef  q \be [X \mid X> F^{-1}(1-q)]$ be the expected reward when rejecting values below the $1-q$ quantile of $F$. Through a change of variable, it can be shown that $R(q)=\int_{0}^q F^{-1}(\theta)d\theta$.

The first step is to express the performance of $\alg_n$ using the fact that the $q_i$ are independently drawn from $\psi_{\ell,n-\ell}$ truncated between $\epsilon_{i-1}$ and $\epsilon_i$. This specific step is the same as in \cite{correa2017}.
The expected probability of not selecting any item up to $i$ is simply $\be [\prod_{j \in [i]} (1-q_j)]$, and using the independence of the $q_i$ it can be expressed as 
\begin{equation*}
\be[\prod_{j \in [i]} (1-q_j)]=\prod_{j \in [i]} \be[(1-q_j)]=\prod_{j \in [i]} \frac{a_j}{\alpha_j}.
\end{equation*}
Then, using that $R(q_i)$ is the expected value of selecting or not item $i$ with threshold $q_i$, we have for $\alg$ the expected performance of the algorithm that
\begin{align*}
\alg&=\sum_{i=1}^n \be[R(q_i)\prod_{j \in [i-1]} (1-q_j)]\\
&=\sum_{i=1}^n \be[R(q_i)]\prod_{j \in [i-1]} \frac{a_j}{\alpha_j}\\
&=\sum_{i=1}^n \int_{\epsilon_{i-1}}^{\epsilon_i} R(q_i)\frac{\psi_{\ell,n-\ell}(q_i)}{\alpha_i}\prod_{j \in [i-1]} \frac{a_j}{\alpha_j}\\
&=\sum_{i=1}^n \rho_i \int_{\epsilon_{i-1}}^{\epsilon_i} R(q_i) \psi_{\ell,n-\ell}(q_i),
\end{align*}
where $\rho_i \eqdef \alpha_i^{-1} \prod_{j \in [i-1]} a_j/\alpha_j$.
Hence, we obtain that 
\begin{align*}
\alg_n &= \sum_{i=1}^n \rho_i \int_{\epsilon_{i-1}}^{\epsilon_i} R(q)\psi_{\ell,n-\ell}(q) dq \\
&\geq \min_{i \in [n]} \rho_i \sum_{i=1}^n \int_{\epsilon_{i-1}}^{\epsilon_i} R(q)\psi_{\ell,n-\ell}(q)dq =\min_{i \in [n]} \rho_i \int_{0}^{1} R(q)\psi_{\ell,n-\ell}(q)dq.
\end{align*}

We now show that this last integral is actually equal to $\opt_{\ell,n}/n$. This was done as a special case in \cite{correa2017}, but for $\ell \geq 2$ this requires the use of the distributions of order statistics. 

We recall the distribution of order statistics: the density of $X_{(i)}$ for $i \in [n]$ is 
\begin{equation*}
\frac{n!}{(n-i)!(i-1)!}f(x)F(x)^{n-i} (1-F(x))^{i-1}.
\end{equation*}
Hence because $\opt_{\ell,n}=\sum_{i=1}^{\ell} \be[X_{(i)}]$ we can express $\opt_{\ell,n}$ using those order statistics distributions. Using the change of variable $q=1-F(t)$ and doing integration by parts, we obtain
\begin{align*}
\opt_{\ell,n} &= \sum_{i \in [\ell]} \int_{0}^{\omega} \frac{n!}{(n-i)!(i-1)!}tf(t)F(t)^{n-i} (1-F(t))^{i-1} dt \\ 
   &=\sum_{i \in [\ell]} \int_{0}^{1} \frac{n!}{(n-i)!(i-1)!}F^{-1}(1-q)(1-q)^{n-i} q^{i-1} dq\\
&=\sum_{i \in [\ell]} \int_{0}^{1} \frac{n!}{(n-i)!(i-1)!}(1-q)^{n-i-1} q^{i-2}( (n-i)q-(i-1)(1-q)) \int_{0}^q F^{-1}(\theta)d\theta dq\\
    &=\sum_{i \in [\ell]} \int_{0}^{1} \frac{n!}{(n-i)!(i-1)!}(1-q)^{n-i-1} q^{i-2}( (n-i)q-(i-1)(1-q)) R(q) dq,
\end{align*}

Exchanging sum and integral, we can observe that the sum is actually telescoping:
\begin{align*}
&\sum_{i \in [\ell]} \frac{n!}{(n-i)!(i-1)!}(1-q)^{n-i-1} q^{i-2}( (n-i)q-(i-1)(1-q)) \\
&= \sum_{i=1}^\ell \frac{n!}{(n-i-1)!(i-1)!} (1-q)^{n-i-1} q^{i-1} - \sum_{i=2}^\ell \frac{n!}{(n-i)!(i-2)!} (1-q)^{n-i} q^{i-2} \\
&= \sum_{i=1}^\ell \frac{n!}{(n-i-1)!(i-1)!} (1-q)^{n-i-1} q^{i-1} - \sum_{i=1}^{\ell-1} \frac{n!}{(n-i-1)!(i-1)!} (1-q)^{n-i-1} q^{i-1}\\
&= \frac{n!}{(n-\ell-1)!(\ell-1)!}(1-q)^{n-\ell-1}q^{\ell-1}\\
&=n \frac{(n-1)!}{(n-\ell-1)!(\ell-1)!}(1-q)^{n-\ell-1}q^{\ell-1}\\
&=n \frac{q^{\ell-1}(1-q)^{n-\ell-1}}{B(\ell,n-\ell)}\\
&=n \psi_{\ell,n-\ell}.
\end{align*}

Therefore, given this algorithm, we can already deduce that for a given $n \in \bb{N}$,
\begin{equation*}
\comp_{\ell}(n)\geq \frac{\min_{i \in [n]} \rho_i}{n}. 
\end{equation*}

\subsection{Proof of \cref{lemma:recurrence_relation}} \label{app:recurrence_relation}

We first recall a property of $\beta_{\ell,n-\ell}$ which can be obtained by integration by parts.
\begin{align*}
\beta_{\ell+1,n-\ell}(z)&=\beta_{\ell,n-\ell}(z)-\frac{z^{\ell}(1-z)^{n-\ell}}{kB(\ell,n-\ell)}\\
\beta_{\ell,n-\ell+1}(z)&=\beta_{\ell,n-\ell}(z)+\frac{z^{\ell}(1-z)^{n-\ell}}{(n-\ell)B(\ell,n-\ell)}.
\end{align*}

First we remark that we can express $\alpha_i$ as $\beta_{\ell,n-\ell}(\epsilon_i)-\beta_{\ell,n-\ell}(\epsilon_{i-1})=b_i-b_{i-1}$. Similarly for $a_i$,
\begin{align*}
a_i&= \int_{\epsilon_{i-1}}^{\epsilon_i} (1-q) \psi_{\ell,n-\ell}(q) dq=   \frac{\int_{\epsilon_{i-1}}^{\epsilon_i} q^{\ell-1}(1-q)^{n+1-\ell} dq}{B(\ell,n-\ell)} \\
&=\frac{B(\ell,n+1-\ell)}{B(\ell,n-\ell)} \frac{\int_{\epsilon_{i-1}}^{\epsilon_i} q^{\ell-1}(1-q)^{n+1-\ell} dq}{B(\ell,n+1-\ell)}\\
&=\frac{\frac{\Gamma(\ell)\Gamma(n+1-\ell)}{\Gamma(n+1)}}{\frac{\Gamma(\ell)\Gamma(n-\ell)}{\Gamma(n)}}(\beta_{\ell,n+1-\ell}(\epsilon_i)-\beta_{\ell,n+1-\ell}(\epsilon_{i-1}))\\
&=\frac{n-\ell}{n}(\beta_{\ell,n+1-\ell}(\epsilon_i)-\beta_{\ell,n+1-\ell}(\epsilon_{i-1})).
\end{align*}
\allowdisplaybreaks

The quantities $\rho_i$ always satisfy a simple recurrence relation, namely that $\rho_{i+1}=a_i \rho_i /\alpha_{i+1}$. So imposing the equality of the $\rho_i$ is equivalent to the relation $\alpha_{i+1}=a_i$.
Now, using this relationship:
\begin{align*}
\alpha_{i+2}&=a_{i+1} \\
\Leftrightarrow \ &\beta_{\ell,n-\ell}(\epsilon_{i+2})-\beta_{\ell,n-\ell}(\epsilon_{i+1})= \frac{n-\ell}{n}(\beta_{\ell,n+1-\ell}(\epsilon_{i+1})-\beta_{\ell,n+1-\ell}(\epsilon_i)) \\
\Leftrightarrow \ &b_{i+2}-b_{i+1}=\frac{n-\ell}{n}\Big( \beta_{\ell,n-\ell}(\epsilon_{i+1})-\beta_{\ell,n-\ell}(\epsilon_i) \\
& + \frac{\epsilon_{i+1}^\ell(1-\epsilon_{i+1})^{n-\ell}}{(n-\ell)B(\ell,n-\ell)} - \frac{\epsilon_{i}^\ell(1-\epsilon_{i})^{n-\ell}}{(n-\ell)B(\ell,n-\ell)}\Big)\\
\Leftrightarrow \ &\Delta[b_{i+1}]=\Delta[b_i]-\Big(\frac{\ell}{n} \beta_{\ell,n-\ell}
(\epsilon_{i+1}) - \frac{\epsilon_{i+1}^\ell(1-\epsilon_{i+1})^{n-\ell}}{nB(\ell,n-\ell)} \\
&- \frac{\ell}{n} \beta_{\ell,n-\ell}
(\epsilon_{i+1}) + \frac{\epsilon_{i+1}^\ell(1-\epsilon_{i+1})^{n-\ell}}{nB(\ell,n-\ell)}\Big) \\
& \Leftrightarrow \ \Delta^2[b_i]=-\frac{\ell}{n}\Delta\left[\beta_{\ell,n-\ell}(\epsilon_i)- \frac{\epsilon_i^\ell (1-\epsilon_i)^{n-\ell}}{\ell B(\ell,n-\ell)}\right]\\
& \Leftrightarrow \ \Delta^{2}[b_i]=-\frac{\ell}{n} \Delta [\beta_{\ell+1,n-\ell}(\epsilon_i)] \\
& \Leftrightarrow \ \Delta^{2}[b_i]=\frac{ \Delta [-\ell\beta_{\ell+1,n-\ell}(\beta_{\ell,n-\ell}^{-1}(b_i))]}{n}.
\end{align*}

By summing those equations, we can hence obtain an explicit recurrence relationship on $b_i$. Indeed, for $i\geq 2$:
\begin{align*}
\sum_{j=0}^{i-2} \Delta^2[b_j] &= \sum_{j=0}^{i-2} \frac{ \Delta [-\ell\beta_{\ell+1,n-\ell}(\beta_{\ell,n-\ell}^{-1}(b_j))]}{n} \\
\Leftrightarrow \ &(b_i-b_{i-1})-(b_1-b_0)=\frac{-\ell(\beta_{\ell+1,n-\ell}(\beta^{-1}_{\ell,n-\ell}(b_i)) - \beta^{-1}_{\ell,n-\ell}(0)) }{n} \\
\Leftrightarrow \ &b_i=b_{i-1} -\frac{\ell}{n} \beta_{\ell+1,n-\ell}\circ \beta^{-1}_{\ell,n-\ell}(b_i) + b_1. 
\end{align*}

\subsection{Proof of \Cref{prop:existence}} \label{app:existence}

Because $\beta_{\ell,n-\ell}(0)=0$ and $\beta_{\ell,n-\ell}(1)=1$, the problem of finding an increasing partition of $\epsilon_i$ which satisfies $\rho_{i+1}=\rho_i$ with $\epsilon_{0}=0$ and $\epsilon_{n}=1$ is equivalent to finding a partition of the $b_i$ which satisfies the recurrence relation in \Cref{eq:recurrence_relation}; this is a specific instance of a discrete boundary value problem. 
We now prove that there exists a $b_1$ such that $b_n=1$, $b_{i+1}\geq b_i$, and it solves the discrete boundary value problem. Remark that these conditions constrain $b_i$ to belong to the interval $[0,1]$: this is crucial to recover from the $b_i$ a valid sequence of $\epsilon_i$ as the domain of $\beta_{\ell,n-\ell}^{-1}$ is $[0,1]$.\\  

First note that for $b_1 \geq \ell/n$ and because $\beta_{\ell+1,n-\ell}(x)\leq 1$, we have that 
\begin{equation*}
b_{i+1}-b_i=-\frac{\ell}{n} \beta_{\ell+1,n-\ell}\circ \beta^{-1}_{\ell,n-\ell} (b_i) +b_1 \geq -\frac{\ell}{n}+\frac{\ell}{n} \geq 0.
\end{equation*}
This implies that any $b_1 \geq \ell/n$ yields an increasing sequence of $b_i$, hence an increasing sequence of $\epsilon_i$.

Second observe that $b_n$ can be expressed as $n-1$ times the composition of the recurrence relation, which is continuous, so $b_n$ is a continuous function of $b_1$. We will show that for $b_1=\ell/n$ and $b_1=(\ell+1)/n$, we end up with $b_n \leq 1$ and $b_n \geq 1$ respectively. Then by intermediate value theorem this proves the existence of a $b_1$ in $[\ell/n,(\ell+1)/n]$ such that $b_n=1$, and $b_{i+1} \geq b_i$ by the first remark because $b_1 \geq \ell/n$. \\

For $b_1\geq (\ell+1)/n$, bounding the difference $b_{i+1}-b_i$ as done above for the monotonicity yields $\Delta[b_i]\geq 1/n$, meaning that $b_n \geq 1$.

The analysis for $b_1 = \ell/n$ uses a simple argument, but needs to be treated carefully. Note that if a $b_1$ is selected such that $b_i$ is increasing and for some $i_0 <n$ we have $b_{i_0}\geq 1$, we can disregard the recurrence after $i_0$ and fix in the algorithm $\epsilon_i=1$ for $i \geq i_0$. This simply corresponds to ignoring any item that comes after $i_0$, and still produces a valid algorithm with competitive ratio lower bounded by $\ell \rho_1 /n= \ell / ( \alpha_1 n) = \ell / ( b_1 n) $. Take $b_1=(\ell-\delta)/n$, with $\delta>0$ small enough such that the sequence remain increasing. Then if the corresponding $b_n$  is greater or equal to $1$, we can recover a valid sequence of $\epsilon_i$ discarding items after some time if necessary. This leads to a lower bound on the competitive ratio of $\ell/(\ell-\delta)>1$. However by considering a constant distribution, it must be that the competitive ratio is smaller than $1$ in the worst case. Therefore it must be that $b_n<1$. By taking the limit as $\delta$ goes to $0$, we have $b_n \leq 1$ for $b_1=\ell/n$. \\

\subsection{Proof of \Cref{lemma:doubling}} \label{app:doubling}
For the detailed proof of $\be[X_{\tau_X}] \geq \be[Y_{\tau_Y}] $ we defer to Appendix A.1 of \cite{Allen2021} (note that their notation of OPT corresponds to the optimal online algorithm).

We recall that the distribution function of the ${\ell}$-th order statistic for a random variable with distribution $F$ is simply $\beta_{n+1-{\ell},{\ell}}\circ F$. If we show that $\beta_{n+1-{\ell},{\ell}}\circ F \geq \beta_{2n+1-{\ell},{\ell}}\circ \sqrt{F}$, then $Y_{({\ell})}$ stochastically dominates $X_{({\ell})}$, which implies the desired result of $\be[Y_{({\ell})}]\geq \be[X_{({\ell})}]$ (this is immediate from writing the expectation of a positive random variable as $\int_0^{\infty} 1-F$).

Let us look at the function $h(x) \eqdef \beta_{n+1-{\ell},{\ell}}(x)-\beta_{2n+1-{\ell},{\ell}}(\sqrt{x})$ for $x \in [0,1]$. Clearly if we show that this function is non-negative over $[0,1]$, then so is $h \circ F$ over $\bb{R}^+$. Let us consider the derivative of $h$ using the density of the beta distribution:
\begin{equation*}
\frac{dh(x)}{dx}= \frac{x^{n-{\ell}}(1-x)^{{\ell}-1}}{B(n+1-{\ell},{\ell})} -\frac{\sqrt{x}^{2n-{\ell}-1}(1-\sqrt{x})^{{\ell}-1}}{2B(2n+1-{\ell},{\ell})}.
\end{equation*}
For simplicity, let $\sqrt{x}=y\in [0,1]$ then we can rewrite the above derivative as 
\begin{align*}
\frac{dh(x)}{dx}&= \frac{y^{2n-2{\ell}}(1-y^2)^{{\ell}-1}}{B(n+1-{\ell},{\ell})} -\frac{y^{2n-{\ell}-1}(1-y)^{{\ell}-1}}{2B(2n+1-{\ell},{\ell})}\\
&=\frac{y^{2n-2{\ell}}(1-y)^{{\ell}-1}(1+y)^{{\ell}-1}}{B(n+1-{\ell},{\ell})} -\frac{y^{2n-{\ell}-1}(1-y)^{{\ell}-1}}{2B(2n+1-{\ell},{\ell})} \\
&=\frac{y^{2n-{\ell}-1}(1-y)^{{\ell}-1}}{B(n+1-{\ell},n)}\left( \frac{(1+y)^{{\ell}-1}}{y^{{\ell}-1}}- \frac{B(n+1-{\ell},{\ell})}{2B(2n+1-{\ell},{\ell})} \right).
\end{align*}
Hence over $[0,1]$, $h'(x)\geq 0$ is equivalent to 
\begin{equation*}
\left(1+\frac{1}{y}\right)^{{\ell}-1} \geq \frac{B(n+1-{\ell},{\ell})}{2B(2n+1-{\ell},{\ell})}.
\end{equation*}
The function on the left is equal to $\infty$ at $y=0$, and is decreasing in $y$: this implies that the derivative is positive until some $y_0=\sqrt{x_0}$, and then possibly negative. If a function is increasing then decreasing over an interval, then its minimum is at the endpoints of the interval. And we have that $h(0)=h(1)=0$, so $h(x)\geq \min_{x' \in [0,1]} h(x')=0$, which concludes the proof. 

\subsection{Proof of \Cref{lemma:unif_conv}} \label{app:unif_conv}

We first recall a limiting relationship between incomplete Gamma and Beta functions. We have for $X_n \sim \mathrm{Beta}(\ell,n)$ that the random variable $Y_n=nX_n$ converges in law to $\mathrm{Gamma}(\ell)$ as $n$ goes to $\infty$.

In particular this means that 
\begin{equation*}
\lim_{n \rightarrow \infty}\beta_{\ell,n-\ell}\left(\frac{z}{n}\right)=\lim_{n \rightarrow \infty}\Pr(X_n\leq \frac{z}{n})=\lim_{n \rightarrow \infty}\Pr( n X_n \leq z)=\gamma_{\ell}(x),
\end{equation*}
where the last equality stems from the equivalence between limit in distribution and point-wise limit of the distribution function (we also indirectly use that $\ell X_n$ goes to $0$ for $k$ fixed).

Let us show the point-wise convergence. The inverse of $\beta_{{\ell},n-{\ell}}(\cdot/n)$ is simply $n \beta_{{\ell},n-{\ell}}^{-1}$. Because $\beta_{{\ell},n-{\ell}}$ is continuous, the point-wise limit of the inverse is the inverse of the point-wise limit. Hence $\lim_{n\rightarrow \infty} \beta_{{\ell},n-{\ell}}^{-1}=\gamma_{\ell}^{-1}$. Rewriting the main function as $-{\ell} \beta_{\ell+1,n-{\ell}}\circ \text{id}/n \circ n \cdot \text{id} \circ \beta_{{\ell},n-{\ell}}^{-1}$ we have by composition that this converges to $-{\ell} \gamma_{\ell+1}\circ \beta_{{\ell}}^{-1}$. 

Now for the uniform convergence. First, note that because of the expansion formula, we have that $\beta_{{\ell},n+1-{\ell}} \geq \beta_{{\ell},n-{\ell}}$ which means that $\beta_{{\ell},n+1-{\ell}}^{-1} \leq \beta_{{\ell},n-{\ell}}^{-1}$ (interpolating over $n$ if necessary) so $\beta_{{\ell},n-{\ell}}^{-1}$ decreasing in $n$, and $\beta_{\ell+1,n+1-{\ell}} \geq \beta_{{\ell},n-{\ell}}$ so $\beta_{\ell+1,n+1-{\ell}}^{-1}$ increasing in $n$. This implies that $\beta_{\ell+1,n-{\ell}}\circ \beta_{{\ell},n-{\ell}}^{-1}$ is decreasing in $n$, and finally $-{\ell} \beta_{\ell+1,n-{\ell}} \circ \beta_{{\ell},n-{\ell}}^{-1}$ is increasing in $n$. Because this function is continuous and takes values in $[0,1]$ which is compact, we can apply Dini's Theorem which guarantees uniform convergence. 

\begin{remark}
Note that we can similarly express this ODE using $\nu=\gamma_{\ell}^{-1}(b) \in [0,\infty)$ to avoid the use of an inverse functions:
\begin{equation*}
\frac{d^2}{dt^2}(\gamma_{\ell}(\nu))= -{\ell} \frac{d}{dt} ( \gamma_{\ell+1}(\nu)).
\end{equation*}
It is not immediately clear what $\nu$ represents compared to the $\epsilon_i$. Indeed, we had that $\beta_{{\ell},n-{\ell}}^{-1}(b_i)=\epsilon_i$, but $\nu$ cannot directly translate into $\epsilon_i$ as $\nu \in [0,\infty)$ whereas $\epsilon_i \in [0,1]$. Actually, we can definite $\nu_i=n\epsilon_i$, which implies that 
\begin{equation*}
\Delta^2 [ \beta_{{\ell},n-{\ell}}(\frac{\nu_i}{n})]=\frac{\Delta[-{\ell}\beta_{\ell+1,n}(\frac{\nu_i}{n})]}{n},
\end{equation*}
which converges using the exact same limiting argument for the pointwise limit as above to the second order ODE which $\nu(t)$ obeys. So $\nu$ is the limit of $n \epsilon_i$. This is to put in perspective to the limit in \cite{correa2017} where the ODE concerned used $y_i=(1-\epsilon_i)^{n-1}$.  
\end{remark}

\subsection{Proof of Theorem \ref{thm:lim}} \label{app:lim}

 Let $\xi_{{\ell},n}(x)=-{\ell} \beta_{\ell+1,n-{\ell}}\circ \beta_{{\ell},n-{\ell}}^{-1} (x)$ and $\xi_{\ell}(x)=-\ell \gamma_{\ell+1}\circ \gamma_{\ell}^{-1}(x)$. First 

\begin{equation*}
b_{i+1}-b_i=\frac{\xi_{{\ell}} (b_i)+c_{1,\ell}}{n}+\frac{c_{1,\ell}(n)-c_{1,\ell}}{n}+\frac{\xi_{{\ell},n}(b_i)-\xi_{\ell}(b_i)}{n}
\end{equation*}
Now let us divide both side by $\xi_{{\ell}}(b_i)+c_{1,\ell}$ (which is strictly positive as $c_{1,\ell}$ must be strictly greater than ${\ell}$)
\begin{equation*}
\frac{c_{1,\ell}(n)-c_{1,\ell}}{n(\xi_{\ell}(b_i)+c_{1,\ell})}=-\frac{1}{n}+\frac{b_{i+1}-b_i}{\xi_{\ell}(b_i)+c_{1,\ell}}-\frac{\xi_{{\ell},n}(b_i)-\xi_{\ell}(b_i)}{n(\xi_{\ell}(b_i)+c_{1,\ell})}
\end{equation*}

Remark also that for $x \in [0,1]$, $x \mapsto -{\ell}x+\zeta_{\ell}(x)+c_{1,\ell}$ is bounded between $c_{1,\ell}-{\ell}>0$ and $c_{1,\ell}$, therefore,

\begin{equation*}
\left \vert \sum_{i=0}^{n-1} \frac{c_{1,\ell}(n)-c_{1,\ell}}{n(\xi_{\ell}(b_i)+c_{1,\ell})}  \right \vert \geq \delta \vert c_{1,\ell}(n)-c_{1,\ell}\vert, \quad \delta >0.
\end{equation*}

Now using the previous equation,
\begin{align*}
\left \vert \sum_{i=0}^{n-1} \frac{c_{1,\ell}(n)-c_{1,\ell}}{n(\xi_{\ell}(b_i)+c_{1,\ell})}  \right \vert  &\leq \left \vert -1 + \sum_{i=0}^{n-1} \frac{b_{i+1}-b_i}{\xi_{\ell}(b_i)+c_{1,\ell}}\right \vert + \frac{1}{n(c_{1,\ell}-{\ell})}\sum_{i=0}^{n-1} \vert \xi_{{\ell},n}(b_i)-\xi_{\ell}(b_i) \vert \\
&\leq \left \vert -1 + \sum_{i=0}^{n-1} \frac{b_{i+1}-b_i}{\xi_{\ell}(b_i)+c_{1,\ell}}\right \vert + \frac{1}{(c_{1,\ell}-{\ell})} \Vert \xi_{{\ell},n}-\xi_{\ell} \Vert_{\infty}.
\end{align*}
The last term goes to $0$ by uniform convergence, and the first one by Riemann sum and the integral equation condition, as due to $c_{1,\ell}(n)$ being bounded independently of $n$ we have $b_{i+1}-b_i=O(1/n)$.

All in all $\lim_{n \rightarrow \infty} c_{1,\ell}(n)=c_{1,\ell}$.

\subsection{Proof of \Cref{prop:convex_comb}} \label{app:convex_comb}

For this proof, we will only detail the parts which need special care compared to the original setting. Most arguments follow through identically.

First we will show that
\begin{equation*}
\frac{1}{p_1} \be[X_{(S)}] = n \be_{q \sim \varphi_{n,\mathbf{p}}}[R(q)],
\end{equation*}
with $q$ a mixture of beta random variables with distribution
\begin{equation*}
\varphi_{n,\mathbf{p}}=\sum_{s \in [\ell]} \frac{p_s-p_{s+1}}{p_1} \cdot \beta_{s,n-s},
\end{equation*}
where $p_{\ell+1}=0$. This is a proper mixture as $\sum_{s \in [\ell]} (p_s-p_{s+1})/p_{1}=p_{1}/p_{1}=1$. This result is immediate from following the final computations of \Cref{prop:alg_perf} and including the $p_i$:

\begin{align*}
&\sum_{s \in [\ell]} \frac{p_s}{p_1} \frac{n!}{(n-s)!(s-1)!}(1-q)^{n-s-1} q^{s-2}( (n-s)q-(s-1)(1-q)) \\
&= \sum_{s=1}^\ell \frac{p_s}{p_1} \frac{n!}{(n-s-1)!(s-1)!} (1-q)^{n-s-1} q^{s-1} - \sum_{s=2}^\ell \frac{p_s}{p_1}\frac{n!}{(n-s)!(s-2)!} (1-q)^{n-i} q^{s-2} \\
&= \sum_{s=1}^\ell \frac{p_s}{p_1} \frac{n!}{(n-s-1)!(s-1)!} (1-q)^{n-s-1} q^{s-1} - \sum_{s=1}^{\ell-1} \frac{p_{s+1}}{p_1} \frac{n!}{(n-s-1)!(s-1)!} (1-q)^{n-s-1} q^{s-1}\\
&= \sum_{s=1}^\ell \frac{p_s-p_{s+1}}{p_1} \frac{n!}{(n-s-1)!(s-1)!} (1-q)^{n-s-1} q^{s-1}\\
&= n  \sum_{s=1}^\ell \frac{p_s-p_{s+1}}{p_1} \cdot \psi_{s,n-s} = n \varphi'_{n,\mathbf{p}},\\
\end{align*}
as we have $\beta_{s,n-s}'=\psi_{s,n-s}$. The condition $p_s \geq p_{s+1}$ is clear from the above computation.\\ 

We now derive the new difference equation. First, $\alpha_i=\varphi_{n,\mathbf{p}}(\epsilon_i)-\varphi_{n,\mathbf{p}}(\epsilon_{i-1})$, and second
\begin{align*}
a_i&= \int_{\epsilon_{i-1}}^{\epsilon_i} (1-q) \varphi_{n,\mathbf{p}}(q) dq \\
&= \sum_{s=1}^{\ell} \frac{p_s-p_{s+1}}{p_1}\int_{\epsilon_{i-1}}^{\epsilon_i} (1-q) \psi_{s,n-s}(q) dq \\
&=\sum_{i=1}^{\ell} \frac{p_s-p_{s+1}}{p_1} \frac{n-s}{n} ( \beta_{s,n+1-s}(\epsilon_i) - \beta_{s,n+1-s}(\epsilon_{i-1}) \\
\end{align*}
Let $b_i=\varphi_{n,\mathbf{p}}(\epsilon_i)$. Because $\beta_{i,n-i}$ is an increasing bijection from $[0,1]$ to $[0,1]$, so is the mixture $\varphi_{n,\mathbf{p}}$, therefore $\varphi_{n,\mathbf{p}}^{-1}$ is well defined, and the boudnary conditions on $\epsilon_i$ translate to the same boundary conditions on $b_i$. The equation $\alpha_{i+2}=a_{i+1}$ which equalizes the $\rho_i$ yields 
\begin{align*}
 \Delta b_{i+1}& =\sum_{i=1}^{\ell} \frac{p_s-p_{s+1}}{p_1} \frac{n-i}{n} ( \beta_{s,n+1-s}(\epsilon_{i+1}) - \beta_{s,n+1-s}(\epsilon_{i})\\
 \Leftrightarrow \Delta b_{i+1} &= \sum_{i=1}^{\ell} \frac{p_s-p_{s+1}}{p_1} \frac{n-s}{n} \Big(\beta_{s,n-s}(\epsilon_{i+1})-\beta_{s,n-s}(\epsilon_i) + \frac{\epsilon_{i+1}^s(1-\epsilon_{i+1})^{n-s}}{(n-s)B(s,n-s)} - \frac{\epsilon_{i}^s(1-\epsilon_{i})^{n-s}}{(n-s)B(s,n-s)}\Big) \\
 \Leftrightarrow \Delta b_{i+1}&= \Delta \varphi_{n,\mathbf{p}}(\epsilon_i) - \sum_{s=1}^{\ell} \frac{p_s-p_{s+1}}{p_1} \frac{s}{n} \Delta \left[ \beta_{s+1,n-s}(\epsilon_i) \right]\\
 \Leftrightarrow \Delta^2 b_i &= - \frac{1}{n} \Delta\left[\sum_{s=1}^\ell \frac{p_s -p_{s+1}}{p_1} \cdot s \cdot \beta_{s+1,n-s}(\varphi_{n,\mathbf{p}}^{-1}(b_i)) \right].
\end{align*}
 This function, and the inverse of $\varphi_{n,\mathbf{p}}$ as a mixture inherit the monotonicity properties of $\beta_{s,n-s}$ in $n$, thus the point-wise convergence towards a mixture of $\mathrm{Gamma}(s)$ and its uniform convergence can be proven all the same. Let $\phi_{\mathbf{p}}$ be the rescaled limit mixture:
\begin{equation*}
\phi_{\mathbf{p}}=\sum_{s=1}^{\ell} \frac{p_s-p_{s+1}}{p_1} \gamma_s.
\end{equation*}
The difference equation yields in the limit as $n \rightarrow \infty$ the differential equation
\begin{equation*}
b'(t)= c - \sum_{s=1}^{\ell} \frac{p_s-p_{s+1}}{p_1} \cdot s \cdot \gamma_{s+1}\circ \phi_{\mathbf{p}}^{-1}(b(t)).
\end{equation*}
The rest of the proof is identical and uses the same arguments as for the case $\mathbf{p}=\mathbf{1}_{\ell}/\ell$. Therefore the $c(\mathbf{p})$ which satisfies the integral equation will yield the lower bound on the competitive ratio of $1/(p_{\ell} \cdot c(\mathbf{p}))$.

\subsection{Proof of \Cref{prop:tightness}} \label{app:tight}

The skeleton of the proof is the same as in Section $3.5$ of \cite{Allen2021}, so we defer to it for some technical details and focus here on the main differences.

We know from \cite{Allen2021} (Equation $23$) that the optimal policy verifies 
\begin{equation}
r'(t)= \int_{r(t)}^{\infty} \log(F(u)) du, \quad  r(1)=0.  
\end{equation}

We will verify that the $r(t)$ proposed above is indeed the optimal policy. For $r(t) \in [q,1]$:
\begin{align*}
\int_{r(t)}^{\infty} \log(F(u))du &= \int_{r(t)}^{r(q)}\log(F(u))du+ \int_{r(q)}^\infty \log(F(u))du \\
&= - \int_q^t \log(F(r(u)) r'(u) du + (H-r(q)) \log(p) \\
&= - \int_q^t \frac{-\gamma_{\ell}^{-1}(1-y(u))}{y'(u)} du + \frac{1}{y'(q)}\\
&= - \int_q^t \frac{y''(u)}{(y'(u))^2} du +\frac{1}{y'(q)}\\
&= \frac{1}{y'(t)} = r'(t),
\end{align*}
where we used that $y''=-\gamma_{\ell}^{-1}(1-y) \cdot y'$. And it is optimal over $[0,q]$ as $r(t)=H$, and $F_q$ has an atom at $H$, which is also the highest value.   \\

Let us now compare $\be[X_{\tau}]$ to $\opt_{\ell,n}$. The probability of a variable being greater than $H$ and arriving in $[0,q]$ is $q \cdot (1-F^{1/n}_q(H))$, so the probability that there is at least one variable greater than $H$ arriving in $[0,q]$ is 
\begin{equation*}
1-(1-q(1-F^{1/n}_q(H)))^n=1-\left(1+q \frac{\log(F_q(H))}{n} +o(n^{-1})\right)^n \xrightarrow[n \rightarrow \infty]{} 1-\exp(q \log(F_q(H))) = 1-p^q. 
\end{equation*}
Therefore 
\begin{equation*}
\alg = (1-p^q) H - p^q \int_q^1 \frac{1}{y'(t)} dt \xrightarrow[q \rightarrow 0]{} - \int_0^1 \frac{1}{y'(t)} dt,
\end{equation*}
where the last inequality follows as when $q \rightarrow 0$ then $p \rightarrow 1$, $H \log(p) \rightarrow 1/y'(0)$, and $(1-p^q)H \approx - \log(1-(1-p^q))H= -q \log(p) H \rightarrow 0$.

Let us now compute $ \lim_{n \rightarrow \infty} \opt_{\ell,n}$. 
\begin{align*}
 & \opt_{\ell,n} = \sum_{i=1}^{\ell} \int_0^{\infty} 1 -  \sum_{j=0}^i \binom{n}{j} (F^{1/n}(u))^{n-j} (1-F^{1/n}(u))^{j} du\\
&\xrightarrow[n \rightarrow \infty]{} \sum_{i=1}^{\ell} \int_0^{\infty} 1-   \sum_{j=0}^{i-1} \frac{(-1)^{j}}{j!} F(u) \cdot  \log^j(F(u)) du \\
&= \int_0^{r(q)} \ell - \sum_{i=1}^{\ell} \sum_{j=0}^{i-1} \frac{(-1)^{j}}{j!} F(u) \cdot  \log^j(F(u)) du + \int_{r(q)}^{\infty} \ell -\sum_{i=1}^{\ell} \sum_{j=0}^{i-1} \frac{(-1)^{j}}{j!} F(u) \cdot  \log^j(F(u)) du\\
&= - \int_q^{1} \left(\ell - \sum_{i=1}^{\ell} \sum_{j=0}^{i-1} \frac{(-1)^{j}}{j!} F(r(t)) \cdot  \log^j(F(r(t))) \right) \cdot r'(t) dt + \left( \ell -\sum_{i=1}^{\ell} \sum_{j=0}^{i-1} \frac{(-1)^{j}}{j!} p \cdot  \log^j(p) \right)(H-r(q)) \\
&= - \int_q^{1} \frac{\ell - \sum_{i=1}^{\ell} \sum_{j=0}^{i-1} \frac{(-1)^{j}}{j!} F(r(t)) \cdot  \log^j(F(r(t)))}{y'(t)} dt + \left( \ell -\sum_{i=1}^{\ell} \sum_{j=0}^{i-1} \frac{(-1)^{j}}{j!} p \cdot  \log^j(p) \right)(H-r(q)) \\
&= -  \int_q^{1} \frac{\ell - \exp(- \gamma_{\ell}^{-1}(1-y(t))) \sum_{i=1}^{\ell} \sum_{j=0}^{i-1} \frac{\left(\gamma_{\ell}^{-1}(1-y(t))\right)^j}{j!} }{y'(t)} dt + \left( \ell -\sum_{i=1}^{\ell} \sum_{j=0}^{i-1} \frac{(-1)^{j}}{j!} p \cdot  \log^j(p) \right)(H-r(q)) 
\end{align*}

We now compute the limit as $q\rightarrow 0$ of the right hand term.
\begin{align*}
\left( \ell -\sum_{i=1}^{\ell} \sum_{j=0}^{i-1} \frac{(-1)^{j}}{j!} p \cdot  \log^j(p) \right)(H-r(q)) &= \left ( \ell - \ell p + (\ell-1) p \log(p) + o(\log(p) \right) (H-r(q))\\
&= \left( - \ell \log(p) + (\ell-1) p \log(p) +o(\log(p)) \right) (H-r(q))\\
&\xrightarrow[q \rightarrow 0]{} -\frac{1}{y'(0)}.
\end{align*}

Before computing the limit, let us re-arrange the integrand of the left-hand term. Recall that $\gamma_{\ell}(x)=1- \exp(-x) \sum_{j=0}^{\ell-1} x^j/j!$.
\begin{align*}
&1-\frac{\exp(-\gamma_{\ell}^{-1}(1-y(t)))}{\ell} \sum_{i=1}^{\ell} \sum_{j=0}^{i-1} \frac{\left(\gamma_{\ell}^{-1}(1-y(t))\right)^j}{j!} =  1-\frac{\exp(-\gamma_{\ell}^{-1}(1-y(t)))}{\ell}\sum_{j=0}^{\ell-1}  \sum_{i=j+1}^{\ell}  \frac{\left(\gamma_{\ell}^{-1}(1-y(t))\right)^j}{j!}\\
&=1-\frac{\exp(-\gamma_{\ell}^{-1}(1-y(t)))}{\ell}\sum_{j=0}^{\ell-1}  (\ell -j) \frac{\left(\gamma_{\ell}^{-1}(1-y(t))\right)^j}{j!} \\
&= \gamma_{\ell+1}\circ \gamma_{\ell}^{-1}(1-y(t)) + \frac{\exp(-\gamma_{\ell}^{-1}(1-y(t)))}{\ell} \left(\frac{\left(\gamma_{\ell}^{-1}(1-y(t))\right)^{\ell}}{(\ell-1)!} + \sum_{j=0}^{\ell-1} j \cdot  \frac{\left(\gamma_{\ell}^{-1}(1-y(t))\right)^j}{j!} \right)\\
&= \gamma_{\ell+1}\circ \gamma_{\ell}^{-1}(1-y(t)) + \frac{\exp(-\gamma_{\ell}^{-1}(1-y(t)))}{\ell}  \sum_{j=1}^{\ell}   \frac{\left(\gamma_{\ell}^{-1}(1-y(t))\right)^j}{(j-1)!} \\
&= \gamma_{\ell+1}\circ \gamma_{\ell}^{-1}(1-y(t)) + \frac{\exp(-\gamma_{\ell}^{-1}(1-y(t)))}{\ell} \gamma_{\ell}^{-1}(1-y(t)) \sum_{j=0}^{\ell-1}   \frac{\left(\gamma_{\ell}^{-1}(1-y(t))\right)^j}{j!}\\
&= \gamma_{\ell+1}\circ \gamma_{\ell}^{-1}(1-y(t))  + \frac{\gamma_{\ell}^{-1}(1-y(t))}{\ell} \left( 1- \gamma_{\ell} \circ \gamma_{\ell}^{-1} (1-y(t))\right)\\
&\gamma_{\ell+1}\circ \gamma_{\ell}^{-1}(1-y(t))  + \frac{\gamma_{\ell}^{-1}(1-y(t))}{\ell} y(t).
\end{align*}

Therefore in the limit, 
\begin{equation*}
\lim_{q \rightarrow 0} \lim_{n \rightarrow \infty} \frac{X_{\tau}}{\opt_{\ell,n}} = \frac{y'(0) \int_0^1 \frac{dt}{y'(t)}}{1+  \ell \cdot y'(0) \int_q^{1} \frac{\gamma_{\ell+1}\circ \gamma_{\ell}^{-1}(1-y(t))  + \frac{\gamma_{\ell}^{-1}(1-y(t))}{\ell} y(t)}{y'(t)} dt}.
\end{equation*}
We want to show that this quantity is equal to $1/c_{\ell}$. This is equivalent, after re-arranging the terms and using that $y'(0)=-c_{\ell}$ to
\begin{equation*}
\int_0^1 \frac{1}{y'(t)} \left( -c_{\ell} -\frac{y'}{c_{\ell}}+ \ell \cdot \gamma_{\ell+1} \circ \gamma_{\ell}^{-1}(1-y(t))  + \gamma_{\ell}^{-1}(1-y(t)) \cdot  y(t)\right) dt =0.
\end{equation*}
Using that $y'(t)=\gamma_{\ell+1} \circ \gamma_{\ell}^{-1}(1-y(t)) -c_{\ell}$, this is equivalent to 
\begin{equation*}
1-\frac{1}{c_{\ell}}= \int_0^1  \frac{-\gamma_{\ell}^{-1}(1-y(t)) \cdot  y(t)}{y'(t)} dt.
\end{equation*}
We now apply the change of variable $y=y(t)$ and $u=\gamma_{\ell}^{-1}(1-y)$ to the integral:
\begin{equation*}
\int_0^1  \frac{-\gamma_{\ell}^{-1}(1-y(t)) \cdot  y(t)}{y'(t)} dt =\int_0^{\infty} \frac{  u \cdot  (1- \gamma_{\ell}(u)) \cdot  \gamma_{\ell}'(u)}{(c_{\ell} - \ell \gamma_{\ell+1}(u))^2} du.
\end{equation*}
Finally, remark that $ u \cdot \gamma_{\ell}'(u)/(c_{\ell}- \ell \gamma_{\ell+1}(u))^2$ is the derivative of $1/(c_{\ell}-\ell \gamma_{\ell+1}(u))$. Integrating by parts, we get 
\begin{equation*}
\int_0^{\infty} \frac{  u \cdot  (1- \gamma_{\ell}(u)) \cdot  \gamma_{\ell}'(u)}{(c_{\ell} - \ell \gamma_{\ell+1}(u))^2} du = \left[ \frac{1-\gamma_{\ell}(u)}{c_{\ell}- \ell \gamma_{\ell+1}(u)} \right]_0^{\infty} + \int_0^{\infty} \frac{\gamma_{\ell}'(u)}{c_{\ell}- \ell \gamma_{\ell+1}(u)} du =-\frac{1}{c_{\ell}} +1,
\end{equation*}
where we used in the last equality the integral characterization of $c_{\ell}$. This concludes the proof.

\subsection{Proof of \Cref{prop:alg_perf2}} \label{app:alg_perf2}

First let us consider the performance of the algorithm, conditionally on the $q_i^j$ being already drawn. Basically, if while waiting to select the $j$-th item the algorithm arrives at step $i$, then the expected reward received taking into account the probability of actually selecting the item is $R(q_i^j)$. However, the probability of arriving at step $i$ while waiting for item $j$ is more complicated to express. Indeed, first the $j-1$ must be selected before time $i$, and then no item must be selected until $i$ while waiting for $j$. Hence the expected reward at step $i$, when selecting the $j$-the item, can be expressed as

\begin{equation*}
\be[R(q^j_i)] \sum_{t_{j-1}=j-1}^{i-1} \sum_{t_{j-2}=j-2}^{t_{j-1}-1} \dots \sum_{t_1=1}^{t_2-1} \left[ \prod_{s=1}^{j-1} \left(1-\frac{a_{t_s}^s}{\alpha_{t_s}^s}\right) \prod_{r_1=1}^{t_1-1} \frac{a^1_{r_1}}{\alpha^1_{r_1}} \dots \prod_{r_j=t_{j-1}+1}^{i-1} \frac{a^j_{r_j}}{\alpha^j_{r_j}} \right],
\end{equation*}
where the $t_m$ correspond to the time when item $m$ is selected, and the sums consider all possible times of selection. While this equation seems complicated, it is merely due to the fact that the thresholds depend on $i$ and $j$, and all possible sequences of selection must be considered. 

The proposition can then be easily obtained by re-using the fact proved in \Cref{prop:alg_perf} that $\opt_{\ell,n}=\be_{q \sim \psi_{\ell,n-\ell}}[R(q)]$ and that for any $j \in [k]$, the $(\epsilon_i^j)_{i \in \{j-1,\dots,n\}}$ are constructed to form a partition of $[0,1]$.

\subsection{Proof of \Cref{lemma:recurrence_relation2}} \label{app:recurrence_relation2}

Let us write down a recursion formula in $i$ on $\rho_i^j$ that might depend on previous values of $j$. We have already dealt with the case $j=1$ when only one item could be selected by the decision maker. As such, we will assume that $j>1$. Due to $\alpha_{i+1}^j \rho_{i+1}^j$ being equal to the probability of reaching time $i+1$ with exactly $j-1$ items already selected, denoting $t_s$ the time when item $s$ is selected, we have 
\begin{align*}
\alpha_{i+1}^j \rho_{i+1}^j& = \sum_{t_{j-1}=j-1}^{i} \sum_{t_{j-2}=j-2}^{t_{j-1}-1} \dots \sum_{t_1=1}^{t_2-1}  \Pr(\text{Selecting item $s$ at time $t_s$ for $s \in [j-1]$})\\
&=\sum_{t_{j-1}=j-1}^{i} \sum_{t_{j-2}=j-2}^{t_{j-1}-1} \dots \sum_{t_1=1}^{t_2-1} \left[ \prod_{s=1}^{j-1} \left(1-\frac{a^s_{t_s}}{\alpha^s_{t_s}}\right) \prod_{r_1=1}^{t_1-1} \frac{a^1_{r_1}}{\alpha^1_{r_1}} \dots \prod_{r_j=t_{j-1}+1}^{i} \frac{a^j_{r_j}}{\alpha^j_{r_j}} \right]\\
&=\sum_{t_{j-1}=j-1}^{i-1} \sum_{t_{j-2}=j-2}^{t_{j-1}-1} \dots \sum_{t_1=1}^{t_2-1} \left[ \prod_{s=1}^{j-1} \left(1-\frac{a^s_{t_s}}{\alpha^s_{t_s}}\right) \prod_{r_1=1}^{t_1-1} \frac{a^1_{r_1}}{\alpha^1_{r_1}} \dots \prod_{r_j=t_{j-1}+1}^{i} \frac{a^j_{r_j}}{\alpha^j_{r_j}} \right] \\
& + \sum_{t_{j-2}=j-2}^{i-1} \dots \sum_{t_1=1}^{t_2-1} \left[ \prod_{s=1}^{j-1} \left(1-\frac{a^s_{t_s}}{\alpha^s_{t_s}}\right) \prod_{r_1=1}^{t_1-1} \frac{a^1_{r_1}}{\alpha^1_{r_1}} \dots \prod_{r_j=i+1}^{i} \frac{a^j_{r_j}}{\alpha^j_{r_j}} \right] \\
&=  \frac{a^j_i}{\alpha_i^j}\sum_{t_{j-1}=j-1}^{i-1} \sum_{t_{j-2}=j-2}^{t_{j-1}-1} \dots \sum_{t_1=1}^{t_2-1} \left[ \prod_{s=1}^{j-1} \left(1-\frac{a^s_{t_s}}{\alpha^s_{t_s}}\right) \prod_{r_1=1}^{t_1-1} \frac{a^1_{r_1}}{\alpha^1_{r_1}} \dots \prod_{r_j=t_{j-1}+1}^{i-1} \frac{a^j_{r_j}}{\alpha^j_{r_j}} \right]  \\
&+ \left( 1- \frac{a^{j-1}_{i}}{\alpha^{j-1}_i} \right) \sum_{t_{j-2}=j-2}^{i-1} \dots \sum_{t_1=1}^{t_2-1} \left[ \prod_{s=1}^{j-2} \left(1-\frac{a^s_{t_s}}{\alpha^s_{t_s}}\right) \prod_{r_1=1}^{t_1-1} \frac{a^1_{r_1}}{\alpha^1_{r_1}} \dots \prod_{r_{j-1}=t_{j-2}+1}^{i-1} \frac{a^j_{r_{j-1}}}{\alpha^j_{r_{j-1}}} \right] \\
& = \alpha_i^j \rho_i^j + (\alpha_i^{j-1}-a_i^{j-1}) \rho_i^{j-1}.
\end{align*}
Notably, if we let the $\epsilon_i^j$ be such that $\rho_i^j=\rho_j^j$ for all $i \in \{j,\dots,n\}$ (starting from $j=1$, and sequentially imposing the boundary values), then dividing both sides by $\rho_j^j$ we obtain
\begin{equation*}
\alpha_{i+1}^j =a_i^j  + (\alpha_i^{j-1}-a_i^{j-1})\frac{\rho_{j-1}^{j-1}}{\rho_j^j}.
\end{equation*}

this overall yields a grid of recurrence equations for $j>1$ and $i\geq j$. Now let us translate this recurrence relationship into one over $b_i^j$. Let us assume that $\rho_{i+1}^{j-1}=\rho_i^j$ and see what this implies for $j$, with $j=1$ being already treated in \Cref{eq:recurrence_relation}.

Following the same computations done in \Cref{lemma:recurrence_relation}, we have that $\alpha_{i+2}^j-a_{i+1}^j=\text{cst}$ is equivalent to $\Delta^2[b_i^j]=\frac{\Delta[-\ell \beta_{\ell+1,n-\ell}\circ \beta_{\ell,n-\ell}^{-1}(b_i^j)]}{n}+\text{cst}$. Then, using that $\alpha_{i+1}^{j-1}-a_{i+1}^{j-1}=\alpha_{i+2}^{j-1}-\alpha_{i+2}^{j-1}+\alpha_{i+1}^{j-1}-a_{i+1}^{j-1}=-(\alpha_{i+2}^{j-1}-\alpha_{i+1}^{j-1})$, we obtain 

\begin{align*}
\Delta^2[b_i^j]&=\frac{\Delta[-\ell \beta_{\ell+1,n-\ell}\circ \beta_{\ell,n-\ell}^{-1}(b_i^j)]}{n}-\frac{\rho_{j-1}^{j-1}}{\rho_j^{j}}\Delta^2[b_i^{j-1}] \\
&=\frac{\Delta[\frac{\rho_{j-1}^{j-1}}{\rho_j^j} \ell \beta_{\ell+1,n-\ell}\circ \beta_{\ell,n-\ell}^{-1}(b_i^{j-1})-\ell \beta_{\ell+1,n-\ell}\circ \beta_{\ell,n-\ell}^{-1}(b_i^j)]}{n}.
\end{align*}
Now summing these equations yields the desired result.

\subsection{Proof of \Cref{prop:existence2}} \label{app:existence2}

We outline here the main steps for the proof: First, it must be that $b_j^j=o(1)$, as otherwise, because we add $b_j^j$ for $n-j$ steps, $b_n^j=\Omega(n)>1$. Then we show that if $\theta_{j,\ell}(n) \eqdef \rho_{j-1}^{j-1}/\rho_{j}^j$ is larger than some positive constant (independent of $n$) then $b_n^j>1$, and if it is smaller than some positive constant then $b_n^j<1$. 
By the intermediate value theorem, as $b^j_n$ is a continuous function of $b_j^j$, there must be a value $b_j^j$ such that $b^j_n=1$, which proves the existence. 
Moreover, by giving an asymptotic expression of $\theta_{j,\ell}(n)$ in terms of $b_j^j$ and $b_{j-1}^{j-1}$, and because $\theta_{j,\ell}(n)$ must remain bounded between two positive constants, we can inductively give an asymptotic expression of $b_j^j$ in terms of $n$ as $b_j^j=\Theta(n^{r_{j,\ell}})$ with $r_{j,\ell}=\ell \cdot (1- (\frac{\ell+1}{\ell})^{j})$. Then we show that the $b_i^j$ remain between $0$ and $1$, allowing us to map back the $b_i^j$ solution to a solution on the $\epsilon_i^j$. Finally, by evaluating the sign of $\rho_i^j$ if any $\epsilon_{i+1}^j$ were to be smaller than an $\epsilon_i^j$, we can obtain a contradiction, thus implying that the $\epsilon_i^j$ must be increasing in $i$, and so does the $b_i^j$. All of the above will be proven inductively for each $j$, so the aforementioned properties will be assumed true for $j-1$, and the initialization to $j=1$ already corresponds to \Cref{prop:existence}.\\

\textbf{Upper bound on $\theta_{j,\ell}(n)$.} Let us inductively show that $\theta_{j,\ell}(n)$ is bounded where we temporarily re-define (only in this proof!) $\theta_{1,\ell}(n) \eqdef (\ell+1)/\ell$. Using the recurrence relation from \Cref{lemma:recurrence_relation2}, we have that
\begin{align*}
 & \Delta[b_i^j] && \geq \frac{1}{n} \left(  \theta_{j,\ell}(n) \ell \beta_{\ell+1} \circ \beta_{\ell}^{-1}(b_i^{j-1}) -\ell \right)  \\
\implies  b_n^j + o(1) = b_n^j -b_j^j =  \frac{1}{n} \sum_{i=j}^{n-1} & \Delta[ b_i^j ]   && \geq  \ell \left(  \theta_{j,\ell}(n) \left( \frac{1}{n}\sum_{i=j}^{n-1} \beta_{\ell+1,n-\ell} \circ \beta_{\ell,n-\ell}^{-1}(b_i^{j-1})\right) -1 \right) \\
& && \geq  \ell \left(  \theta_{j,\ell}(n) \left( \frac{1}{n}\sum_{i=j}^{n-1} \gamma_{\ell+1} \circ \gamma_{\ell}^{-1}(b_i^{j-1}) + o(1)\right) -1 \right),
\end{align*}
where the last equality is due to the uniform convergence in \Cref{lemma:unif_conv}.
We now need to lower bound the above sum. Due to the induction hypothesis, we have that $b_n^{j-1}=1$, $b_j^j =o(1/n)$ for $j>1$ and thus $\Delta[b_i^{j-1}] \leq \theta_{j-1,\ell}(n) \ell /n +o(1/n)$ (this is why we define $\theta_{1,\ell}(n)=(\ell+1)/\ell$ to make sure that this expression remains true for $j=1$ as $\Delta[b_i^1]\leq (\ell+1)/n$). Therefore it takes some time to go from any value $y \in (0,1)$ to $1=b_{n-1}^j$. More specifically, if we denote $t$ the first time where $b_{t}^{j-1} \geq y$ we obtain an inequality on $t$:
\begin{align*}
1=b_t^{j-1}+\sum_{i=t}^{n-1} \Delta[b_i^{j-1}] & \leq y +\frac{1}{n} \theta_{j-1,\ell}(n)  \ell + \frac{n-t-1}{n} \theta_{j-1,\ell}(n)  \ell +o(1) \\
\implies n-t \geq \frac{1-y}{\theta_{j-1,\ell}(n)  \ell } +o(1).
\end{align*}
Due to the monotonicity of $b_i^{j-1}$ in $i$, for any $i \geq t$, $b_i^{j-1} \geq y$. We can now lower bound the sum:
\begin{equation}
b_n^j \geq \ell \left( \theta_{j,\ell}(n) \frac{(1-y)\gamma_{\ell+1} \circ \gamma_{\ell} ^{-1} (y)}{\ell \theta_{j-1,\ell}(n)  }  -1 \right) +o(\theta_{j,\ell}(n) ).
\end{equation}
We could get tighter bounds on $\theta_{j,\ell}(n)$ if we maximize this inequality in $y$, but let us simply take $y=1/2$, which numerically is not so bad for low values of $\ell$ and looks to be the maximum for large values of $\ell$ anyway. All in all, for $b_n^j$ to remain below $1$, it must be that 
\begin{equation*}
\theta_{j,\ell}(n)  +o(\theta_{j,\ell}(n)) \leq \frac{2(\ell+1)}{\gamma_{\ell+1}\circ \gamma_{\ell}^{-1}(1/2)} \theta_{j-1,\ell}(n).
\end{equation*}
Iterating this inequality in $j$ with $\theta_{1,\ell}(n)$ bounded shows that for $n$ large enough all the $\theta_{j,\ell}(n)$ indeed remain bounded. \\

\textbf{Size estimate of $b_j^j$.} Before showing that $\theta_{j,\ell}(n)$ is bounded below by some positive constant, we will first show that $b_j^j$ must be very small in front of $1/n$ due to the previous upper bound on $\theta_{j,\ell}(n)$. Let us express $\rho_j^j$ in terms of $\rho_{j-1}^{j-1}$:
\begin{align*}
\rho_j^j&=\frac{1}{\alpha_j^j}\prod_{r \in [j-1]} \left(1-\frac{a_r^r}{\alpha_r^r}\right)=\frac{1}{\alpha_j^j}\left(1-\frac{a^{j-1}_{j-1}}{\alpha_{j-1}^{j-1}}\right)\alpha_{j-1}^{j-1}\rho_{j-1}^{j-1}  \\
&=\frac{1}{\alpha_j^j}\left(\alpha_{j-1}^{j-1}-a^{j-1}_{j-1}\right)\rho_{j-1}^{j-1} \\
&=\frac{\ell \beta_{\ell+1,n-\ell}\circ \beta_{\ell,n-\ell}^{-1}(b_{j-1}^{j-1})}{n b_j^j}\rho_{j-1}^{j-1},
\end{align*}
where the last equality can be obtained following the same computations done in \Cref{app:recurrence_relation}. 
This implies that $b_j^j=O((\ell \beta_{\ell+1,n-\ell} \circ \beta_{\ell,n-\ell}(b_{j-1}^{j-1}))/n)$. \\

The expansion of $\beta_{\ell,n-\ell}^{-1}(x)$ around $0$ is $(\ell \cdot x B(\ell,n-\ell))^{1/\ell}+o(x^{1/\ell})=(\ell \cdot x)^{1/\ell}/n+o(1)$. Using the combinatorial formula for $\beta_{\ell+1,n-\ell}$,
we have 
\begin{align}
\ell \beta_{\ell+1,n-\ell} \circ \beta_{\ell,n-\ell}(b_j^j)&= \ell \sum_{t=\ell+1}^n \binom{n}{t} \beta_{\ell,n-\ell}^{-1}(b_j^j)^{t}\cdot (1-\beta_{\ell,n-\ell}^{-1}(b_j^j))^{n-t} \nonumber \\
&=\ell \frac{n^{\ell+1}}{(\ell+1)!} \frac{(\ell!\cdot b_j^j)^{1+1/\ell}}{n^{\ell+1}}+o(( b_j^j)^{1+1/\ell}) \nonumber \\
&=\frac{\ell}{\ell+1}(\ell!)^{1/\ell} (b_j^j)^{1+1/\ell}+o(( b_j^j)^{1+1/\ell})  \label{app:eq:rho_eq}.
\end{align}
The second equality is because $(1-\beta_{\ell,n-\ell}^{-1})^{n-t} \rightarrow_{n \rightarrow \infty} 1$ as long as $b_j^j=o(1)$, and because 
\begin{equation*}
\binom{n}{t+1} \beta_{\ell,n-\ell}^{-1}(b_j^j)^{t+1}/(\binom{n}{t} \beta_{\ell,n-\ell}^{-1}(b_j^j)^{t}) =O((b_j^j)^{\ell}) =o(1),
\end{equation*}
so the first term of the sum dominates the other ones.
Hence, using the growth rate induction hypothesis $b_{j-1}^{j-1}=\Theta(n^{r_{j-1},\ell})$, we can upper bound $b_{j-1}^{j-1}$ by $O(n^{(1+1/\ell)r_{j-1,\ell}-1})$. We now solve the recurrence $r_{j,\ell}= ((\ell+1)/ \ell) r_{j-1,\ell} -1 $ which will allow us to conclude that $b_j^j \leq O(n^{r_{j,\ell}})$. The first term and the fixed point of this recurrence are respectively $r_{1,\ell}=-1$ and $\ell$, a classic exercise shows that $r_{j,\ell}=-((\ell+1)/\ell)^{j-1}(1+\ell)+\ell=- \ell( ((\ell+1)/\ell)^j -1)$. Moreover, for $j>1$, $r_{j,\ell}<-1$, meaning that $b_j^j=o(n^{-1})$. \\

\textbf{Lower bound on $\theta_{j,\ell}(n)$.} We can now proceed to lower bound $\theta_{j,\ell}(n)$. For $\Delta[b_i^j]$ to be big enough and for $b_n^j$ to reach $1$, $\theta_{j,\ell}(n)$ must be large enough. Indeed
\begin{equation*}
\Delta[b_i^j] \leq \frac{\theta_{j,\ell}(n)\ell}{n} +b_j^j= \frac{\theta_{j,\ell}(n)\ell}{n} +o(\frac{1}{n}),
\end{equation*}
using the previous upper bound on $b_j^j$ and that $0 \leq \beta_{\ell+1,n-\ell} \leq 1$. Therefore $b_n^j \leq \theta_{j,\ell}(n) \ell +o(1)$, which implies that if $\theta_{j,\ell}(n)$ is strictly smaller than $1/\ell+o(1)$, then $b_n^j$ is smaller than $1$. We can then apply the intermediate value theorem to the function $b_n^j(b_j^j)$ to obtain the existence, and for the value $b_j^j$ which satisfies the boundary value condition, it must be that $\theta_{j,\ell}(n) \geq 1/\ell +o(1)$. Using once again the expression of $\theta_{j,\ell}(n)$ and induction hypothesis on $b_{j-1}^{j-1}$, we can conclude that $b_j^j= \Omega(n^{r_{j,\ell}})$ and therefore that $b_j^j = c_{j,\ell} n^{r_{j,\ell}}$ with $c_{j,\ell} \in [ 1/\ell, ( 2(\ell+1)/(\gamma_{\ell+1} \circ \gamma_{\ell}^{-1}(1/2)))^{j-1} (\ell+1)/\ell]+o(1)$. The actual constants could be tightened, and this would immediately yield a lower bound on the competitive ratio, akin to using the bound $c_{1,\ell} \leq \ell+1$ to prove that $\comp_{1,\ell} \geq \ell/(\ell+1)$. \\

\textbf{Mapping $b_i^j$ back to $\epsilon_i^j$.} To ensure that the solution to the discrete boundary value problem in $b_i^j$ translates into a solution to the discrete boundary value problem in $\epsilon_i^j$, we must ensure that the $b_i^j$ remain in $[0,1]$ for $\beta_{\ell,n-\ell}^{-1}(b_i^j)$ to be well defined. 
For $b_i^j \leq 1$, one way to see this is to define a continuous extension of $\beta_{\ell+1,n-\ell} \circ \beta_{\ell,n-\ell}(x)$ by $1$ for any $x>1$, and by $0$ whenever $x<0$. 
This ensures that if for some $i$, $b_i^j >1$ then it remains strictly greater than $1$. Indeed for $t_1$ the first time it crosses $1$ the difference $\Delta[b_{t_1}^j]$ must be positive, and for any $t>t_1$ $\Delta[b_t^j] \geq \Delta[b_{t_1}^j] \geq 0$ due to the monotonicity of $b_i^{j-1}$ by the induction hypothesis and due to the continuous extension which remains fixed at $1$. This entails that $b_n^j > 1$. This is a contradiction with $b_n^j=1$.\\

Now let us show that the $b_i^j$ that solves the boundary value problem always remain positive. The main idea is that, due to the relative size of the $b_j^j$ compared to the $b_j^{j-1}$, the sequence must be increasing at the beginning and hence positive. After some time due to the `discrete Lipschitzness' the sequence i.e. the difference between two consecutive terms is bounded, and because the $b_i^{j-1}$ are increasing, it cannot go below a certain positive threshold without encountering previous values of $b_i^j$ which were increasing and thus the sequence must go back up. First, when $b_i^j=o(1)$, we can always approximate $\ell \beta_{\ell+1,n-\ell} \circ \beta_{\ell,n-\ell}(b_i^j) = L (b_i^j)^{1+1/\ell} + o( (b_i^j)^{1+1/\ell})$, where $L=\frac{\ell}{\ell+1}(\ell!)^{1/\ell}$. From a high level, what this means is that whenever $x$ is small when we compare $x$ and $\beta_{\ell+1,n-\ell} \circ \beta_{\ell,n-\ell}^{-1}(x)\approx L \cdot x  \cdot x^{1/\ell}$ the first will dominate the second. So, for any $i \leq n/\log(n)$, $\Delta[b_i^1]=c_{1,\ell}/n+o(1/n)$, and $b_i^1=c_{1,\ell} (i/n)+o(i/n)$. From there we can verify by induction that for $i \leq n/\log(n)$, $b_i^j=\kappa_j \cdot (n/i)^{r_{j,\ell}}+o( (n/i)^{r_{j,\ell}})$ with $\kappa_j$ bounded between two positive constant independent of $n$, and $r_{j,\ell}=-\ell(((\ell+1)/\ell)^{j}-1)$. We can start by upper bounding $\Delta[b_i^j]$ using  $i \leq n/\log(n)$ by
\begin{align*}
\Delta[b_i^j] &\leq \theta_{j,\ell}(n) L \frac{1}{n} (b_i^{j-1})^{1/\ell+1}+c_{j,\ell}(n) n^{r_{j,\ell}} \\
&\leq \theta_{j,\ell}(n) L \kappa_{j-1}^{1+1/\ell} \frac{1}{n} \left(\frac{n}{i}\right)^{(1+1/\ell)\cdot r_{j-1,\ell}} +c_{j,\ell}(n) n^{r_{j,\ell}} \\
&\leq \kappa_{j} \frac{\log(n)^{(1+1/\ell) \cdot r_{j-1,\ell}}}{n} + o(\frac{\log(n)^{(1+1/\ell) \cdot r_{j,\ell}}}{n}),
\end{align*}
with $\kappa_j=\theta_{j,\ell}(n)L \kappa_{j-1}^{1+1/\ell}+c_{j,\ell}(n)$. This implies that for $i \leq n/\log(n)$:
\begin{equation*}
b_i^j \leq \kappa_j (i\log(n)^{(1+1/\ell) \cdot r_{j-1,\ell}})/n +o( (i\log(n)^{(1+1/\ell) \cdot r_{j,\ell}})/n) \leq \kappa_j \log(n)^{r_{j,\ell}}+ o(\log(n)^{r_{j,\ell}}).
\end{equation*}
Therefore $b_i^j$ is negligible in front of $b_i^{j-1}$, so we can redo the same computations by replacing above the inequality by an equality as $\beta_{\ell+1,n-\ell}\circ \beta_{\ell,n-\ell}^{-1}(b_i^{j})$ is negligible in front of $\beta_{\ell+1,n-\ell}\circ \beta_{\ell,n-\ell}^{-1}(b_i^{j-1})$. This also implies that $\Delta[b_i^j] \geq 0$ as the only negative term is negligible. We obtain $b_i^j=\kappa_j \cdot (n/i)^{r_{j,\ell}}+o( (n/i)^{r_{j,\ell}})$. Now for $i= \lfloor \log(n)/n \rfloor$, we have that $b_i^j= \kappa_j \log(n)^{r_{j,\ell}}+o(\log(n)^{r_{j,\ell}})$, which is strictly greater than $\ell/n$, an upper bound on the minimum value of $\Delta[b_i^j]$ derived from \Cref{lemma:recurrence_relation2}. Because $\Delta[b_i^j] \geq 0$ for any $i \leq n/\log(n)$ and $b_i^{j-1}$ is non decreasing, for any $t$ such that $b_i^j \leq b_{n/\log(n)}$ we have that $\Delta [b_t^{j}] \geq \Delta[b_{n/\log(n)}^{j}] \geq 0$. Thus the sequence $b_i^j$ cannot keep on decreasing when going below $b_{n/\log(n)}^j$, and $b_{n/\log(n)}^j - \ell/n >0$ which overall yields the non-negativity of the $b_i^j$. \\

\textbf{Monotonicity of the $b_i^j$.} It remains to show the monotonicity. The quantity $\rho_i^j$, while it stems from a probabilistic event that had assumed that the $\epsilon_i^j$ were increasing, can be defined for any $\epsilon_i^j$ using integrals with no further requirements on the $\epsilon_i^j$. For now, we have shown that there exists $\epsilon_i^j$ such that all the $\rho_i^j$ are equal, $\epsilon_{j-1}^j=0$ and $\epsilon_n^j=1$. First, the sign of $\alpha_i^j \rho_i^j$ is always positive. Indeed, $\alpha_i^j$ and $a_i^j$ are of the same sign, positive if $\epsilon_i^j \geq \epsilon_{i-1}^j$, and negative otherwise. In both cases, the ratio is positive and smaller than $1$, which also implies the positivity of $1-a_i^j/\alpha_i^j$. As a sum of products of positive terms, $\alpha_i^j \rho_i^j$ is always positive. Because all the $\rho_i^j$ are equal, they have the same sign, so either all the $\rho_i^j$ and $\alpha_i^j$ are positive, or they are all negative. Because $\epsilon_{j-1}^j=0$ and $\epsilon_n^j=1$, there must be some $t \geq j$ such that $\epsilon_t^j \geq \epsilon_{t-1}^j$, implying that $\alpha_t^j$ is positive. Therefore all the $\alpha_i^j=\beta_{\ell,n-\ell}(\epsilon_i^j)-\beta_{\ell,n-\ell}(\epsilon_{i-1}^j)$ are positive, which means that the $\epsilon_i^j$ are non-decreasing. Finally, because $\epsilon_{j-1}^n = 0 \neq 1 = \epsilon_n^j$, then at least one of the $\rho_i^j$ is finite, which by equality of the $\rho_i^j$ means that all of them must be finite, and so all the $\epsilon_i^j$ are distinct. Hence the $\epsilon_i^j$ are non-decreasing and distinct, so are increasing.

\subsection{Proof of \Cref{prop:convergence} } \label{app:convergence} 

 Due to the uniform convergence in \Cref{lemma:unif_conv}, it seems intuitive that we do have the convergence from the discrete boundary value problem to the continuous one. However, there are many technical difficulties that make proving this convergence especially challenging, in particular the fact that the limit function $\gamma_{\ell+1} \circ \gamma_{\ell}^{-1}$ is not Lipschitz over $[0,1]$ having an infinite derivative at $1$. The convergence of $c_{1,\ell}(n)$ is already proven in Theorem \ref{thm:lim}, so it remains to prove the convergence of $\theta_{j,\ell}(n)$ towards $\theta_{j,\ell}$ (the solution to the continuous boundary value problem). \\

 Let $\xi_{{\ell},n}(x)=-{\ell} \beta_{\ell+1,n-{\ell}}\circ \beta_{{\ell},n-{\ell}}^{-1} (x)$ and $\xi_{\ell}(x)=-\ell \gamma_{\ell+1}\circ \gamma_{\ell}^{-1}(x)$. First due to the boundedness of the sequence $\theta_{j,\ell}(n)$, by the Bolzano-Weierstrass theorem there is at least one subsequence with an accumulation point $\theta$, and we will work with such a subsequence. \\
 
 \textbf{Existence of solution to ODE.} We now consider the ODE $y'(t)=\xi_{\ell}(y)-\theta \xi_{\ell}(b^{j-1}(t)) $ with initial condition $y(0)=0$. as $\xi_{\ell}$ and $\xi_{\ell}(b^{j-1}(t))$ are continuous, there exists a solution $b^j$ over $[0,1]$. We also have that \begin{equation*}
\xi_{\ell}'(x)= - \ell \cdot (\gamma_{\ell}^{-1})'(x) \cdot (\gamma_{\ell+1})'(\gamma_{\ell}^{-1}(x))= - \ell \cdot \frac{\gamma_{\ell+1}' \circ \gamma_{\ell}^{-1}(x)}{\gamma_{\ell}' \circ \gamma_{\ell}^{-1}(x)} = - \gamma_{\ell}^{-1}(x),
 \end{equation*} 
 so as long as $x<1$ then $\xi_{\ell}'$ is bounded and $\gamma^{-1}_{\ell}(x)$ Lipschitz over the interval $[0,x]$. This means that as long as $b^j$ is strictly smaller than $1$, then by the Cauchy Lipschitz theorem the solution must be unique. The case when $\max_{[0,1]} b_j <1$ is easier to treat, so we focus on when $\max_{[0,1]} b_j \geq 1$. Because $\xi_{\ell}$ is bounded, $b^j$ itself is Lipschitz over $[0,1]$, so denoting $t_1$ the first time $b^j=1$ the solution is unique over $[0,t_1]$ due to the Lipschitzness of $b^j$. Note that over $(t_1,1]$ there can be potentially multiple solutions satisfying the initial condition $b^j(0)=0$. \\
 
 We now wish to prove the convergence of $b_{\lfloor tn \rfloor}^j$ towards $b^j(t)$ for any $t \in [0,t_1]$. We will prove it first on $[0,t_1-\epsilon]$ for any $\epsilon>0$. The main idea is that $b_i^j$ is almost an Euler discretization of the continuous solution, and the same ideas used in the convergence of the Euler method can be modified to take into account that the discretization uses $\xi_{\ell,n}$ and not $\xi_{\ell}$. \\
 
\textbf{$b_{\lfloor tn \rfloor}$ and $b(t)$ are different from $1$.} Let $t \in [0,t_1-\epsilon]$, in which case $b_j(t)<1$ by continuity as $t<t_1$ and $t_1$ is the first time for which $b^j(t)=1$. Similarly, we now show that for $n$ large enough, $b_{\lfloor tn \rfloor} \leq 1-c$ for some $c>0$. The quantity $b_{\lfloor tn \rfloor}$ is bounded between $(0,1)$ so has a non-empty set of accumulation points. For $n$ large enough, the distance between $b_{\lfloor tn \rfloor}$ and the set of accumulation points will go to $0$. If not we can look at the sub-sequence of points which do not converge to an accumulation point and apply Bolzano-Weierstrass again to exhibit a new accumulation point towards which at least some of the points converge, showing a contradiction. Therefore, if all the accumulation points are strictly smaller than $1$, then there exists some constant $c$ such that for $n$ large enough $b_{\lfloor tn \rfloor} \leq 1-c$. All the accumulation points must be smaller than $1$ as the $b_i^j$ are smaller than $1$. Suppose that $1$ belongs to the set of accumulation points. Because $b_n^j=1$, this means that $b_n^j-b_{\lfloor tn \rfloor}^j = n^{-1} \sum_{i=\lfloor tn \rfloor}^n \Delta[b_i^j] \rightarrow 0$. So, due to the monotonicity of $b^j_i$ and the convergence of $b^{j-1}_i$ by the induction hypothesis, for any $t'$ in $[t,1]$ we have $b_{\lfloor t'n \rfloor}^j \rightarrow  1$, and $n\Delta[b_{\lfloor t'n \rfloor}^j] \rightarrow \theta f(b^{j-1}(t')) - \ell$. Using that $\xi_{\ell,n}$ converges uniformly towards $\xi_{\ell}$, the Riemann sum approximation tells us that $n^{-1} \sum_{i=\lfloor t' n \rfloor}^n \Delta[b_i^j] \rightarrow \int_{t'}^1 \theta \xi_{\ell}(b^{j-1}(u))du - \ell =0$. This equation is valid for any $t'>t$, which is not possible as $b^{j-1}$ is increasing and therefore the integral value must be different. Overall, we have proven that the $b_{\lfloor tn \rfloor}^j$ must remain far from $1$ as long as $t <1$. \\

\textbf{Convergence by Euler's method.} Instead of working with the discrete sequence $b_i^j$, we work with the affine by parts function $b_{(n)}^j$ which takes value $b_i^j$ at times $i/n$ and each of those values are interpolated through linear segments. We will prove the uniform convergence of this affine by part function to the continuous limit. One way to prove this could be to use the Arzela-Ascoli theorem, as we have now a sequence of functions with approximately the same Lipschitz constant, and are thus equicontinuous. Instead we will apply Euler's method. Let $\varphi(y,t)= \xi_{\ell}(y)-\theta \xi_{\ell}(b^{j-1}(t))$ and $\varphi_n(y,t)= \xi_{\ell,n}(y)-\theta_{j,\ell}(n) \xi_{\ell,n}(b^{j-1}_{(n)}(t))$. The function $\varphi_n$ converges uniformly to $\varphi$ due to \Cref{lemma:unif_conv}, that $\theta_{j,\ell}(n)$ converges to some $\theta$ for the subsequence considered, and $b^{j-1}_{(n)}$ converges uniformly to its limit solution. Let $t_i=i/n$, and $\delta_i=b^j(t_i)-b^j_i$ be the global truncation error up to time $t_i$ (the notation is omitted but $b^j_i$ depends on $n$). We only consider time $t_i$ with $t_i \leq t_1-\epsilon$, so that $\varphi(y,t)$ is $B$-Lipschitz in $y$ for $B>0$ and $b^j(t)''=(b^j)'(t)\xi_{\ell}'(b^j(t))-\theta \cdot (b^{j-1})'(t) \xi_{\ell}'(b^{j-1}(t))$ is also bounded by some $A>0$ due to the boundedness of $(b^{j-1})'$ and $(b^j)'$. Looking at the global truncation error, denoting by $\omega_n=\Vert \varphi_n - \varphi \Vert_{\infty}$, we have
 \begin{align*}
& \vert \delta_{i+1} \vert =\vert b^j(t_{i+1})-b_{i+1}^j = b^j(t_{i+1})-b_i^j-\frac{\varphi_n(b^j_i,t_i)}{n} \vert\\
&=\vert b^j(t_{i+1})-b_i^j-\frac{\varphi(b^j_i,t_i)}{n}+\frac{\varphi(b^j_i,t_i)-\varphi_n(b^j_i,t_i)}{n}\vert\\
&=\vert b^j(t_{i})-b_i^j+\frac{\varphi(b^j(t_i),t_i)-\varphi(b^j_i,t_i)}{n}+(b^j(t_{i+1})-b^j(t_i)-\frac{1}{n}\varphi(b^j(t_i),t_i))+\frac{\varphi(b^j_i,t_i)-\varphi_n(b^j_i,t_i)}{n}\vert \\
&\leq \vert \delta_i \vert + \frac{1}{n} B \vert \delta_i \vert + \frac{1}{2} A \frac{1}{n^2}+\frac{\omega_n}{n} \leq (1+\frac{B}{n}) \delta_i + \frac{\omega'_n}{n},
 \end{align*}
 where $\omega'(n)=\max(A/(2n),\omega_n)$ which goes to zero as both $A/n$ and $\omega_n$ do. Using that $\delta_0=0$, we can apply this inequality iteratively leading to 
 \begin{equation*}
    \delta_i \leq \sum_{j=1}^{i-1} \left(1+\frac{B}{n}\right)^j \frac{\omega'_n}{n} = \frac{\omega'_n}{n} \frac{(1+B/n)^{i}-1}{1+B/n-1} \leq \omega'_n \exp((t_1-\epsilon)B) \xrightarrow[n \longrightarrow \infty]{} 0.
 \end{equation*}
 Because $t_{i+1}-t_i \rightarrow 0$ and by Lipschitzness of $b^j_{(n)}$ we have the convergence towards $b^j$ for any $t \in [0,t_1-\epsilon]$. Finally 
 \begin{align*}
\vert b^j(t_1) - b^j_{(n)}(t_1) \vert & \leq \vert b^j(t_1) - b^j(t_1-\epsilon) \vert + \vert b^j(t_1-\epsilon)-b^j_{(n)}(t_1-\epsilon) \vert + \vert b^j_{(n)}(t_1-\epsilon) - b^j_{(n)}(t_1) \vert \\
& \leq M \epsilon + \delta_{\lfloor (t_1 -\epsilon) n \rfloor}(\epsilon) + M \epsilon, 
 \end{align*}
 with $M$ a common upper bound on $\varphi_n$ and $\varphi$ for $n$ large enough. We can take the limit of this inequality over $n$ for any fixed $n$, and then take the limit over $\epsilon$. This implies that $\lim_{n \rightarrow \infty} b^j_{(n)}(t_1)=b^j(t_1)=1$, which is impossible unless $t_1=1$ as we have already proven that this limit is different from $1$ as long as $t<1$. This implies that $\theta$ is a solution to the continuous boundary value problem. This also proves the existence of a solution to the continuous boundary value problem.  \\

We now prove that the solution of the continuous boundary value problem must be unique. We will show that $b^j(1)$ is strictly increasing in the parameter $\theta \geq$ of the ODE. Let $\theta_1 > \theta_2$, $b_1$ and $b_2$ the respective solutions, and $d=b_2-b_1$ their difference. First we show that $b_1 \geq b_2$. Let $M=\max_{[0,t]} d(t)$, $t_0$ the point at which the maximum is reached, and $M \geq d(0)=b_2(0)-b_1(0)=0$. If $t_0 >0$, we have at $t_0$ that $d'(t_0)=b'_2(t_0)-b'_1(t_0)= \xi_{\ell}(b_2(t_0))-\xi_{\ell}(b_1(t_0))-(\theta_2 - \theta_1) \xi_{\ell} (b^{j-1}(t_0)<0 $ using that $b_2(t_0)\geq b_1(t_0)$, $\xi_{\ell}$ is decreasing, and $\xi_{\ell}(b^{j-1}(t_0))<0$ as  $t_0>0$. By continuity $d$ is strictly decreasing in a neighborhood of $t_0$, and therefore for $\delta>0$ small enough  $d(t_0-\delta) > d(t_0)=M$ which is impossible by definition of $M$. Thus $t_0=0$ and $b_1 \geq b_0$. Finally if there is some $t_0>0$ such that $b_1(t_0)=b_2(t_0)$, the same argument yields $d'<0$ which with $d(t_0)=0$ contradicts $d \geq 0$. Hence $b_1(t) > b_2(t)$ over $(0,1]$ and in particular $b_1(1)>b_2(1)$. \\
 
 Because there is a unique possible value for the limit, there is only one possible accumulation point for $\theta_{j,\ell}(n)$, implying that $\theta_{j,\ell}(n)$ does converge to the unique solution of the continuous boundary value problem in \Cref{eq:continuousBVP}.\\
 
 To finish, $b^j$ is non-decreasing as $\Delta[b_i^j]\geq 0$ and it must remain so in the limit. Moreover because $b^{j-1}$ is strictly increasing, so is $b^j$. The initialization for this property is that $b^1$ is strictly increasing as $c_{1,\ell} >1$. Additionally the convergence of $\theta_{j,\ell}(n)$ implies the convergence of $c_{j,\ell}(n)$, and the relation between these two quantities is immediate from taking the limit in \Cref{app:eq:rho_eq}. The monotonicity of $b^j$ immediately implies that $\theta_{j,\ell}\geq 1$, as $(b^j)'(1)=\theta_{j,\ell} \cdot \ell - \ell=\ell(\theta_{j,\ell}-1)\geq 0$.

 \subsection{Proof of \Cref{prop:static_LB}} \label{app:static_LB}

 The proof consists of two steps, using Jensen's inequality on the reward of the single threshold algorithm similarly done in \cite{Correa2019} to prove in a simple way the $1-1/e$ performance of $F^{-1}(1-1/n)$ in the \ac{iid}  single item setting, and algebraic manipulations as well as inequalities to obtain the desired lower bound.\\

 We can start by noting that the expression of the online algorithm when only a single threshold is used is much simpler. Let $q \in [0,1]$ be the quantile corresponding to the selected threshold, e.g. $T=F^{-1}(1-q)$. The expected reward given by the $j$-th item at time $i$ is simply $R(q)$ times the probability of having selected exactly $j-1$ item up to time $i-1$, which corresponds to a random variable distributed according to $\mathrm{Binomial}(i-1,q)$ to be equal to $j-1$. The total expected reward obtained through the $j$-th item is thus 
 \begin{equation*}
\sum_{i=j}^n R(q) \binom{i-1}{j-1} q^{j-1} (1-q)^{i-j}.
 \end{equation*}
 Moreover, we know through the proof of \Cref{prop:alg_perf} that for $Q$ distributed according to $\mathrm{Beta}(\ell,n-\ell)$, $\opt_{\ell,n}=n \be[R(Q)]$. The expectation of $Q$ is $\be[Q]=\ell/n$, and using the concavity of $R$ we obtain 
 \begin{equation}
\opt_{\ell,n} = n \be[R(Q)] \leq n R\left( \frac{\ell}{n} \right).
 \end{equation}
 Due to this inequality, we set the deterministic quantile to be $q=\ell/n$, which immediately implies that 
 \begin{equation*}
\sum_{i=j}^n R(q) \binom{i-1}{j-1} q^{j-1} (1-q)^{i-j} \geq \left( \sum_{i=j}^n \frac{1}{n} \binom{i-1}{j-1} \left( \frac{\ell}{n} \right)^{j-1} \left(1-\frac{\ell}{n} \right)^{i-j} \right) \opt_{\ell,n}.
 \end{equation*}
 To obtain a lower bound on this competitive ratio, it remains to lower bound this sum, which we will denote by $S_{j,\ell,n}$.
 \begin{equation*}
S_{j,\ell,n}= \sum_{i=0}^{n-j} \frac{1}{n} \binom{i+j-1}{j-1} \left( \frac{\ell}{n} \right)^{j-1} \left(1- \frac{\ell}{n} \right)^{i} = \frac{\ell^{j-1}}{n^j (j-1)!} \sum_{i=0}^{n-j} \frac{(i+j-1)!}{i!} \left(1- \frac{\ell}{n} \right)^i.
 \end{equation*}
 We recognize that $\sum_{i=0}^{n-j} (i+j-1) \times \dots \times (i+1) (1-\ell/n)^i$ is the $j-1$-th derivative of the geometric sum $\sum_{i=0}^{n-1} (1-\ell/n)^i=(1-(1-\ell/n)^n)/(1-(1-\ell/n))$. the $t$-th derivative of $1/(1-x)$ is $t!/(1-x)^{t+1}$ and the $t$-th derivative of $1-x^{n}$ for $t\geq 1$ is $-n!/(n-t)! x^{n-t} \bb{1}[t \leq n]$. Using Leibniz rule for derivation, 
 \begin{align*}
\frac{d^{j-1}}{dx} \frac{1-x^n}{1-x}&=\sum_{t=0}^{j-1} \binom{t-1}{t} \frac{d^{j-1-t}}{dx}\left( 1-x^n \right)  \frac{d^{t}}{dx} \left( \frac{1}{1-x} \right)\\
&= (j-1)!\frac{1-x^{n}}{(1-x)^j}-\sum_{t=0}^{j-2} \binom{j-1}{t} \cdot \frac{t!}{(1-x)^{t+1}} \cdot \frac{n!}{(n+1+t-j)!} x^{n+1+t-j} \\
&=(j-1)!\frac{1}{(1-x)^j}-\sum_{t=0}^{j-1} \binom{j-1}{t} \cdot \frac{t!}{(1-x)^{t+1}} \cdot \frac{n!}{(n+1+t-j)!} x^{n+1+t-j} \\
 \end{align*}
 For $x=(1-\ell/n)\leq 1$, $x^{n+1+t-j} = (1-\ell/n)^{n+1+t-j} \leq \exp(-\ell) (1-\ell/n)^{1+t-j} $, $(1-x)^{t+1}=(\ell/n)^{t+1}$, and $n!/(n+1+t-j)! \leq n^{j-t-1}$. Therefore 
 \begin{align*}
S_{j,\ell,n} &\geq \frac{\ell^{j-1}}{n^j (j-1)!} \left[ (j-1)!  \frac{1}{(\ell/n)^j}  - \sum_{t=0}^{j-1} \frac{(j-1)!}{t!(j-1-t)!} \cdot \frac{t!}{(\ell/n)^{t+1}} \cdot n^{j-t-1} \exp(-\ell) (1-\frac{\ell}{n})^{1+t-j} \right] \\
&= \frac{1}{\ell} \left(1  -  e^{-\ell} \sum_{t=0}^{j-1} \frac{(\ell/(1-\ell/n))^{j-1-t}}{(j-1-t)!} \right) \\
&= \frac{1}{\ell} \left(1  -  e^{-\ell} \sum_{t=0}^{j-1} \frac{(\ell /(1-\ell/n) )^{t}}{t!} \right) \\
&= \frac{1}{\ell} \left(1 - e^{\ell^2/(n-\ell)} + e^{\ell^2/(n-\ell)} -  e^{-\ell} \cdot e^{\ell/(1-\ell/n)} \cdot e^{-\ell/(1-n/\ell)} \sum_{t=0}^{j-1} \frac{(\ell / (1-\ell/n))^{t}}{t!} \right) \\
&= \frac{\gamma_j(\ell / (1-\ell/n)) e^{\ell^2/(n-\ell)}+(1-e^{\ell^2/(n-\ell)})}{\ell} \geq \frac{\gamma_j(\ell)}{\ell}-\frac{1}{n}\left(\ell-\gamma_j(\ell) \cdot \ell - \frac{\ell^{j} e^{-\ell}}{(j-1)!}\right) -o\left( \frac{1}{n^2} \right),
 \end{align*}
 where we used Taylor approximations to get estimates of $\gamma_j(\ell/(1-\ell/n))$, and $e^{\ell^2/(n-\ell)}$. Summing the contribution of every item $j \in [k]$, we immediately obtain the desired lower bound 

 \begin{equation*}
\comp^S_{k,\ell}(n) \geq \frac{\sum_{j=1}^k \gamma_{\ell}(j)}{k} -\frac{1}{n}\left(1-\gamma_j(\ell) - \frac{\ell^{j-1} e^{-\ell}}{(j-1)!}\right) -o\left( \frac{1}{n^2} \right) .
 \end{equation*}

 \subsection{Proof of \Cref{prop:static_UB}} \label{app:static_UB}

 Once the worst case instance from \cite{Arnosti2021} is correctly modified, their proof almost entirely follows through. First of all, they show that for any quantity $W$ independent of $n$, the prophet's expected reward is at least 
 \begin{equation*}
\ell+W-\frac{1+W}{n+1},
 \end{equation*}
 and the decision maker's expected reward is at most
 \begin{equation*}
\be[\min(\mathrm{Poisson}(nq),k)]\left( 1+ \frac{W}{nq}\right)  +2kWn^{-2/3} +2kn^{-1/3},
 \end{equation*}
 with $q$ the probability of accepting any item which is a function of the random tie-break probability. They further show that the derivative of $\be[\min(\mathrm{Poisson}(nq),k)]\left( 1+ W/nq\right)$ in $\lambda=nq$ is equal to 
 \begin{equation*}
\frac{d}{d\lambda}\be[\min(\mathrm{Poisson}(\lambda),k)]\left( 1+ \frac{W}{\lambda}\right)=\Pr(\mathrm{Poisson}(\lambda)<k)\left( 1- W\frac{k}{\lambda^2} \frac{\Pr(\mathrm{Poisson}(\lambda)>k)}{\Pr(\mathrm{Poisson}(\lambda)<k)}\right).
 \end{equation*}
 To have a simple expression of the competitive ratio, we pick $W=W_{k,\ell}$ such that the above derivative cancels at exactly $\lambda=\ell$. Hence 
 \begin{equation*}
\frac{d}{d\lambda}\be[\min(\mathrm{Poisson}(\lambda),k)]\left( 1 \! + \! \frac{W_{k,\ell}}{\lambda}\right) \! = \! \Pr(\mathrm{Poisson}(\lambda)<k)\left( \! 1 \! - \frac{\ell^2}{\lambda^2}\frac{\Pr(\mathrm{Poisson}(\ell)\! <\! k)}{\Pr(\mathrm{Poisson}(\ell)\! >\! k)} \frac{\Pr(\mathrm{Poisson}(\lambda)\! > \! k)}{\Pr(\mathrm{Poisson}(\lambda)\! <\! k)}\right)  
 \end{equation*}
  It remains to show that this critical point corresponds to a maximum. The computations will be almost identical to \cite{Arnosti2021}. \\

 For $\lambda<\ell$ we have 
 \begin{align*}
\frac{\ell^2}{\lambda^2}\frac{\Pr(\mathrm{Poisson}(\ell)<k)}{\Pr(\mathrm{Poisson}(\ell)>k)} \frac{\Pr(\mathrm{Poisson}(\lambda)>k)}{\Pr(\mathrm{Poisson}(\lambda)<k)} &= \frac{\sum_{j>k} \lambda^{j-2}/j!}{\sum_{j>k} \ell^{j-2}/j!} \cdot \frac{\sum_{j<k} \ell^{j}/j!}{\sum_{j<k} \lambda^{j}/j!}\\
&<\left( \frac{\lambda}{\ell} \right)^{k-1} \frac{\sum_{j<k} \ell^{j}/j!}{\sum_{j<k} \lambda^{j}/j!} \\
&= \frac{\sum_{j<k} (\frac{\lambda}{\ell})^{k-1-j} \lambda^{j}/j!}{\sum_{j<k} \lambda^{j}/j!} \leq 1.
 \end{align*}
 The same can be done for $\lambda>\ell$, which shows that $nq=\ell$ indeed yields the optimal static rule with tie-break for $F^*$. 

 In the limit as $n\rightarrow \infty$, this implies that 
 \begin{equation*}
\comp_{k,\ell}^{S} \leq \frac{\ell}{k} \cdot \frac{\be[\min(\mathrm{Poisson}(\ell),k)] \left(1+\frac{W_{k,\ell}}{\ell}\right)}{\ell+W_{k,\ell}}= \frac{\be[\min(\mathrm{Poisson}(\ell),k)]}{k}.
 \end{equation*}
 This last quantity can then be related to $\gamma_j(\ell)$, as 
 \begin{align*}
\be[\min(\mathrm{Poisson}(\ell),k)]=\sum_{j=0}^{k-1} \Pr(\mathrm{Poisson}(\ell)>j) &= k - \sum_{j=0}^{k-1} \Pr(\mathrm{Poisson}(\ell) \leq j) \\
& =  \sum_{j=0}^{k-1} \left( 1 -   \sum_{i=0}^j \frac{\ell^i}{i!} e^{-\ell} \right) \\
& = \sum_{j=0}^{k-1} \gamma_{j+1}(\ell)\\
&= \sum_{j=1}^k \gamma_j(\ell).
 \end{align*}

\subsection{Proof on finite dimension reduction} \label{app:balayage}

In this section, we take care of proving the reduction procedure, for the general $(k,\ell)$ setting, from a general distribution $F$ to a discrete distribution in $[0,1]$, with a smaller competitive ratio $\comp_{k,\ell}(n)$. This immediately implies the result of \Cref{prop:reduction2} for general $(k,\ell)$.

Let us first define the technique of balayage.

\begin{definition}
For a random variable $X$ and constants $0\leq a< b <\infty$ we denote by $X_{a:b}$ the random variable which takes the same value as $X$ when $X \not \in [a,b]$, takes value $a$ with probability $p_a=\be[(b-X)\bb{1}[X \in [a,b]]]/(b-a)$ and takes value $b$ with probability $p_b=\be[(X-a)\bb{1}[X \in [a,b]]]/(b-a)$.
\end{definition}

This new random variable conserves some characteristic of the original one: $X$ and $X_{a,b}$ have the same probability of taking values outside $[a,b]$ thus $\mathbb{E}[X \bb{1}_{[X \not \in [a,b]]}]=\mathbb{E}[X_{a,b} \bb{1}_{[X \not \in [a,b]]}]$, and by definition of $p_a$ and $p_b$ we have $\mathbb{E}[X \bb{1}_{[X  \in [a,b]]}]=\mathbb{E}[X_{a,b} \bb{1}_{[X  \in [a,b]]}]$. Both properties imply that $\be[X]=\be[X_{a,b}]$.

We can derive that $\opt_{\ell,n}$ is increasing with balayage, which generalize the proof of \cite{Kertz1982} for $\ell=1$.

\begin{lemma} \label{lemma:balayage}
For $Y=X_{a:b}$, 
\begin{equation}
\be[\sum_{i \in [\ell]} X_{(i)}] \leq \be[\sum_{i \in [\ell]} Y_{(i)}].
\end{equation}
\end{lemma}

\begin{proof}
We denote by $\opt_{\ell}(X_1,\dots,X_n)$ the function which takes into input the variables $\bx=(X_1,\dots,X_n)$ and outputs the sum of the top $\ell$ variables. Clearly, $\be[\opt_{\ell}(\bx)]=\opt_{\ell,n}$. We first show that for all $i \in [n]$, $\be[\opt_{\ell} (\bx)] \leq \be[\opt_{\ell} (X_{a,b},\bx_{-i})] $, the statement of the proposition then follows by applying multiple times this inequality. 

 First, let us remark that $\opt_{\ell}({\bx})$ can be rewritten as the value of the following linear (integer) program: the objective is $\bf{S}^{\top}\bx$ with $S_i \in \{0,1\}$ and $\sum_{i \in [n]} S_i =\ell$. Because the objective is convex, and as the supremum of a family of convex functions, $\opt_{\ell}$ is convex in $\bx$. In particular, for some $i \in [n]$, $\varphi(x)\eqdef \be[\opt_{\ell} (\bx) \mid X_i=x]$ is convex in $x$.
By convexity and independence of the $X_i$, we have that \begin{align*}
\varphi(x)&=\varphi( b \cdot \frac{x-a}{b-a}+ a \cdot \frac{b-x}{b-a}) \leq \frac{x-a}{b-a}\varphi(b)+ \frac{b-x}{b-a} \varphi(a) \\
&= \frac{x-a}{b-a} \be[\opt_{\ell} (b,\bx_{-i})]+\frac{b-x}{b-a}\be[\opt_{\ell} (a,\bx_{-i})].
\end{align*}
Therefore, we have 
\begin{align*}
\be[\opt_{\ell} (\bx)] &=\be[\opt_{\ell} (\bx) \bb{1}_{[X_i \in [a,b]]}]+\be[\opt_{\ell} (\bx) \bb{1}_{[X_i \not \in [a,b]]}]\\
&= \be [\opt_{\ell} (\bx) \bb{1}_{[X_i \in [a,b]]}] + \be[\opt_{\ell} (X_{a,b},\bx_{-i})) \bb{1}_{[X_i \not \in [a,b]]}] \\
&= \be [\varphi(X_i) \bb{1}_{[X_i \in [a,b]]}]+\be[\opt_{\ell} (X_{a,b},\bx_{-i})) \bb{1}_{[X_i \not \in [a,b]]}] \\
& \leq \be\left[ \left(\frac{X_i-a}{b-a} \be[\opt_{\ell} (b,\bx_{-i})]+\frac{b-X_i}{b-a}\be[\opt_{\ell} (a,\bx_{-i})]\right) \bb{1}_{[X_i \in [a,b]]} \right]\\ 
&+\be[\opt_{\ell} (X_{a,b},\bx_{-i})) \bb{1}_{[X_i \not \in [a,b]]}] \\
&=p_b \be[\opt_{\ell} (b,\bx_{-i})]+ p_a \be[\opt_{\ell} (a,\bx_{-i})]+\be[\opt_{\ell} (X_{a,b},\bx_{-i})) \bb{1}_{[X_i \not \in [a,b]]}]\\
&=\be[\opt_{\ell} (X_{a,b},\bx_{-i})) \bb{1}_{[X_i  \in [a,b]]}]+\be[\opt_{\ell} (X_{a,b},\bx_{-i})) \bb{1}_{[X_i \not \in [a,b]]}]\\
&= \be[\opt_{\ell} (X_{a,b},\bx_{-i})) ]. \qedhere
\end{align*}
\end{proof}

If the distribution is bounded by some constant $v$, then we can simply consider the variable $X_i/v$ which is in $[0,1]$, and this does not change the value of the competitive ratio. If the distribution is unbounded, we can do a balayage to infinity which will recover the same property in \Cref{lemma:balayage}. All the mass above $a$ is put into either $a$ with probability $p_a=\be[(b-X_i)\bb{1}[X_i \geq a]]/(b-a)$ and into $b$ with probability $p_b=\be[(a-X_i)\bb{1}[X_i \geq a]]/(b-a)$. The only requirement is that $p_a\geq 0$ which can be guaranteed for $b$ large enough as $X_i$ was assumed unbounded and therefore $\Pr(Y\geq a)>0$. See Lemma 2.7 in \cite{Kertz1982} for more details. From now on we consider that the support of $F$ is in $[0,1]$.\\

We are now be able to show that for well chosen constants $v_0,\dots,v_m$, we obtain a new distribution such that the value of the optimal algorithm remains the same, while the value of the prophet must be bigger due, hence the competitive ratio smaller, due to \Cref{lemma:balayage}. 

The problem of finding the optimal sequence of stopping times $(\tau_j)_{j \in [k]}$ is directly related to the theory of optimal stopping \cite{Chow71}, and it well known that the \ac{bdp} stopping rule is optimal. For $k$ items to select we define the following \ac{bdp} rule:

\begin{definition}
For $i\in [n]$, let $\val{i}{j} $ be the \ac{bdp} optimal expected reward of sequentially selecting $j$ among $i$ items, defined by $\val{0}{j}=0$, $\val{i}{0}=0$ and by the recurrence relation 
\begin{equation*}
\val{i}{j} =\be[(\val{i-1}{j-1}+X) \vee \val{i-1}{j}].
\end{equation*}
The \ac{bdp} stopping rule is to select an item at time $i$ when selecting the $j$-th item if  $X_i+\val{i-1}{j-1} \geq \val{i-1}{j}$.
\end{definition}
This sequence of stopping rules is as mentioned above optimal, and therefore the competitive ratio can be rewritten as the problem of minimizing $(\ell/k) \val {n}{k}(F)/\opt_{\ell}(F)$. 

We will use the following convenient notation for $i \in [n-1]$ and $j \in [k]$:
\begin{equation*}
\gap i j \eqdef \val i{j} -\val i{j-1} .
\end{equation*}

For some finite set $B=\{x_i \in \bb{R}, i \in [m]\}$ of $m$ real values , we denote by $X_B$ the successive balayage from left to right of $X$ over $B$. For instance for $B=\{a,b,c\}$ with $a<b<c$, we have $X_B=(X_{a,b})_{b,c}$. 

\begin{remark}
It is crucial for the balayage to be done on the ordered values: if we consider $(X_{a,c})_{a,b}$, then $p_b=\be [(X_{a,c}-a)\bb{1}_{[X_{a,c} \in [a,b]}]/(b-a) $. Because $X_{a,c}$ is already balayed, and $b<c$, $X_{a,b}$ only takes the value $a$ over the interval $[a,b]$, which implies that  $p_b=0$. Whereas for $X_B$, we can have $p_b \neq 0$. Actually what really matters is for the balayage to be done always with the closest value, but doing it from increasing values gives a proper process to follow.
\end{remark}

\begin{lemma}\label{lemma:succ_balayage}
For $B_{\Delta}=\{0\} \cup \{1\} \cup \{ \gap {i}j, (i,j) \in [n-1] \times [k]\} $, we have $\val nk (X_{B_{\Delta}}) = \val nk (X)$ and $\opt (X_{B_{\Delta}}) \geq \opt (X)$.
\end{lemma}

\begin{proof}
The second part of the proposition is clear from the fact that $\opt$ is increasing when applying balayage (\Cref{lemma:balayage}), and that $X_B$ stems from successive balayage. It remains to show the first part.

We denote the ordered elements of $B_{\Delta}$ with $0=x_1<x_2\dots x_{m-1}<x_m=1$ and consider the sets $B_r=\{ x_s, s \in [r]\}$. We show by induction that $\val nk (X_{B_r})=\val nk (X)$. The initialization for $\{x_1,x_2\}$ can be proved almost identically to the second induction step, see below.
Let us assume that the property is true for $r-1$, with $r \geq 3$.

We have $X_{B_{r}}=(X_{B_{r-1}})_{x_{r-1},x_r}$. Hence, we need to show that for  $Y=X_{B_r}$, we have that $Y_{x_{r-1},x_r}$ preserves $\val nk (Y)$.

We do a second induction to show that for all $i \in [n] $ the following property is true: for all $j \in [k]$, $\val ij (Y)=\val ij (Y_{x_{r-1},x_r})$. The initialization is true as $\val 1j (Y)=\be [Y] =\be [Y_{x_{r-1},x_r}]=\val 1j (Y_{x_{r-1},x_r})$ where the second inequality comes from balayage preserving expectation. Let us assume that the property is true for $i-1$. 
We have by the recurrence relation and the induction hypothesis that 
\begin{align*}
&\val ij(Y_{x_{r-1},x_r})=\be [ (Y_{x_{r-1},x_r}+\val {i-1}{j-1}(Y_{x_{r-1},x_r})) \vee \val {i-1}{j}(Y_{x_{r-1},x_r}) ]\\
&=\be [ (Y_{x_{r-1},x_r}+\val {i-1}{j-1}(Y)) \vee \val {i-1}{j}(Y) ]\\
&=\be \left[ \left((Y_{x_{r-1},x_r}+\val {i-1}{j-1}(Y)) \vee \val {i-1}{j}(Y) \right) \bb{1}_{[Y_{x_{r-1},x_{r}} < \gap {i-1}j]}\right]\\
&+\be \left[ \left((Y_{x_{r-1},x_r}+\val {i-1}{j-1}(Y)) \vee \val {i-1}{j}(Y) \right) \bb{1}_{[Y_{x_{r-1},x_{r}} \geq \gap {i-1}j]}\right]\\
&=\val {i-1}j (Y)\Pr(Y_{x_{r-1},x_r} < \gap {i-1}j ) + \be [ ((Y_{x_{r-1},x_r}+ \val {i-1}j (Y)) \bb{1}_{[Y_{x_{r-1},x_{r}} \geq \gap {i-1}j]}]\\
&\text{or} \quad \val {i-1}j (Y)\Pr(Y_{x_{r-1},x_r} \leq \gap {i-1}j ) + \be [ ((Y_{x_{r-1},x_r}+ \val {i-1}j (Y)) \bb{1}_{[Y_{x_{r-1},x_{r}} > \gap {i-1}j]}]
\end{align*}
If $\gap {i-1}j \not \in [x_{r-1},x_{r}]$, then we directly have the desired equality. Otherwise $\gap {i-1}j \in [x_{r-1},x_{r}]$. In this case because of the construction of $B_{\Delta}$, we have that $\gap {i-1}j$ is either $x_r$ or $x_{r-1}$. 

If $\gap {i-1}j=x_{r-1}$, by the balayage being equal outside $[\gap {i-1}j,x_r]$ we have that $\Pr(X_{x_{r-1},x_r} < \gap {i-1}j)=\Pr(X < \gap {i-1}j)$ and $\Pr(X_{x_{r-1},x_r} \geq \gap {i-1}j)=\Pr(X_{x_{r-1},x_r} \geq \gap {i-1}j)$, and in addition with the expectation being equal over $[\gap {i-1}j,x_r]$ we can deduce that $\be [Y_{x_{r-1},x_r} \bb{1}_{[Y_{x_{r-1},x_r} \geq \gap {i-1}j]}]=\be [Y  \bb{1}_{[Y\geq \gap {i-1}j]}]$. Using the first alternate formula described above we obtain $\val ij(Y_{x_{r-1},x_r}) =\val ij (Y)$. If $\gap {i-1}j=x_r$ we obtain similar properties and use the second alternate formula. In all case we have the equality. We can conclude the second induction, and also conclude the first induction as well. 
\end{proof}

Using the above proposition, we know that we can lower the competitive ratio by applying this specific $B_{\Delta}$ balayage on $X$. All those distributions are supported on the values described by $B_{\Delta}$. Notice that whenever $j \geq i$, then $\gap ij = \gap ii$, so those two values are not distinct. We prove now prove \Cref{prop:reduction2}.

\begin{proposition}
The value of $\comp_{k,\ell}$ is attained by a discrete distribution with a support of $2+k(k-1)/2+k(n-k)$ points on $[0,1]$. 
\end{proposition}
\begin{proof}
This is immediate by applying \Cref{lemma:succ_balayage} and \Cref{lemma:balayage}, and because 
\begin{equation*}
\vert B_{\Delta} \vert =2+ \sum_{i \in [n-1]} \sum_{1 \leq j \leq \max(i,k)} 1=2+k(k-1)/2+k(n-k).  \qedhere
\end{equation*}
\end{proof}
\noindent It is possible that other reductions are more efficient in terms of numbers of values.

Hence we can consider an optimization problem over $2(2+k(k-1)/2+k(n-k))$ parameters instead (to take into account different possible values with different associated distributions).\\

Interestingly, the gaps respect some monotonicity property:
\begin{proposition}
The $\gap ij$ are increasing in $i$ and decreasing in $j$.
\end{proposition}

Let us first show that it is increasing in $j$. We have that $\gap ij -\gap i{j-1}=\val ij -2 \val i{j-1} + \val i{j-2}$. Let us compare $\val ij + \val i{j-2}$ to $2\val i{j-1}$. The first quantity correspond to the supremum of stop rules, where it is allowed to select $2$ times the same item for the $j-2$ first items, and then is allowed to select $2$ more items at different times. The second quantity correspond to stop rules allowed to select $2$ times the same item for the first $j-1$ items encountered. This is strictly more lax in terms of constraints compared to the first quantity, hence we have that $\gap ij$ is decreasing in $j$.

We now show that it is increasing in $i$. 
\begin{align*}
\gap {i-1}j &= \val {i-1}j- \val {i-1}{j-1} = \be[(X+\val i{j-1})\vee \val ij] - \be[(X+\val i{j-2})\vee \val i{j-1}] \\
&= \val ij - \val i{j-1} + \be[ (X+ \val i{j-1} -\val ij)_+ - (X+\val i{j-2}-\val i{j-1})_+]\\
&= \gap ij + \be[ (X+ \val i{j-1} -\val ij)_+ - (X+\val i{j-2}-\val i{j-1})_+].
\end{align*}
Using that if $z>y$, then $z_+>y_+$, and because 
\begin{equation*}
X+ \val i{j-1} -\val ij -X -\val i{j-2}+\val i{j-1}= \gap i{j-1} - \gap ij \geq 0,
\end{equation*}
we can conclude. The inequality comes from the monotonicity of $\gap ij$ in $j$.

\section{Further related works} 
\label{sec:related-work}

While the \ac{iid} version of the prophet inequality has received significant attention, other variants have been studied extensively. If  $1/2$ is the best competitive ratio when the values are not distributed identically and arrive in a fixed sequence, \cite{Esfandiari2017,Ehsani2017} show that when the $X_i$ are presented in a random order, named prophet secretary problem, a competitive ratio of at least $1-1/e$ can be achieved. \cite{Chawla2010, Beyhaghi2021} study the free order prophet where the order of arrival of the $X_i$ can be freely chosen. Recently, \cite{Bubna2022,Giambartolomei2023} have shown that both of these variants are intrinsically different, in that their worst-case competitive ratio are distinct. An important remaining question, is whether the free order variant is as hard as the \ac{iid} case. This is related to our work, as any upper bound on the \ac{iid} case directly translates into an upper bound on the free order prophet. 

In an orthogonal direction, it is possible to examine prophet settings with increasingly complex combinatorial constraints or payoffs. There has been a rich stream of literature on the multi-unit prophet, which assumes that the decision maker and the prophet both actually have a budget of $k \in \bb{N}$ items, which was initiated by \cite{Hajiaghayi2007}. Lower bounds for the competitive ratio explicit in $k$ of order $1-O(1/\sqrt{k})$ were subsequently given by \cite{Alaei2011} for an adaptive algorithm, and \cite{Chawla2020} then proved that $1-O(\log(k)/\sqrt{k})$ can be reached using only a single threshold. More recently \cite{Jiang2021} gave tight constants that are solutions to a limiting ODE. \cite{Jiang2022} also proposes optimization problems that compute the competitive ratio for any $k$ but only for a given $n$. Different types of constraints are also studied such as \cite{Kleinberg2012} which assumes that the allocation must respect matroid constraints, or \cite{Correa2023} who proved competitive ratio guarantees for an online combinatorial auction. \\

The idea of considering weaker benchmarks, as proposed by \cite{Kennedy1985} and our paper, can be readily considered for any of these different combinatorial or distributional assumptions.  The more general framework where the decision maker and the prophet can respectively select  $k$ and $\ell$ items was introduced by \cite{Kennedy1987} in the non \ac{iid} case, but significant results were only proven for $\ell=1$. This is of the same flavor as the $(J,K)$-secretary problem introduced by \cite{Buchbinder2010}, where the goal is to find an element in the top $K$ with only $J$ tries. Selecting one item among the top $k$ has also been studied for the prophet setting in \cite{Esfandiari2019}, where the probability of getting an item from the top $k$ is of order $1-O(\exp(-k))$. A recent work by \cite{Elfarouk2025} proposes a new sharding technique to obtain better prophet inequalities, and in particular achieves a $O(1-k^{k/5})$ lower bound in the \ac{iid} setting for the problem where the decision maker recovers as a value the maximum of $k$ selected items. Another setting where a rate of $1-e^{-k}$ is achievable is in \cite{Alaei2022}, where $k$ different thresholds are fixed in advance, and the decision maker is allowed to make $k$ passes over the data.

Finally, we mention that there has been a recent nice concurrent work by \cite{Brustle2024} on the $k$ multi-unit \ac{iid} prophet who, using a complementary approach through a linear program characterization of $\comp_{k,k}(n)$, achieve similar results. They obtain the same limit system of ODE for $k=\ell$, and additionally provide an estimate of the error between the asymptotic value $\liminf_n \comp_{k,k}(n)$ and $\comp_{k,k}(n)$. They also leverage this result to prove a tight approximation ratio for the stochastic sequential assignment problem.\\

There has also been a lot of focus  \cite{Azar2014, Correa2019b} on sample prophet inequalities, where decision makers do not have access to the distribution themselves, but only samples of the distribution. A remarkable result from \cite{Rubinstein2020} is that a single sample per distribution is enough to achieve the $1/2$ competitive ratio in the original prophet setting. They also show how to use the quantile strategies from \cite{correa2017} to obtain sample prophet inequalities in the \ac{iid} case. This is especially relevant for this work, as the strategies we propose are also quantile algorithms, and therefore the proof from \cite{Rubinstein2020} can likely be extended by using \Cref{alg:algorithm}.

\section*{Acknowledgements}

This work has been partially supported by MIAI @ Grenoble Alpes (ANR-19-P3IA-0003), by the Hi! PARIS Center, by the French National Research Agency (ANR) through grant ANR-20-CE23-0007,  ANR-19-CE23-0015, ANR-23-CE23-0002, ANR-19-CE23-0026 and PEPR IA FOUNDRY project (ANR-23-PEIA-0003). This work has also been funded by the European Union through the ERC OCEAN 101071601 grant. Views and opinions expressed are however those of the author(s) only and do not necessarily reflect those of the European Union or the European Research Council Executive Agency. Neither the European Union nor the granting authority can be held responsible.


\begin{thebibliography}{99}

\bibitem{Alaei2011}
Saeed Alaei.
\newblock Bayesian combinatorial auctions: Expanding single buyer mechanisms to many buyers.
\newblock \emph{IEEE Annual Symposium on Foundations of Computer Science}, 2011.

\bibitem{Alaei2022}
Saeed Alaei, Ali Makhdoumi, Azarakhsh Malekian, and Rad Niazadeh.
\newblock Descending price auctions with bounded number of price levels and batched prophet inequality.
\newblock In \emph{ACM Conference on Economics and Computation}, 2022.

\bibitem{Arnosti2021}
Nick Arnosti and Will Ma.
\newblock Tight guarantees for static threshold policies in the prophet secretary problem.
\newblock \emph{Conference on Economics and Computation}, 2021.

\bibitem{Azar2014}
Pablo~D. Azar, Robert Kleinberg, and S.~Matthew Weinberg.
\newblock Prophet inequalities with limited information.
\newblock In \emph{ACM-SIAM Symposium on Discrete Algorithms}, 2014.

\bibitem{Borodin1998}
Allan Borodin and Ran El-Yaniv.
\newblock \emph{Online computation and competitive analysis}.
\newblock Cambridge University Press, 1998.

\bibitem{Brustle2024}
Johannes Brustle, Sebastian Perez-Salazar, and Victor Verdugo.
\newblock Splitting guarantees for prophet inequalities via nonlinear systems
\newblock In \emph{Workshop on Internet and Network Economics}, 2024.

\bibitem{Bubeck2018}
S{\'e}bastien Bubeck, Michael~B. Cohen, James~R. Lee, and Yin~Tat Lee.
\newblock Metrical task systems on trees via mirror descent and unfair gluing.
\newblock In \emph{ACM-SIAM Symposium on Discrete Algorithms}, 2018.

\bibitem{Bubna2022}
Archit Bubna and Ashish Chiplunkar.
\newblock Prophet inequality: Order selection beats random order.
\newblock In \emph{ACM Conference on Economics and Computation}, 2022.

\bibitem{Buchbinder2010}
Niv Buchbinder, Kamal~Kumar Jain, and Mohit Singh.
\newblock Secretary problems via linear programming.
\newblock In \emph{Mathematics of Operations Research}, 2010.

\bibitem{Chawla2010}
Shuchi Chawla, Jason~D. Hartline, David~L. Malec, and Balasubramanian Sivan.
\newblock Multi-parameter mechanism design and sequential posted pricing.
\newblock In \emph{Behavioral and Quantitative Game Theory}, 2010.

\bibitem{Chawla2020}
Shuchi Chawla, Nikhil~R. Devanur, and Thodoris Lykouris.
\newblock Static pricing for multi-unit prophet inequalities.
\newblock In \emph{Workshop on Internet and Network Economics}, 2020.

\bibitem{Chow71}
Y.S. Chow, Herbert Robbins, Siegmund~David Herbert, and Siegmund David.
\newblock {Great Expectations: The Theory of Optimal Stopping}.
\newblock \emph{Royal Statistical Society. Journal. Series A: General}, 1971.

\bibitem{correa2017}
Jos{\'e} Correa, Patricio Foncea, Ruben Hoeksma, Tim Oosterwijk, and Tjark Vredeveld.
\newblock Posted price mechanisms for a random stream of customers.
\newblock In \emph{ACM Conference on Economics and Computation}, 2017.

\bibitem{Correa2019}
Jos{\'e} Correa, Patricio Foncea, Ruben Hoeksma, Tim Oosterwijk, and Tjark Vredeveld.
\newblock Recent developments in prophet inequalities.
\newblock \emph{SIGecom Exch.}, 2019.

\bibitem{Correa2019b}
Jos{\'e} Correa, Paul D{\"u}tting, Felix~A. Fischer, and Kevin Schewior.
\newblock Prophet inequalities for i.i.d. random variables from an unknown distribution.
\newblock In \emph{ACM Conference on Economics and Computation}, 2019.

\bibitem{Correa2023}
Jos{\'e} Correa and Andr{\'e}s Cristi.
\newblock A constant factor prophet inequality for online combinatorial auctions.
\newblock \emph{ACM Symposium on Theory of Computing}, 2023.

\bibitem{Elfarouk2025}
Elfarouk Harb
\newblock New Prophet Inequalities via Poissonization and Sharding
\newblock In \emph{ACM-SIAM Symposium on Discrete Algorithms}, 2025.


\bibitem{Ehsani2017}
Soheil Ehsani, Mohammad~Taghi Hajiaghayi, Thomas Kesselheim, and Sahil Singla.
\newblock Prophet secretary for combinatorial auctions and matroids.
\newblock In \emph{ACM-SIAM Symposium on Discrete Algorithms}, 2017.

\bibitem{Esfandiari2017}
Hossein Esfandiari, MohammadTaghi Hajiaghayi, Vahid Liaghat, and Morteza Monemizadeh.
\newblock Prophet secretary.
\newblock \emph{SIAM Journal on Discrete Mathematics}, 2017.

\bibitem{Esfandiari2019}
Hossein Esfandiari, Mohammad~Taghi Hajiaghayi, Brendan Lucier, and Michael Mitzenmacher.
\newblock Prophets, secretaries, and maximizing the probability of choosing the best.
\newblock In \emph{International Conference on Artificial Intelligence and Statistics}, 2019.

\bibitem{Ferguson1989}
Thomas~S. Ferguson.
\newblock Who solved the secretary problem.
\newblock \emph{Statistical Science}, 1989.

\bibitem{Giambartolomei2023}
Giordano Giambartolomei, Frederik Mallmann-Trenn, and Raimundo Saona.
\newblock Prophet inequalities: Separating random order from order selection.
\newblock \emph{ArXiv}, 2023.

\bibitem{Hajiaghayi2007}
Mohammad~Taghi Hajiaghayi, Robert~D. Kleinberg, and Tuomas Sandholm.
\newblock Automated online mechanism design and prophet inequalities.
\newblock In \emph{AAAI Conference on Artificial Intelligence}, 2007.

\bibitem{Hill83}
T.~P. Hill.
\newblock Prophet inequalities and order selection in optimal stopping problems.
\newblock \emph{Proceedings of the American Mathematical Society}, 1983.

\bibitem{Kertz1982}
T.~P. Hill and Robert~P. Kertz.
\newblock Comparisons of stop rule and supremum expectations of i.i.d. random variables.
\newblock \emph{The Annals of Probability}, 1982.

\bibitem{Hill83b}
T.~P. Hill and Robert~P. Kertz.
\newblock Stop rule inequalities for uniformly bounded sequences of random variables.
\newblock \emph{Transactions of the American Mathematical Society}, 1983.

\bibitem{Jiang2021}
Jiashuo Jiang, Will Ma, and Jiawei Zhang.
\newblock Tight guarantees for multi-unit prophet inequalities and online stochastic knapsack.
\newblock In \emph{ACM-SIAM Symposium on Discrete Algorithms}, 2021.

\bibitem{Jiang2022}
Jiashuo Jiang, Will Ma, and Jiawei Zhang.
\newblock Tightness without counterexamples: A new approach and new results for prophet inequalities.
\newblock In \emph{ACM Conference on Economics and Computation}, 2022.

\bibitem{Kennedy1985}
Douglas~P. Kennedy.
\newblock Optimal stopping of independent random variables and maximizing prophets.
\newblock \emph{Annals of Probability}, 1985.

\bibitem{Kennedy1987}
Douglas~P. Kennedy.
\newblock Prophet-type inequalities for multi-choice optimal stopping.
\newblock \emph{Stochastic Processes and their Applications}, 1987.

\bibitem{Kertz1986}
Robert~P. Kertz.
\newblock Stop rule and supremum expectations of i.i.d. random variables: a complete comparison by conjugate duality.
\newblock \emph{J. Multivar. Anal.}, 1986.

\bibitem{Kertz1986b}
Robert~P. Kertz.
\newblock Comparison of optimal value and constrained maxima expectations for independent random variables.
\newblock \emph{Advances in Applied Probability}, 1986.

\bibitem{Kleinberg2012}
Robert Kleinberg and Seth~Matthew Weinberg.
\newblock Matroid prophet inequalities.
\newblock In \emph{ACM Symposium on Theory of Computing}, 2012.

\bibitem{Krengel1977}
Ulrich Krengel and Louis Sucheston.
\newblock Semiamarts and finite values.
\newblock \emph{Bulletin of the American Mathematical Society}, 1977.

\bibitem{Allen2021}
Allen Liu, Renato~Paes Leme, Martin P\'{a}l, Jon Schneider, and Balasubramanian Sivan.
\newblock Variable decomposition for prophet inequalities and optimal ordering.
\newblock In \emph{ACM Conference on Economics and Computation}, 2021.

\bibitem{Mehta2013}
Aranyak Mehta.
\newblock Online matching and ad allocation.
\newblock \emph{Found. Trends Theor. Comput. Sci.}, 2013.

\bibitem{Motwani1994}
Rajeev Motwani, Steven Phillips, and Eric Torng.
\newblock Nonclairvoyant scheduling.
\newblock \emph{Theoretical Computer Science}, 1994.

\bibitem{PerezSalazar2022}
Sebastian Perez-Salazar, Mohit Singh, and Alejandro Toriello.
\newblock The iid prophet inequality with limited flexibility.
\newblock \emph{ArXiv}, 2022.

\bibitem{Rubinstein2020}
Aviad Rubinstein, Jack~Z. Wang, and S.~Matthew Weinberg.
\newblock Optimal single-choice prophet inequalities from samples.
\newblock In \emph{Innovations in Theoretical Computer Science Conference}, 2020.

\bibitem{SamuelCahn1984}
Ester Samuel-Cahn.
\newblock Comparison of threshold stop rules and maximum for independent nonnegative random variables.
\newblock \emph{Annals of Probability}, 1984.

\bibitem{Beyhaghi2021}
Balasubramanian Sivan, Hedyeh Beyhaghi, Martin P{\'a}l, Negin Golrezaei, and Renato~Paes Leme.
\newblock Improved approximations for posted price and second-price mechanisms.
\newblock \emph{Operations Research}, 2021.


\end{thebibliography}
\end{document}